%% file: isi_farr_ringdowns.tex
\pdfoutput=1
\documentclass[aps,prd,twocolumn,superscriptaddress,preprintnumbers,floatfix,nofootinbib]{revtex4-2}

\usepackage{graphicx}
\usepackage{amsmath}
\usepackage[caption=false]{subfig}
\usepackage{placeins}
\usepackage{color}
\usepackage{standalone}
\usepackage{dcolumn}
\usepackage{microtype}
\usepackage{etoolbox}
\usepackage{amssymb}
\usepackage{mathrsfs}
\usepackage[dvipsnames]{xcolor}
\usepackage[colorlinks,urlcolor=NavyBlue,citecolor=NavyBlue,linkcolor=NavyBlue,pdfusetitle]{hyperref}
\usepackage[all]{hypcap}
\usepackage[inline]{enumitem}
\usepackage[utf8]{inputenc}
\usepackage{csquotes}
\usepackage{array}
\usepackage{booktabs}
\usepackage{xspace}

\newcommand{\mnras}{Mon.~Not.~Royal Ast.~Soc.}

\DeclareMathOperator{\atantwo}{atan2}
\newcommand{\ts}{\textsuperscript}

\newcommand{\beq}{\begin{equation}}
\newcommand{\eeq}{\end{equation}}

\newtoggle{commentsoff}
\togglefalse{commentsoff}

\ifdefined\nocomments
    \toggletrue{commentsoff}
\fi

\iftoggle{commentsoff}{
  \newcommand*{\mi}[1]{}
  \newcommand*{\mg}[1]{}
  \newcommand*{\wf}[1]{}
  \newcommand*{\comment}[1]{}
  
  \newcommand*{\todo}[1]{}
  \newcommand*{\warn}[1]{}

  \newcommand*{\commentmark}[1]{}
}{
  \newcommand*{\mi}[1]{{\color{magenta} [{\bf MAX}: #1]}}
  \newcommand*{\wf}[1]{\textcolor{green}{[\textbf{WILL}: #1]}}
  
  \newcommand*{\comment}[1]{{\color{blue} [{\bf NOTE}: #1]}}
  \newcommand*{\warn}[1]{{\color{red} [{\bf WARNING}: #1]}}
  \newcommand*{\todo}[1]{{\color{red} [{\bf TODO}: #1]}}

}

\graphicspath{{./fig/}}

\newcommand{\dcc}{LIGO-P2100227}

\newcommand{\infd}{\mathrm{d}}
\newcommand{\white}{\bar}

\newcommand{\snropt}{\mathrm{SNR}}
\newcommand{\snrmf}{\mathrm{SNR}_{\rm mf}}
\newcommand{\cov}{C}
\newcommand{\acf}{\rho}
\newcommand{\nmode}{D}

\newcommand{\nrsim}{\textsc{SXS:BBH:0305}\xspace}
\newcommand{\unaryminus}{\scalebox{0.55}{\( - \)}}

\begin{document}

\title{Analyzing black-hole ringdowns}

\author{Maximiliano Isi}
\email[]{maxisi@mit.edu}
\thanks{NHFP Einstein fellow}
\affiliation{
LIGO Laboratory, Massachusetts Institute of Technology, Cambridge, Massachusetts 02139, USA
}%
\affiliation{Center for Computational Astrophysics, Flatiron Institute, 162 5th Ave, New York, NY 10010}

\author{Will M. Farr}
\email{will.farr@stonybrook.edu}
\affiliation{Center for Computational Astrophysics, Flatiron Institute, 162 5th Ave, New York, NY 10010}
\affiliation{Department of Physics and Astronomy, Stony Brook University, Stony Brook NY 11794, USA}

\hypersetup{pdfauthor={Isi, Farr}}

\date{\today}

\begin{abstract}
A perturbed black hole rings down by emitting gravitational waves in tones with specific frequencies and durations.
Such tones encode prized information about the geometry of the source spacetime and the fundamental nature of gravity, making the measurement of black hole ringdowns a key goal of gravitational wave astronomy.
However, this task is plagued by technical challenges that invalidate the naive application of standard data analysis methods and complicate sensitivity projections.
In this paper, we provide a comprehensive account of the formalism required to properly carry out ringdown analyses, examining in detail the foundations of recent observational results, and providing a framework for future measurements.
We build on those insights to clarify the concepts of ringdown detectability and resolvability---touching on the drawbacks of both Bayes factors and naive Fisher matrix approaches---and find that overly pessimistic heuristics have led previous works to underestimate the role of ringdown overtones for black hole spectroscopy.
We put our framework to work on the analysis of a variety of simulated signals in colored noise, including analytic injections and a numerical relativity simulation consistent with GW150914.
We demonstrate that we can use tones of the quadrupolar angular harmonic to test the no-hair theorem at current sensitivity, with precision comparable to published constraints from real data.
Finally, we assess the role of modeling systematics, and project measurements for future, louder signals.  We release \textsc{ringdown}, a \textsc{Python} library for analyzing black hole ringdowns using the the methods discussed in this paper, under a permissive open-source license.
\end{abstract}

\maketitle

\section{Introduction}
\label{sec:intro}

Black-hole (BH) ringdowns provide an exceptional observational handle on the nature of gravity: a perturbed BH radiates gravitational waves (GWs) in the form of damped sinusoids that cleanly encode information about the structure of the BH spacetime \cite{Vishveshwara:1970cc, Press:1971wr, Teukolsky:1973ha, Chandrasekhar:1975zza}, and make it accessible to LIGO \cite{TheLIGOScientific:2014jea}, Virgo \cite{TheVirgo:2014hva} and future GW detectors \cite{Evans:2016mbw,Sathyaprakash:2012jk,Audley:2017drz}.
The research program known as ``BH spectroscopy'' \cite{Detweiler:1980gk,Dreyer:2003bv,Berti:2005ys} aims to analyze such quasinormal modes (QNMs) in order to test general relativity (GR) and the BH paradigm \cite{Will:2014kxa,Cardoso:2019rvt}, as embodied by the Kerr metric for astrophysical BHs \cite{Kerr:1963ud,Teukolsky:2014vca}.
This includes basic tenets such as the no-hair theorem and the area law---respectively, the statements that astrophysical BHs are fully characterized by their mass and spin \cite{Doroshkevich:1966,Israel:1967wq,Carter:1971zc,Hawking:1971vc,Robinson:1975bv,Chrusciel:2012jk}, and that the total area of classical BH horizons may not decrease over time \cite{Hawking:1971tu,Wald:1999vt,Chrusciel:2000cu}.

Although the literature on BH spectroscopy is vast (e.g., \cite{Dreyer:2003bv,Berti:2005ys,Kamaretsos:2011um,Gossan:2011ha,Meidam:2014jpa,Nagar:2016iwa,Berti:2016lat,Cabero:2017avf,Thrane:2017lqn,Baibhav:2017jhs,Baibhav:2018rfk,Brito:2018rfr,Carullo:2018sfu,Carullo:2019flw,Giesler:2019uxc,Isi:2019aib,Bhagwat:2019dtm,Ota:2019bzl,Forteza:2020hbw,CalderonBustillo:2020tjf,Abbott:2020jks,Isi:2020tac,Ghosh:2021mrv,Capano2021}), sensitivity projections have often relied on simplistic data analysis treatments that reduce the scope of their conclusions.
For instance, analyses based on Fisher matrices tend to assume that a test of the Kerr hypothesis would require \emph{independently} analyzing two or more QNMs in order to obtain separate estimates of the BH parameters to be compared for consistency, e.g., visually on a plot \cite{Dreyer:2003bv,Berti:2005ys,Kamaretsos:2011um,Berti:2016lat,Bhagwat:2019dtm,Forteza:2020hbw}.
Although useful for projections in some regimes, such a setup does not reflect how an optimal measurement would be carried out:
in order to verify that a given ringdown signal is consistent with a Kerr spectrum, we should self-consistently fit for multiple modes \emph{simultaneously} under a Bayesian framework \cite{Gossan:2011ha,Meidam:2014jpa,Nagar:2016iwa,Brito:2018rfr,Carullo:2018sfu,Carullo:2019flw,Giesler:2019uxc,Isi:2019aib}.
The isolated-mode treatment, together with overly pessimistic heuristics, has led to misconceptions---for example, that BH overtones (fast-decaying modes with similar frequencies) should only be spectroscopically useful for uncommonly loud detections \cite{Bhagwat:2019dtm,Forteza:2020hbw}, in contradiction with observational results obtained with modest signal amplitudes \cite{Isi:2019aib,Abbott:2020jks}.

The prevalence of oversimplifications in ringdown studies is hardly surprising: formulating a robust analysis to extract QNMs from GW data is an unexpectedly challenging task.
Currently, our best chance for gaining access to a BH ringdown is to target the late stage of GWs from binary BH coalescences, which consists of emission by the perturbed remnant BHs born in the merger.
Operationally, the ringdown portion of such a signal is defined to begin at some time $t_0$, after which the strain can be accurately described as a superposition of damped sinusoids with complex frequencies corresponding to the QNMs of the final BH \cite{Baibhav:2018rfk}.%
\footnote{In other words, we \emph{define} ``ringdown'' as the portion of the signal that can be accurately described as a superposition of damped sinusoids (we do not consider polynomial tails \cite{Leaver:1986gd}). This is a phenomenological statement that establishes our data analysis target, not necessarily a statement about BH physics.}
Thus, studies aiming to isolate the ringdown should be able to probe data at times $t \geq t_0$ irrespective of what preceded it.
The presence of this time-domain distinction introduces subtle complications to ringdown studies in the presence of ``colored'' (i.e.\ time-correlated) noise that invalidate the naive application of standard data analysis strategies \cite{Nagar:2016iwa,Cabero:2017avf,Carullo:2019flw,Isi:2019aib,Capano2021}.

There are two broad avenues for addressing this challenge: (i) enhance regular inspiral-merger-ringdown (IMR) waveform models with additional freedom in the ringdown stage \cite{Brito:2018rfr,Meidam:2017dgf,Ghosh:2021mrv}; or (ii) define a model only for the late data, and discard the rest \cite{Nagar:2016iwa,Carullo:2018sfu,Carullo:2019flw,Isi:2019aib,Capano2021}.
The former is simpler to accommodate within the standard infrastructure for LIGO-Virgo analyses, especially when starting from a waveform whose ringdown model is explicitly based on perturbation theory \cite{Brito:2018rfr,Ghosh:2021mrv}.
However, it has the disadvantage of necessarily coupling the ringdown measurement to the inspiral-merger regime, thus becoming susceptible to systematics induced by the attachment procedure, or shortcomings of the pre-ringdown model.

On the other hand, having an independent model for the post-merger data is conceptually cleaner, and allows for a ringdown analysis fully agnostic about the inspiral dynamics.
However, this approach demands truncating the GW signal at a specific time, which is difficult to handle with the usual LIGO-Virgo analysis techniques \cite{Veitch2015}.
Instead, it calls for special treatment in the time domain \cite{Isi:2019aib,Carullo:2019flw}, or an equivalent nontrivial procedure in the frequency domain \cite{Capano2021}.
There is also potential uncertainty about the optimal truncation point, although this choice can be reasonably informed by numerical studies \cite{Giesler:2019uxc}.\footnote{In fact, this issue is also implicitly present in the enhanced-waveform strategy mentioned above, in the choice of ringdown start time within the IMR waveform model itself \cite{Brito:2018rfr,Pan:2011gk}.}

The main goal of this paper is to lay down, and then empirically validate, the formalism for properly carrying out this second type of analysis, examining in detail the foundations of recent observational results \cite{Isi:2019aib,Abbott:2020tfl,Abbott:2020mjq,Abbott:2020jks,Isi:2020tac}.
This involves a thorough description of how to construct a suitable ringdown-only template for BH spectroscopy, as well as of the Bayesian infrastructure and data manipulations required to use it without bias.
The end result is a comprehensive, self-contained account of ringdown analyses that expands, clarifies and, when necessary, corrects previous literature, under a uniform notation and set of conventions.
Consequently, this paper synthesizes the conceptual and methodological backbone of past observations in \cite{Isi:2019aib} and others like it, elucidating the role of different assumptions and data analysis choices made in the literature.
More importantly, it provides a flexible and robust platform for future, richer measurements.

The second goal is to build on those insights to revisit the concepts of ringdown detectability and resolvability.
To that end, we formulate a procedure for operationally establishing the number of ringdown modes required by certain data, in a way that is not strongly sensitive to arbitrary choices of prior bounds (in contrast with, e.g., approaches based exclusively on Bayes factors).
With this in hand, we examine what it means to ``resolve'' two ringdown modes, and the conditions required to spectrally characterized them.
We show that Rayleigh-like spectral separation criteria can be easily misapplied to obtain overly pessimistic projections, as we argue has been the case in some previous studies of BH spectroscopy.

The third and final goal of the paper is to showcase our overall conceptual framework, as well as our specific implementation, by applying it to the question of BH spectroscopy with overtones.
Studying GW150914-like ringdowns in colored noise, we show how our method is successful at reconstructing the spectral and polarization properties of the signal.
We then evaluate the detectability of the first overtone, and demonstrate its potential to improve measurements of BH parameters and, more crucially, to constrain deviations away from the Kerr spectrum.
Through a numerical relativity injection, we confirm once more that ringdowns are detectable and resolvable with current detectors, and that overtones offer a valuable resource for tests of the no-hair theorem.

This paper is intended as a comprehensive reference that can be read in its entirety or by selecting specific sections.
We begin in Sec.~\ref{sec:model} by showing that the most generic ringdown template suitable for data analysis consists of a superposition of elliptically polarized damped sinusoids, and that nuances like spherical-spheroidal mixing are overall inconsequential; we also examine key properties of the Kerr spectrum, with a focus on tones of the (usually dominant) quadrupolar harmonic.
In Sec.~\ref{sec:inference}, we discuss the technical challenges intrinsic to ringdown-only analyses and show how to circumvent them by operating fully in the time domain; we show how to compute the matched-filter signal-to-noise ratio (SNR) of a signal in noisy data.
In Sec.~\ref{sec:ds}, we demonstrate the efficacy of the method and show how to use it to detect and characterize damped sinusoids; importantly, we show modes need not be separated in frequency to be resolvable.
In Sec.~\ref{sec:analysis}, we turn to more realistic simulations of BH ringdowns, including several tones of the quadrupolar angular harmonic of a GW150914-like system \cite{Abbott:2016blz}; we demonstrate the recovery of analytic Kerr and non-Kerr injections at different SNRs, and replicate some of those results with a numerical-relativity injection.
In Sec.~\ref{sec:conclusion}, we offer an itemized summary of the main conclusions in the paper, and outline future work.

As an important companion to this paper, we simultaneously release a software package, which we name \textsc{ringdown} \cite{ringdown}.
This software implements our analysis framework within a flexible platform that users can use to easily develop their own ringdown analyses.
The code is written in the \textsc{Python} language, with a Bayesian sampling component based in \textsc{Stan} \cite{Stan,Carpenter:2017}.

\section{Ringdown model}
\label{sec:model}

\subsection{Generic template}
\label{sec:template}

There exist two qualitatively distinct families of Kerr QNMs: prograde modes, corotating with the hole, and retrograde modes, counterotating with the hole \cite{Leaver:1985ax,Berti:2009kk}.%
\footnote{It is standard for theoretical QNM studies to disregard modes with negative frequency, in which case negative (positive) $m$ becomes synonymous with ``retrograde'' (``prograde'') modes; we do not adopt that terminology.}
Taking both types into account and setting $t_0=0$, the most generic template based on Kerr perturbation theory for the complex-valued ringdown strain, $h\equiv h_+ - i h_\times$, is
\begin{equation} \label{eq:h}
h
= \sum C_{[p]\ell m n}\, e^{-i \tilde{\omega}_{[p]\ell m n} t}\, {}_{\unaryminus 2} S_{[p]\ell m n}(\iota, \varphi)\, ,
\end{equation}
summing over $p=+1$ for prograde modes and $p=-1$ for retrograde modes, azimuthal number $\ell \geq 2$, magnetic number $-\ell \leq m \leq \ell$, and overtone number $0 \leq n$.
Each QNM carries a complex frequency $\tilde{\omega}_{[p]\ell m n} \equiv \omega_{[p]\ell m n} - i \gamma_{[p]\ell m n}$, with damping time $\tau_{[p]\ell m n} \equiv 1 / \gamma_{[p]\ell m n}$.
The overtone number $n$ is defined to order modes of a given $p$, $\ell$ and $m$ by decreasing damping time, such that $\tau_{[p]\ell m n} > \tau_{[p]\ell m (n+1)}$.
The mode amplitudes and phases are encoded in the complex amplitudes $C_{[p]\ell m n}$, which are set by the initial state of the perturbation and are not known a priori \cite{Berti:2006wq,Kamaretsos:2011um}.

The angular structure of the strain is given by the spin-weighted spheroidal harmonics ${}_{\unaryminus 2} S_{[p]\ell m n}$, as a function of polar and azimuthal angles $\iota$ and $\phi$ relative to the BH spin direction,
\begin{equation} \label{eq:swsh}
{}_{\unaryminus 2} S_{[p]\ell m n}(\iota, \varphi) \equiv e^{im\varphi} {}_{\unaryminus 2} S_{\ell m}(\chi\, t_M \tilde{\omega}_{[p]\ell m n}, \cos\iota)\, ,
\end{equation}
with a nontrivial dependence on the product of the dimensionless spin magnitude, $\chi$, and the dimensionless complex frequency, $t_M \tilde{\omega}_{[p]\ell m n} \equiv (G M/c^3)\, \tilde{\omega}_{[p]\ell m n}$, for BH mass $M$
\cite{Teukolsky:1973ha,Press:1973zz,Leaver:1985ax,Berti:2005gp,Cook:2014cta}.
These functions are not orthogonal over the sphere, causing the modes to mix: the projection of the strain of mode $\{[p]\ell m n\}$ onto the corresponding spherical harmonic ${}_{\unaryminus 2} Y_{\ell m}(\iota,\varphi)$ will, in principle, pick up contributions from an infinite set of modes $\{[p']\ell' m n'\}$ sharing the same magnetic number $m$ \cite{Teukolsky:1973ha,Berti:2014fga}.
This can be important in the study of numerical relativity waveforms (e.g., \cite{Buonanno:2006ui,Giesler:2019uxc}), wherein the strain is extracted from the simulation through a projection into the ${}_{\unaryminus 2} Y_{\ell m}$'s (which are themselves complete and orthogonal over the sphere, unlike the ${}_{\unaryminus 2} S_{\ell m}$'s) \cite{Boyle:2019kee}.
One can address this by expanding the ${}_{\unaryminus 2} S_{\ell m}$ as a series of ${}_{\unaryminus 2} Y_{\ell m}$'s,  most often finding that ${}_{\unaryminus 2} S_{\ell m} \approx {}_{\unaryminus 2} Y_{\ell m}$ is a sufficiently good approximation \cite{Berti:2005gp}.
Either way, as will become apparent below, none of this is relevant for our purposes because we will not be interested in measuring the $C_{[p]\ell m n}$'s.

By definition, the prograde versus retrograde distinction implies $\mathrm{sgn}(m) = p ~ \mathrm{sgn}(\omega_{[p]\ell m n})$ for $m\neq 0$.%
\footnote{For $m=0$, there are still two families of modes indexed by $p=\pm1$, but the prograde vs retrograde terminology loses meaning.}
Although both prograde ($p=+1$) and retrograde ($p=-1$) modes are equally fundamental in principle, studies have suggested the latter to be suppressed for regular binaries in which the BHs rotate in the same sense as the orbit \cite{Berti:2005ys,Berti:2006wq,Buonanno:2006ui,London:2014cma,Lim:2019xrb} (however, see \cite{Dhani:2020nik,Finch:2021iip}).
Based on this, it is standard to focus on prograde modes and set $C_{[-1]\ell m n}=0$.
We will do so here and drop the $p$ index below, even though our formalism can trivially accommodate retrograde modes.

We can further narrow the scope of Eq.~\eqref{eq:h} by appealing to symmetry.
Parity-time (PT) symmetry of the baseline (unperturbed) metric implies $\omega_{\ell m n } = - \omega_{\ell \unaryminus m n}$;
indeed, in the Kerr case, $\tilde{\omega}^{\rm (GR)}_{\ell m n } = - \tilde{\omega}^{\rm (GR)*}_{\ell \unaryminus m n}$, where ${}^*$ indicates complex conjugation \cite{Teukolsky:1973ha}.
Assuming this holds and implicitly setting $p=1$ everywhere, we may rewrite the sum in Eq.~\eqref{eq:h} as\footnote{This double counts $m=0$ modes, which can be solved by redefining $C_{\ell 0 n} \to C_{\ell 0 n}/2$.}
\begin{align} \label{eq:h_pt}
h = \sum_{\ell}\sum_{0 \leq m \leq \ell} \sum_{n} &\left[ C_{\ell m n} e^{-i \tilde{\omega}_{\ell m n} t} {}_{\unaryminus 2} S_{\ell m n}(\iota, \varphi) \, +  \right. \nonumber \\
&\phantom{[}\left. C_{\ell \unaryminus m n} e^{i \tilde{\omega}_{\ell m n}^* t} {}_{\unaryminus 2} S_{\ell \unaryminus m n}(\iota, \varphi) \right] ,
\end{align}
where ${}_{\unaryminus 2} S_{\ell \unaryminus m n}(\iota, \varphi) =  (-1)^\ell {}_{\unaryminus 2} S^*_{\ell m n}(\pi-\iota, \varphi)$ by symmetry \cite{Press:1973zz,Cook:2014cta}.%
\footnote{This property is sometimes presented without writing out the angular dependence (e.g., \cite{Berti:2005ys,Buonanno:2006ui}), which could be incorrectly taken to mean that ${}_{\unaryminus 2} S_{\ell \unaryminus m n}$ and ${}_{\unaryminus 2} S^*_{\ell m n}$ are directly interchangeable (App.~\ref{app:swsh_id}).}
Following this reasoning, we will always consider the two $\pm m$ terms in the summand of Eq.~\eqref{eq:h_pt} as a set, denoting them jointly by $h_{\ell |m| n}$.
(See also App.~D of \cite{Berti:2007fi}.)

In the absence of a physical model for the $C_{\ell m n}$'s, we can simply absorb the ${}_{\unaryminus 2} S_{\ell m n}$ factors into redefined complex amplitudes $C_{\ell m n}' \equiv {}_{\unaryminus 2} S_{\ell m n}(\iota, \varphi) C_{\ell m n}$, so that the summand in Eq.~\eqref{eq:h_pt} becomes
\begin{align} \label{eq:h_pt_lmn}
h_{\ell |m| n} &=  C_{\ell m n}' e^{-i \tilde{\omega}_{\ell m n} t} +
C_{\ell \unaryminus m n}' e^{i \tilde{\omega}^*_{\ell m n} t} \nonumber \\
&=  \left[C_{\ell m n}' e^{-i \omega_{\ell m n} t } +
C_{\ell \unaryminus m n}' e^{i \omega_{\ell m n} t} \right] e^{-t/\tau_{\ell m n}} .
\end{align}
This expression makes it clear that assuming $\tilde{\omega}_{\ell m n } = - \tilde{\omega}^{*}_{\ell \unaryminus m n}$ was equivalent to requiring that each pair of prograde $\{\ell, \pm m, n\}$ modes add up to a single elliptically polarized contribution, to which we refer as ``the $\{\ell, |m|, n\}$ mode.''

To make the elliptical character of $h_{\ell|m|n}$ explicit, factor the new complex amplitudes $C'$ as
\begin{equation} \label{eq:ellip}
C_{\ell \pm |m| n}' = \frac{1}{2} \left(1 \pm \epsilon_{\ell |m| n} \right) A_{\ell |m| n} \, e^{i\phi_{\ell \pm|m| n}} ,
\end{equation}
for some arbitrary real amplitude $A_{\ell |m| n}$, an ellipticity $-1 \leq \epsilon_{\ell |m| n} \leq 1$, and two independent phases $\phi_{\ell \pm|m| n}$.
The inverse transformation is simply
\begin{equation} \label{eq:amplitude}
A_{\ell |m| n} = |C_{\ell +|m| n}'| + |C_{\ell -|m| n}'|\, ,
\end{equation}
\begin{equation} \label{eq:ellipticity}
\epsilon_{\ell |m| n} = \frac{|C_{\ell +|m| n}'| - |C_{\ell -|m| n}'|}{|C_{\ell +|m| n}'| + |C_{\ell -|m| n}'|} ,
\end{equation}
and $\phi_{\ell \pm|m| n} = \arg ({C_{\ell \pm |m| n}'} )$, so there is no loss of generality.
To simplify the indices, define $j \equiv \{\ell, |m|, n\}$ and $\phi_{\pm j} \equiv \phi_{\ell \pm|m| n}$, so that Eq.~\eqref{eq:h_pt_lmn} becomes
\begin{align} \label{eq:h_pt_lmn_LR}
h_{j} =  \frac{1}{2} A_{j}\,  e^{-t/\tau_{j}} &\left[\left(1 + \epsilon_{j}\right)e^{-i (\omega_{j} t - \phi_{+j})}~ + \right. \nonumber \\
&\phantom{[}\left. \left(1 - \epsilon_{j}\right) e^{i (\omega_j t + \phi_{-j})} \right] .
\end{align}
It is now clear that this is an elliptical GW: $\epsilon_j = 1$ yields a right-hand circularly polarized mode; $\epsilon_j = -1$ yields a left-hand circularly polarized mode; and intermediate values yield generic, elliptically polarized waves, including linear polarizations for $\epsilon_j = 0$.

When analyzing GW detector data, it is convenient to work in the linear polarization basis.
The plus and cross polarizations corresponding to Eq.~\eqref{eq:h_pt_lmn_LR}, $h^{(+)}_{j} = \Re \left[h_j\right]$ and $h^{(\times)}_{j} = - \Im \left[h_j\right]$, can be written
\begin{equation} \label{eq:h_ellip_lmn_plus}
h^{(+)}_{j} = h^c_{j}\, \cos \theta_{j} - \epsilon_{j} h^s_{j}\, \sin\theta_{j}\, ,
\end{equation}
\begin{equation} \label{eq:h_ellip_lmn_cross}
h^{(\times)}_j = h^c_{j}\, \sin \theta_j + \epsilon_j h^s_{j}\, \cos\theta_j\, ,
\end{equation}
for cosine and sine quadratures
\begin{equation}  \label{eq:h_ellip_lmn_sin}
h^c_j \equiv A_j\, e^{-t/\tau_j} \cos(\omega_j t - \phi_j) \, ,
\end{equation}
\begin{equation} \label{eq:h_ellip_lmn_cos}
h^s_j \equiv A_j\, e^{-t/\tau_j} \sin(\omega_j t - \phi_j) \, .
\end{equation}
Here, we have defined two new angles: $\theta_j \equiv - (\phi_{+j} + \phi_{-j})/2$ and $\phi_j \equiv (\phi_{+j} - \phi_{-j})/2$.

\begin{figure}
\includegraphics[width=0.6\columnwidth]{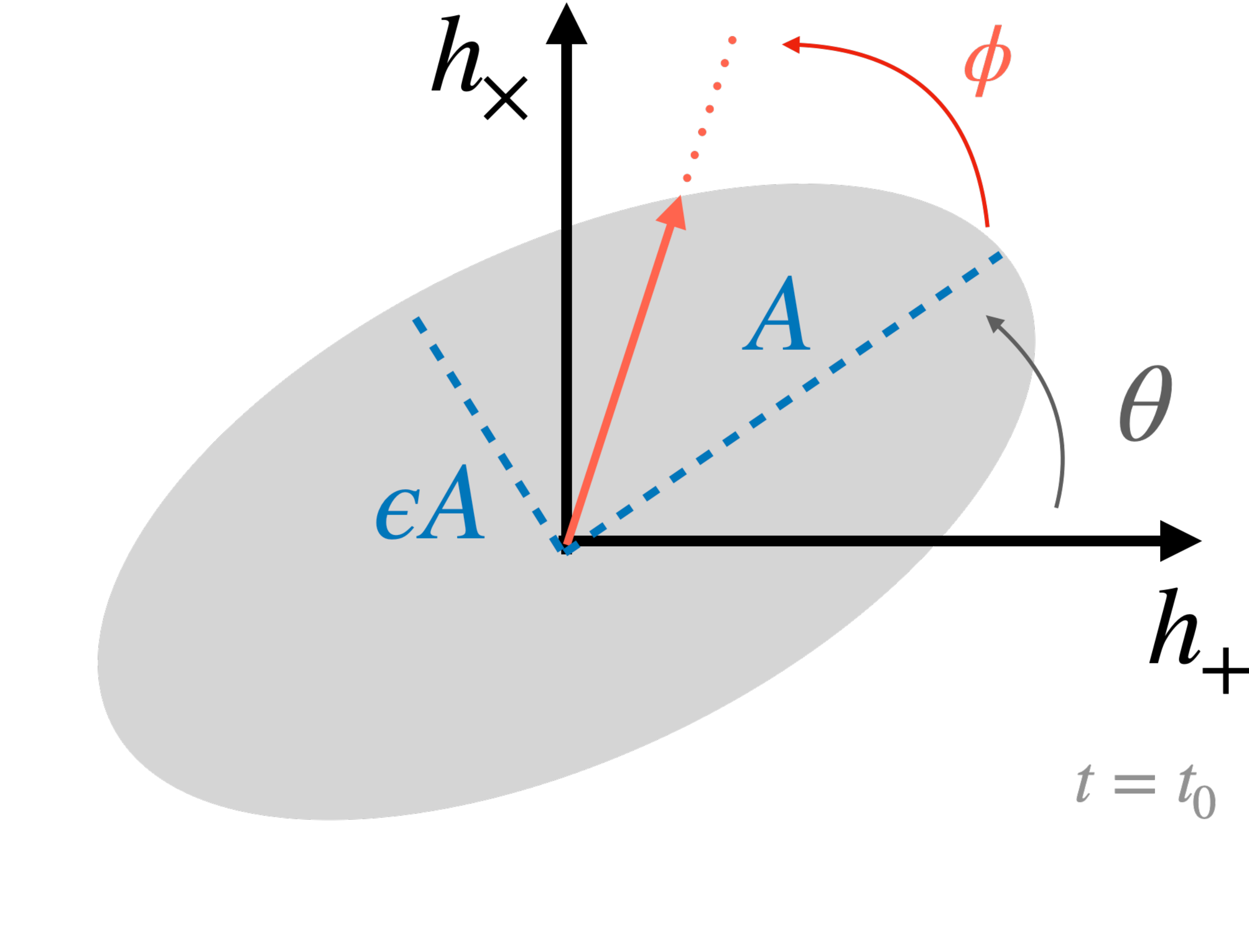}
\caption{Polarization ellipse. At any given time, the phasor of a given $j\equiv\{\ell,|m|,n\}$ ringdown mode lies on an ellipse with amplitude $A_j$ and ellipticity $\epsilon_j$, with semimajor axis tilted by an angle $\theta_j$ with respect to the plus-polarization axis (abscissa); a second angle, $\phi_j$, determines the initial location of the phasor within the ellipse. The phasor circles the ellipse with angular frequency $\omega_j$, while the ellipse itself shrinks with $e$-folding time $\tau_j$.}
\label{fig:ellipse}
\end{figure}

The role of these quantities is illustrated in Fig.~\ref{fig:ellipse}.
At time $t=0$, the polarization state of $h_j$ is given by a phasor lying somewhere along an ellipse in the $h^{(+)}_{j}$, $h^{(\times)}_j$ plane.
The ellipse has semimajor and semiminor axes $A_j$ and $\epsilon A_j$, respectively, oriented such that the semimajor axis defines an angle $\theta_j$ with the $h^{(+)}_{j}$ axis.
Finally, $\phi_j$ is the angle between the initial state and the semimajor axis within the ellipse.
As time passes, the phasor circles the ellipse with angular velocity $\omega_j$, while the size of the ellipse itself shrinks at an exponential rate with $e$-folding time $\tau_j$.

Equations (\ref{eq:h_ellip_lmn_plus}--\ref{eq:h_ellip_lmn_cos}) refer to a single mode $j = \{\ell, |m|, n\}$.
To construct a template for the signal $s_I$ recorded by GW detector $I$, we need only add all modes under consideration and project each polarization onto the detector through the corresponding antenna patterns $F_{+/\times}^I$,
\begin{equation} \label{eq:s}
s_I = \sum_j\left[F_+^I h^{(+)}_j(t - \delta t_I) + F_\times^I h^{(\times)}_j(t - \delta t_I) \right] .
\end{equation}
The $F^I_{+/\times}$ encode the relative orientation of the detector with respect to the GW strain tensor, and so depend on the source right ascension $\alpha$, declination $\delta$ and polarization angle $\psi$.%
\footnote{These response functions vary only on the timescale of a sidereal day, so we can ignore their time dependence when analyzing ringdowns.}
However, for fixed $\alpha$ and $\delta$, the effect of $\psi$ is fully degenerate with a rotation of the polarization ellipse through $\theta_j$, so we can evaluate the $F^I_{+/\times}$ at an arbitrary value of $\psi$ without loss of generality (see, e.g., App. A in \cite{Isi:2017equ}).
The $\delta t_I$ represent time-of-flight delays accounting for the different signal arrival times at each detector, and are usually defined with respect to the geocenter; these are also functions of $\alpha$ and $\delta$, but not $\psi$.

From Eq.~\eqref{eq:s}, it should be clear why we need not worry about mode mixing.
Even though we are using $j = \{\ell, |m|, n\}$ as a label, the polarization functions $h_j^{(+/\times)}$ know nothing about the meaning of those indices: that information has been absorbed by $A_j$, $\epsilon_j$, $\theta_j$ and $\phi_j$, which for us will be nuisance parameters [see Eq.~\eqref{eq:h_pt_lmn}].
In fact, the functions
\begin{equation} \label{eq:h_j}
h^{(+/\times)}_j \equiv h_{+/\times}(t;\, \omega_j, \tau_j, A_j, \epsilon_j, \theta_j, \phi_j)
\end{equation}
also ignore that $\omega_j$ and $\tau_j$ originate in BH perturbation theory: we can construct templates $h_{+/\times}(t;\, \omega, \tau, A, \epsilon, \theta, \phi)$ with any generic set of arguments.
If we wanted to incorporate a physical model for the amplitudes (e.g., in terms of the progenitor parameters in a binary coalescence, as in \cite{London:2014cma}), we could always do so by restoring the spheroidal harmonics in Eq.~\eqref{eq:h_pt_lmn}, as we discuss in App.~\ref{app:slms}; in the absence of such a model, the spheroidal harmonics just introduce redundant parameters to the model, and are thus detrimental.

In spite of the several assumptions we have used to motivate it, the template in Eq.~\eqref{eq:s} is extremely general.
It can easily be made to accommodate retrograde modes by passing the corresponding $(\omega_{[-1]\ell m n},\, \tau_{[-1]\ell m n})$ as input to Eq.~\eqref{eq:h_j}.
Although this would be inappropriate when perturbing around the Kerr solution, it can also circumvent the assumption of PT symmetry by simultaneously including pairs of modes like
\begin{equation}
h^{(+/\times)}_{\pm j} = h_{+/\times}(t;\, \omega_{\pm j}, \tau_{\pm j}, A_{\pm j}, \pm 1, \theta_{\pm j}, \phi_{\pm j})\, ,
\end{equation}
defining $\pm j = \{\ell, \pm |m|, n\}$, and allowing $\omega_{\pm j}$ and $\tau_{\pm j}$ to take independent values unrestricted by the PT condition, $\tilde{\omega}_{+j} = - \tilde{\omega}^*_{-j}$.
In that general case, the two circularly-polarized modes, $+|m|$ and $-|m|$, will not combine into a single elliptical contribution.

\begin{figure*}
  \includegraphics[height=0.31\textwidth]{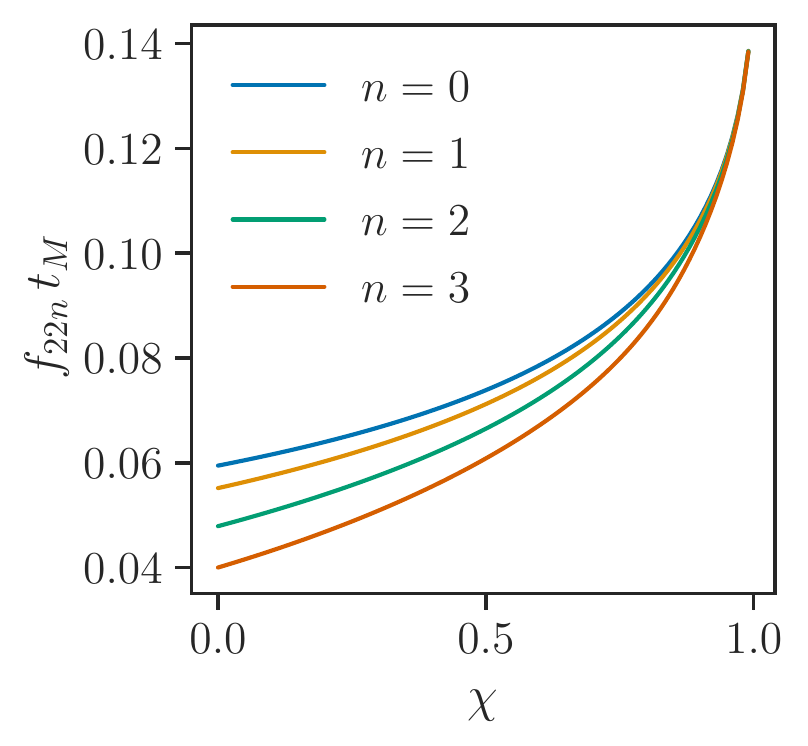}
  \includegraphics[height=0.31\textwidth]{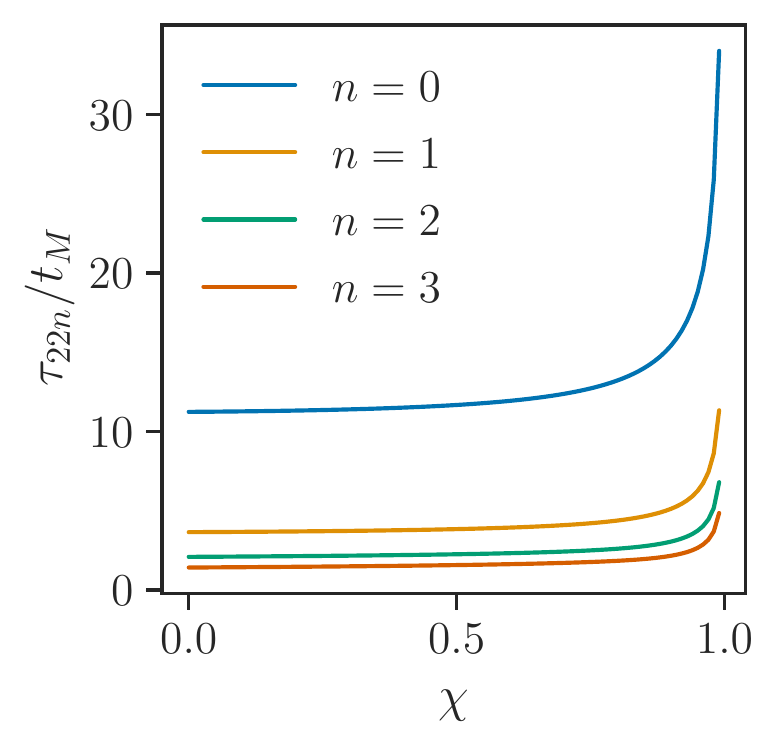}
  \includegraphics[height=0.31\textwidth]{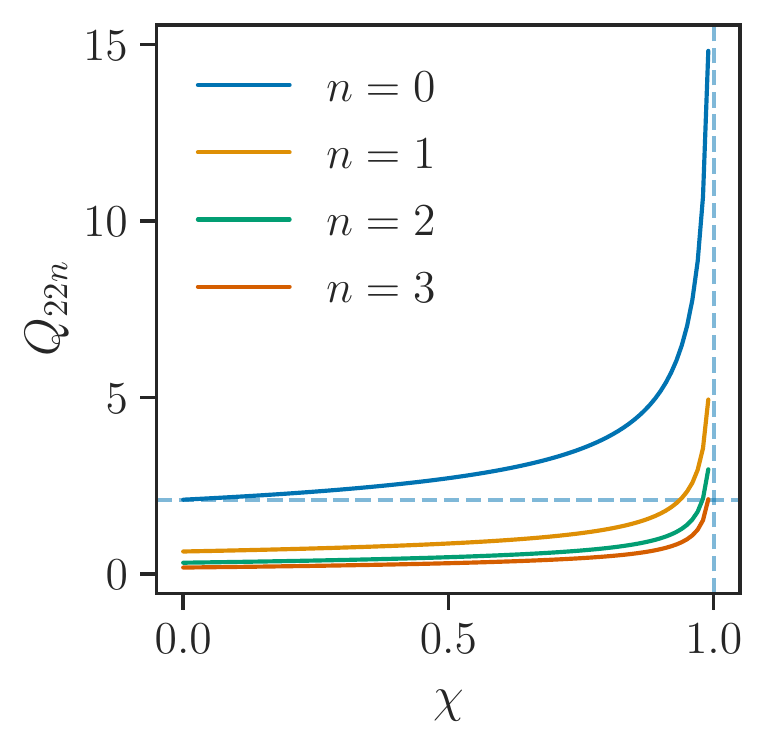}
  \caption{\label{fig:ftau-chi} Frequency $f_{22n}$ (left),
  damping time $\tau_{22n}$ (middle) and quality factor $Q_{22n}=\pi f_{22n} \tau_{22n}$ (right) for different $\ell=m=2$ tones, as a function of of dimensionless BH spin $\chi$.
  Times are measured in units of $t_M \equiv G M /c^3$ for BH mass $M$.
  The fundamental mode ($n=0$) is defined to be the mode with the largest $\tau_{22n}$; the overtones ($n\geq 1$) are ordered with decreasing $\tau_{22n}$ for increasing $n$.
  For the fundamental mode (blue), $Q_{220}$ asymptotes to $Q_\mathrm{min} \simeq 2.1$ as $\chi \to 0$, while it diverges to infinity as $\chi \to 1$ (right, dashed lines).
  }
\end{figure*}

Under most circumstances, we expect the signal to be dominated by modes with $\ell = |m| = 2$.
If so, we may wish to construct a template including $N$ overtones of the $\ell = |m| = 2$ angular mode in Eq.~\eqref{eq:s}, i.e.
\begin{equation} \label{eq:s_N}
s_I^{(22N)} = \sum_{n=0}^N \left[F_+^I h^{(+)}_{22n} + F_\times^I h^{(\times)}_{22n} \right] ,
\end{equation}
suppressing time dependence.
This is a template with $N+1$ modes, for a total of $6\left(N+1\right)$ damped-sinusoid parameters in the general case (fewer if we restrict to a Kerr spectrum, as discussed below).
The choice of $N$ is, in principle, arbitrary.
If $\ell > 2$ modes are also expected, we may incorporate them with the same or different number of respective overtones.

\subsection{Kerr ringdowns}
\label{sec:model:kerr}

In the case of a Kerr BH, all QNM frequencies and damping rates are determined fully by the hole's mass $M$ and dimensionless spin $\chi$.
We may thus write $\tilde{\omega}_{\ell m n}=\tilde{\omega}_{\ell m n}^{\rm (GR)}(M,\chi)$, so that the set of free parameters reduces to
\begin{equation} \label{eq:params_gr}
  \left\{ M, \chi, A_{j}, \epsilon_{j}, \theta_{j}, \phi_{j} \right\},
\end{equation}
for a total of $2(2N + 3)$ degrees of freedom for an $N$-overtone model including only $\ell=|m|=2$.
Throughout, we use the \textsc{qnm} Python package to calculate Kerr ringdown frequencies and damping times \citep{Stein:2019mop}.

The Kerr metric imposes some notable restrictions on the allowed QNM frequencies $f_{\ell m n} \equiv \omega_{\ell m n} / (2\pi)$ and damping rates $\gamma_{\ell m n} \equiv 1/\tau_{\ell m n}$.
As we mentioned above, the symmetries of the perturbation equations imply $\tilde{\omega}_{\ell m n}^{\rm (GR)} = - \tilde{\omega}_{\ell \unaryminus m n}^{\rm (GR)*}$, or, equivalently,
\begin{equation}
f_{\ell m n}^{\rm (GR)} = - f_{\ell \unaryminus m n}^{\rm (GR)} \, ,
\end{equation}
\begin{equation}
\tau_{\ell m n}^{\rm (GR)} = \tau_{\ell \unaryminus m n}^{\rm (GR)} \, .
\end{equation}
Further structure arises when we consider the functional dependence of these quantities on the BH parameters.
For example, Fig.~\ref{fig:ftau-chi} shows the frequency and damping times for different
prograde tones with $\ell=|m|=2$ as a function of BH spin. The mass $M$ of the
Kerr BH acts simply as an overall scale on the frequency and damping times,
so that dimensionless numbers like the product $f \tau$ or ratio $f_{j}/f_{j'}$ are
functions of the spin parameter $\chi$ only.

\begin{figure}
  \includegraphics[width=\columnwidth]{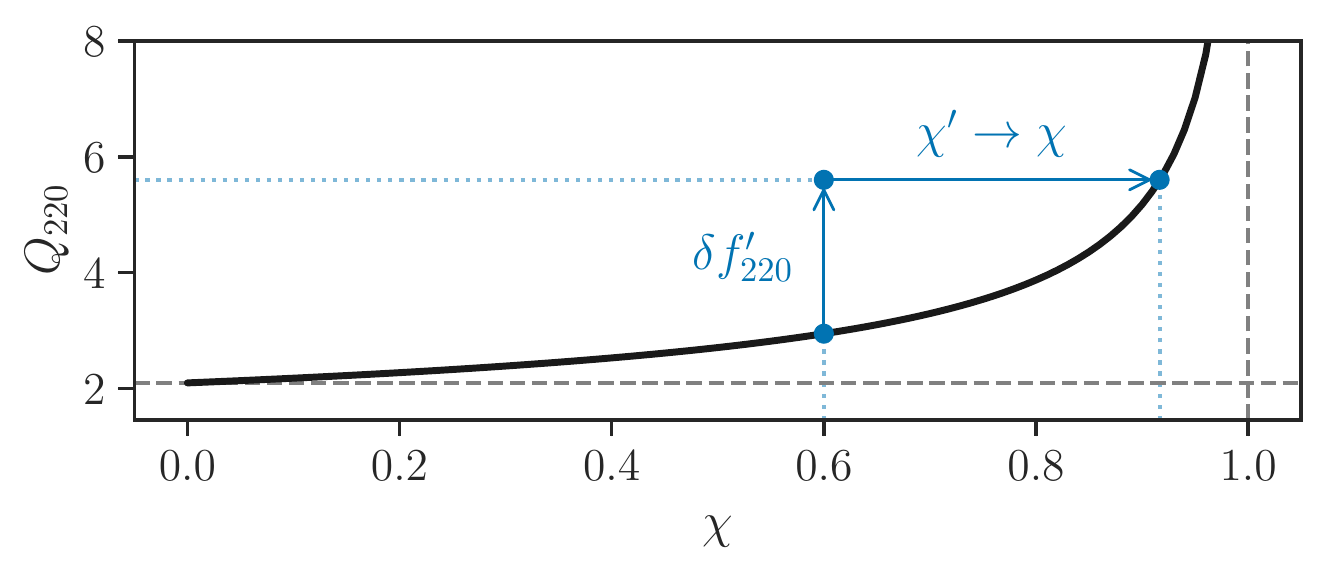}
  \caption{\label{fig:Q-chi} Quality factor versus spin magnitude for the
  fundamental $\ell = |m| =2,\, n =0$ Kerr ringdown mode (cf., Fig.~\ref{fig:ftau-chi}).
  A fractional deviation $\delta f_{220}'$ around the Kerr
  fundamental frequency results in an equal perturbation to $Q_{220}$; a model
  that assumes GR ($\delta f_{220}'=0$) can account for this by inferring a BH
  spin ($\chi$) different than the truth ($\chi'$).  The figure illustrates this
  for $\delta f_{220}'=0.9$ and true spin $\chi'=0.6$, resulting in an inferred
  value $\chi \approx 0.92$. The correspondence between $\delta f_{220}'$ and
  $\chi$ breaks down for negative values of $\delta f_{220}'$ that would yield
  $Q_{220} < Q_\mathrm{min}$ (horizontal dashed line).
  }
\end{figure}

Although the frequencies and damping rates can be made to take a broad range of values by varying $M$ and $\chi$, there are some restrictions.
For the fundamental $\ell = |m| = 2$ mode ($n=0$), the quality factor (suppressing indices)
\begin{equation}
  Q \equiv \pi f \tau
\end{equation}
is required to be greater than $Q_\mathrm{min} = Q_{220}\left( \chi = 0 \right)
\simeq 2.1$, so it is not possible to match \emph{every} underdamped ($Q > 1/2$)
mode with some combination of $M$ and $\chi$, nor is it possible to match
\emph{any} critically damped or overdamped mode (see Fig.~\ref{fig:Q-chi}).

We have found that the spin-dependence of the $f$ and $\tau$ parameters for low
order Kerr modes in GR can be well approximated by a linear combination of $\log
(1 - \chi)$, and powers from $\chi^0$ to $\chi^4$: for each mode the dimensionless frequency $f\, t_M$ satisfies
\begin{equation}
  f\, t_M \simeq c_l \log \left( 1 - \chi \right) + \sum_{i = 0}^4 c_i \chi^i,
\end{equation}
and similarly for the dimensionless damping rate $\gamma\, t_M$.
The approximating coefficients $c$ can be found by, e.g., least-squares fits to the complex mode frequencies computed by the \textsc{qnm} package \citep{Stein:2019mop}.
Our \textsc{ringdown} package does this automatically for any requested $\{\ell,|m|,n\}$, but we tabulate the values for a number of tones of the quadrupolar harmonic in App.~\ref{app:coeffs} as an example.
Implementing the relation between the Kerr parameters and the mode frequencies and damping times in this way is advantageous in contexts where it is required to differentiate mode parameters with respect to Kerr parameters (e.g., when using Hamiltonian Monte Carlo to sample from a posterior density defined in the \textsc{Stan} language \citep{Carpenter:2017}, as we will do below).

\subsection{Deviations from Kerr}
\label{sec:model:pert}

In order to verify that a given ringdown signal is consistent with a Kerr spectrum, one could imagine applying a fully-general model by which all frequencies and damping rates are allowed to vary freely in Eq.~\eqref{eq:h_j}.
However, such a generic model provides no straightforward way of quantifying agreement with the Kerr hypothesis: it can produce a $(2\times D)$-dimensional posterior for the frequencies and damping times of $D$ modes, but cannot in itself evaluate the degree of consistency with combinations allowed by the Kerr model.
Furthermore, parameter degeneracies render such a model impractical if more than one or two modes are included, presenting challenges even for high SNRs.
It is also useful for practical reasons (specifying priors, sampling efficiently, etc.) to be able to make direct connections between the observed mode properties and physically relevant variables in GR, such as masses and spins.

While a fully-general analysis can match any modes observed in the data, under most circumstances we expect that the modes that do appear will be at least close to the predictions for a Kerr BH in GR.
Thus, instead of allowing all frequencies and damping rates to vary freely, we may replace at least two of those parameters with the Kerr values derived from some $M$ and $\chi$, while allowing other modes to float around their corresponding Kerr values.  In other words, $M$ and $\chi$ are standing in for $f$ and $\tau$ of one of the spectroscopic modes, upon which we therefore implicitly impose a minimum allowed $Q$ (see Fig.~\ref{fig:Q-chi}); in a sufficiently modified theory these parameters need not have, even approximately, the usual Kerr interpretation.
In that way, the set of free parameters becomes
\begin{equation} \label{eq:params_nongr}
  \left\{ M, \chi, \delta {f}_j, \delta {\tau}_j, A_{j}, \epsilon_{j}, \theta_{j}, \phi_{j} \right\},
\end{equation}
where $\delta {f}_j$ and $\delta {\tau}_j$ are fractional deviations away from the Kerr prediction, i.e.,
\begin{equation} \label{eq:delta_f}
f_j = f_j^{\rm (GR)}(M,\chi) \left(1 + \delta{f}_j\right),
\end{equation}
\begin{equation} \label{eq:delta_tau}
\tau_j = \tau_j^{\rm (GR)} (M,\chi) \left(1 + \delta{\tau}_j\right) ,
\end{equation}
for all but two of the included $j$'s; for those two designated modes, we set $\delta f_j = \delta \tau_j = 0$, so that $f_j=f_j^{\rm (GR)}(M,\chi)$ and $\tau_j = \tau_j^{\rm (GR)}(M,\chi)$.
Equivalently, we could chose to work with fractional deviations on the damping rate $\gamma_j$, with $\delta\gamma_j \approx - \delta \tau_j$ for small deviations; although we do not do this here, taking $\delta\gamma_j$ as the primary quantity would have the advantage of avoiding technical issues that arise when $\delta \tau_j \to -1$ (see discussion toward the end of Sec.~\ref{sec:analysis:pert:kerr}).  Finally, another alternative is to work with deviation quantities that exhibit no singularities for finite parameter values via
\begin{align}
f_j &= f_j^{\rm (GR)}\left( M, \chi \right) \exp\left( \delta f_j \right) \\
\tau_j &= \tau_j^{\rm (GR)} \left( M, \chi \right) \exp\left( \delta \tau_j \right).
\end{align}
In all cases, the Kerr prediction is recovered when $\delta f_j = \delta \tau_j = 0$, and all parameterizations agree in the limit $\delta f_j, \delta \tau_j \ll 1$.

When the modes in question are overtones of a given $(\ell,|m|)$, we will additionally require that the $\delta \tau_j$ values be bounded so as to preserve tone ordering, i.e.~$\tau_{22,n+1} < \tau_{22,n}$ (a similar condition could be imposed on the frequencies when dealing with modes of different $|m|$).
For an $N$-overtone template, Eq.~\eqref{eq:s_N}, this model has $6(N + 1)$ degrees of freedom.

The fractional deviations $\delta f_j$ and $\delta \tau_j$ encapsulate all relevant information about the agreement of a given signal with the Kerr hypothesis.
If the signal is well described by a Kerr spectrum for any valid combination of $M$ and $\chi$, then we should find the posterior to be consistent with $\delta f_j = \delta \tau_j = 0$ for all modes under consideration.
Moreover, the characteristic width of the $\delta f_j$ or $\delta \tau_j$ posteriors (say, the 90\%-credible interval) quantifies the degree to which we can establish agreement with the Kerr scenario, providing a natural measure of the test's precision.

In generic beyond-Kerr models, it is reasonable to expect the $\delta f_j$ and $\delta \tau_j$ to themselves be functions of the mass and spin, plus any other parameters intrinsic to the model (e.g., the value of a scalar field).
Although this has little bearing on the analysis of individual signals, it does complicate the pooling of results across catalogs of sources, which would demand a hierarchical Bayesian treatment \cite{Zimmerman:2019wzo,Isi:2019asy}.
Alternatively, it is possible to write the $\delta f_j$ and $\delta \tau_j$ as a power series expansion on the BH spin $\chi$, with source-independent coefficients to be specified by any given theory \cite{Maselli:2019mjd,Carullo:2021dui}.
Under that framework, the $\delta f_j$ and $\delta \tau_j$ would be replaced by the new set of coefficients (whose number depends on the order of the expansion) as free parameters, potentially facilitating the combination of observations\footnote{This is only true if we assume that all sources analyzed belong to the same population; hierarchical treatments would still be needed if we allowed for the possibility of mixtures (e.g., a population composed of both Kerr BHs, and some exotic BH mimicker).} and the derivation of theoretical implications.
Although we do not adopt such a parameterization here, it would be trivial to do so, without altering the qualitative nature of our conclusions.

\subsubsection{Four-parameter models}
\label{sec:4D-parameterizations}

We now turn to the question of how to best assign the deviation parameters in Eqs.~\eqref{eq:delta_f} and \eqref{eq:delta_tau} to different QNMs.
As implied above, the spectroscopic analysis requires that at least two modes be detected: for a single-mode measurement, $\delta f$ and $\delta \tau$ would be fully degenerate with the mass and spin, modulo the $Q \geq Q_{\rm min}$ constraints imposed by the Kerr spectrum (Sec.~\ref{sec:model:kerr}).
In a model with $D$ modes, we may consider up to $2D-2$ deviation parameters;
in the simplest case, $D=2$ and we are left with four spectroscopically relevant parameters: $M$ and $\chi$, plus a single $\delta f$ and a single $\delta\tau$.
As we show below, even when considering angular harmonics other than $\ell=|m|=2$, it is generally preferable to assign $M$ and $\chi$ to the best measured mode.%
\footnote{Readers not interested in parameterizations of the two-mode model may skip the rest of this section with impunity; we continue the discussion of the formalism in Sec.~\ref{sec:inference}.}

While keeping our focus on the $\ell=|m|=2$ angular harmonic, the most natural two-mode model to consider is the one made up of the two longest lived tones, i.e., the fundamental and the first overtone \cite{Isi:2019aib,Abbott:2020jks}.
In that case, the Kerr-deviation parameters may be assigned to any two of $\{f_{220}, \tau_{220}, f_{221}, \tau_{221}\}$.
Since, we expect the fundamental mode to be more easily measured, it is reasonable to leave its frequency and damping time unperturbed, and assign both deviation parameters to the first overtone.
The spectral model would thus become
\begin{subequations}
\label{eq:nongr_df1_dtau1}
\begin{align}
  f_{220} & =f^{\rm (GR)}_{220}\left( M, \chi \right) \label{eq:f220-default}, \\
  \tau_{220} & = \tau^{\rm (GR)}_{220}\left( M, \chi \right) \label{eq:tau220-default}, \\
  f_{221} & = f^{\rm (GR)}_{221}\left( M, \chi \right) \left( 1 + \delta f_{221} \right), \label{eq:f221-default} \\
  \tau_{221} & = \tau^{\rm (GR)}_{221}\left( M, \chi \right) \left( 1 + \delta \tau_{221} \right). \label{eq:tau221-default}
\end{align}
\end{subequations}
Whenever the fundamental mode dominates the measurement (which we expect to be the case essentially always), it will pin down $M$ and $\chi$ in this model, while the overtone is allowed to explore alternative values around Kerr solution.
Except for the restriction to $Q_{220} > Q_\mathrm{min}$ (Fig.~\ref{fig:ftau-chi}), this parameterization is fully general for the two-mode case, in that it can fit (almost) any pair of QNMs.
Equations \eqref{eq:nongr_df1_dtau1} were put to use in \cite{Isi:2019aib} and, subsequently, \cite{Abbott:2020jks} to produce observational results.

We will adopt the above as our default parameterization.  It is instructive to see why this choice is a good one by exploring inferior alternatives.
For example, another option for the $N = 1$ model would be to allow deviations in the two frequencies by writing
\begin{subequations}
\label{eq:nongr_df0_df1}
\begin{align}
  f_{220} &= f_{220,\mathrm{Kerr}}\left( M', \chi' \right) \left( 1 + \delta f_{220}' \right) \label{eq:two-df-f0}\\
  \tau_{220} &= \tau_{220,\mathrm{Kerr}}\left( M', \chi' \right) \label{eq:two-df-tau0} \\
  f_{221} &= f_{221,\mathrm{Kerr}}\left( M', \chi' \right) \left( 1 + \delta f_{221}' \right) \label{eq:two-df-f1}\\
  \tau_{221} & = \tau_{221,\mathrm{Kerr}}\left( M', \chi' \right) \label{eq:two-df-tau1},
\end{align}
\end{subequations}
using primes to distinguish this from our default parameterization.
While not strictly degenerate, Eqs.~\eqref{eq:nongr_df0_df1} are \emph{practically} degenerate: although they appear to involve four degrees of freedom, there are only three in practice.

We can see this by studying the relation between $\{ M',\, \chi',\, \delta f_{220}',\, \delta f_{221}'\}$ and $\{M,\, \chi,\, \delta f_{221},\, \delta \tau_{221}\}$ through Eqs.~\eqref{eq:nongr_df0_df1} and Eqs.~\eqref{eq:nongr_df1_dtau1}.
To obtain a mapping between the two sets of coordinates, first note that the quality factor $Q_{220} = \pi f_{220} \tau_{220}$ can be written in both parameterizations as
\begin{align} \label{eq:Q220-transf}
  Q_{220} &= f_{220}^{\rm (GR)}\left( M, \chi \right) \tau_{220}^{\rm (GR)}\left( M, \chi \right) \\ \nonumber
  &= f_{220}^{\rm (GR)}\left( M', \chi' \right) \tau_{220}^{\rm (GR)}\left( M', \chi' \right) \left( 1 + \delta f_{220}' \right).
\end{align}
Since $f_{220}^{\rm (GR)}\tau_{220}^{\rm (GR)}$ is independent of $M$, we can numerically solve the above expression for $\chi$ as a function of the primed quantities.
In other words, as long as $\delta f_{220}'$ is not so small as to render $Q_{220} < Q_\mathrm{min}$, we can always find a value $\chi$ that yields the same quality factor as a different value $\chi'$ plus the deviation $\delta f_{220}'$ (Fig.~\ref{fig:Q-chi}).
The restriction to $Q_{220} > Q_\mathrm{min}$ is not worrisome because, since we can assume that the $n=0$ mode dominates the signal, a ringdown with $Q_{220} < Q_\mathrm{min}$ would be easily identified as anomalous by regular parameter estimation of the full IMR signal within GR.

\begin{figure}
  \includegraphics[width=\columnwidth]{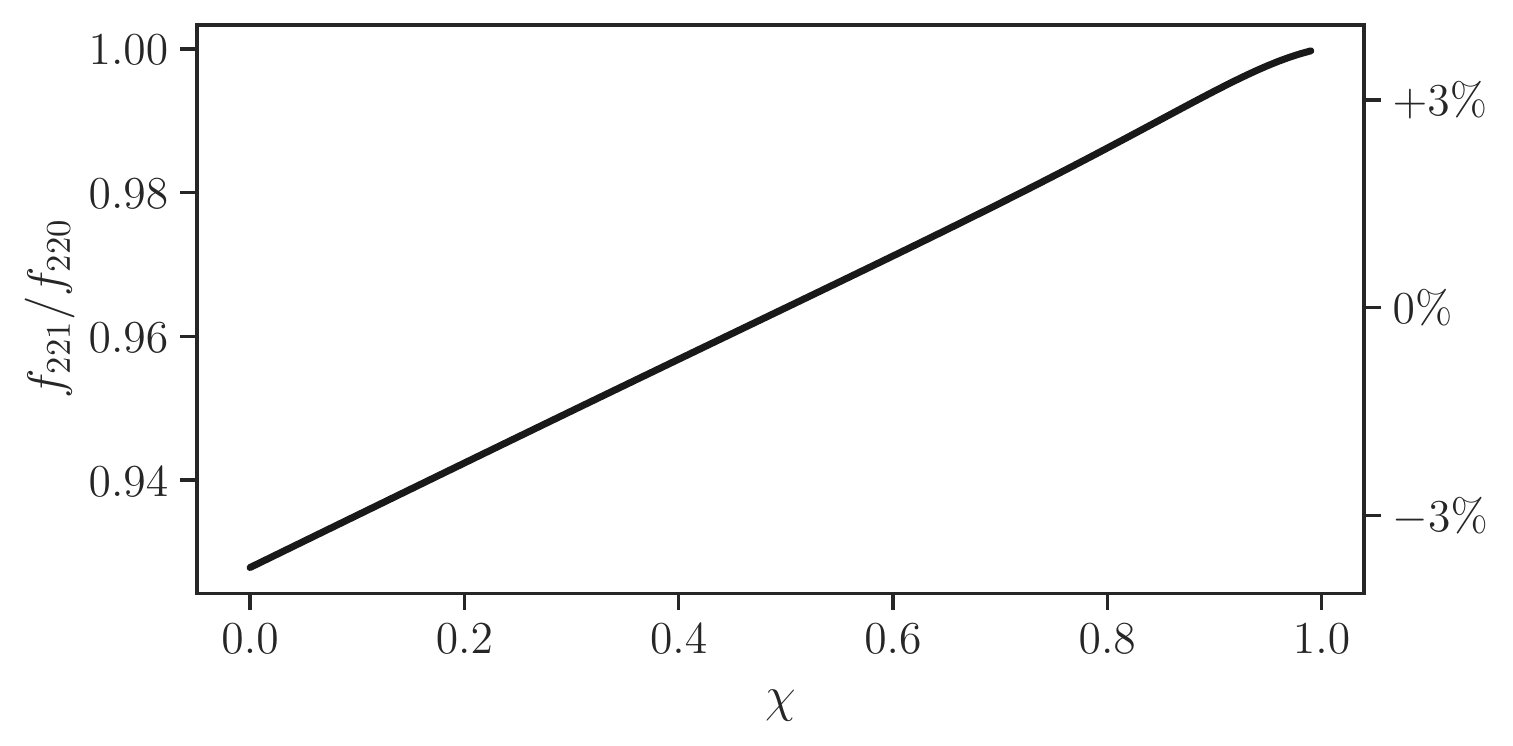}
  \includegraphics[width=\columnwidth]{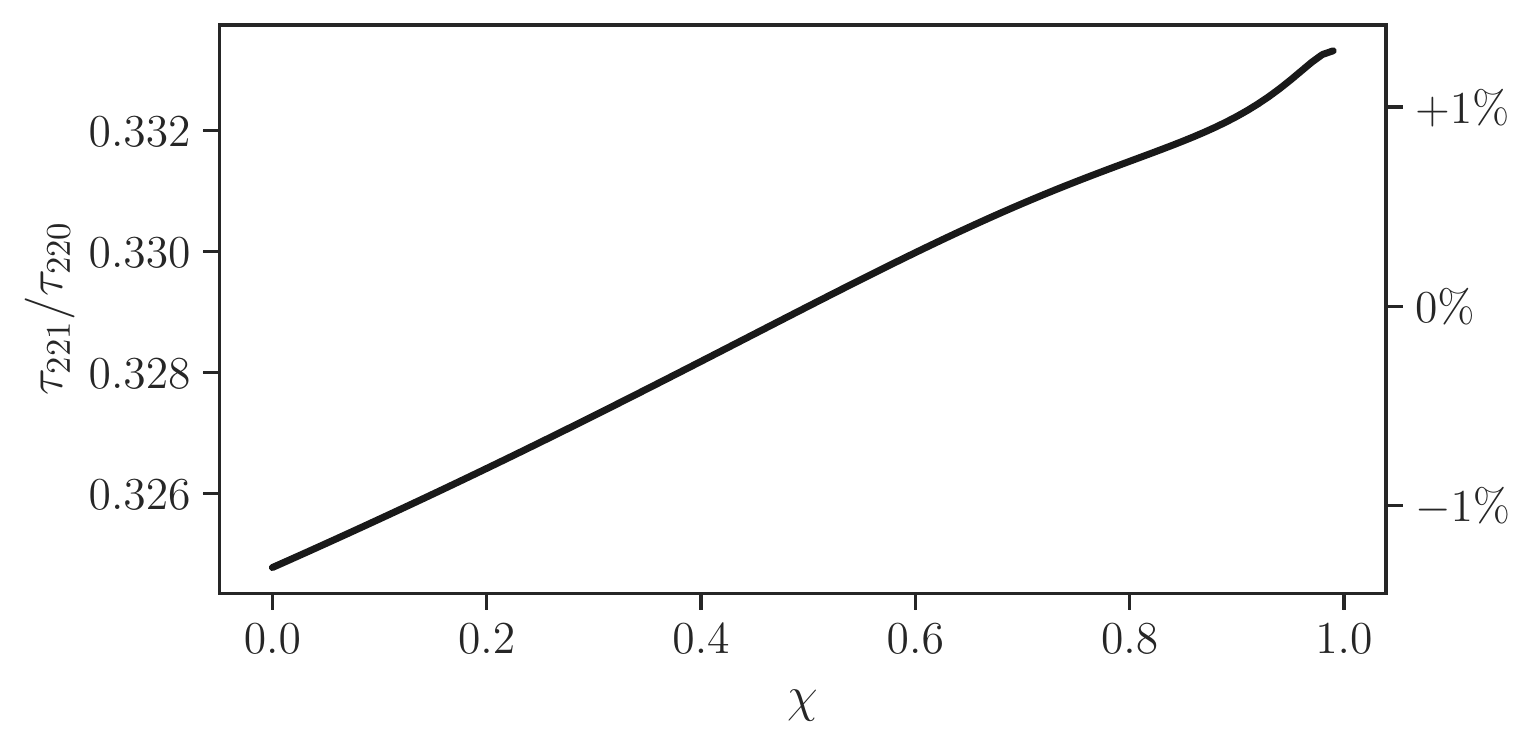}
  \caption{\label{fig:ftau-ratios-chi} Ratio of the first-overtone frequency
  (top) and damping time (bottom) to the fundamental mode values, as a function
  of BH spin $\chi$, assuming a Kerr BH. The labels on the right of each plot
  present the ratios as fractional deviations of the value at $\chi=0.5$. The
  ratios only vary by a few percent over the full range of $\chi$. (See also
  Fig.~\ref{fig:ftau-chi}.)}
\end{figure}

\begin{figure*}
  \includegraphics[width=1.5\columnwidth]{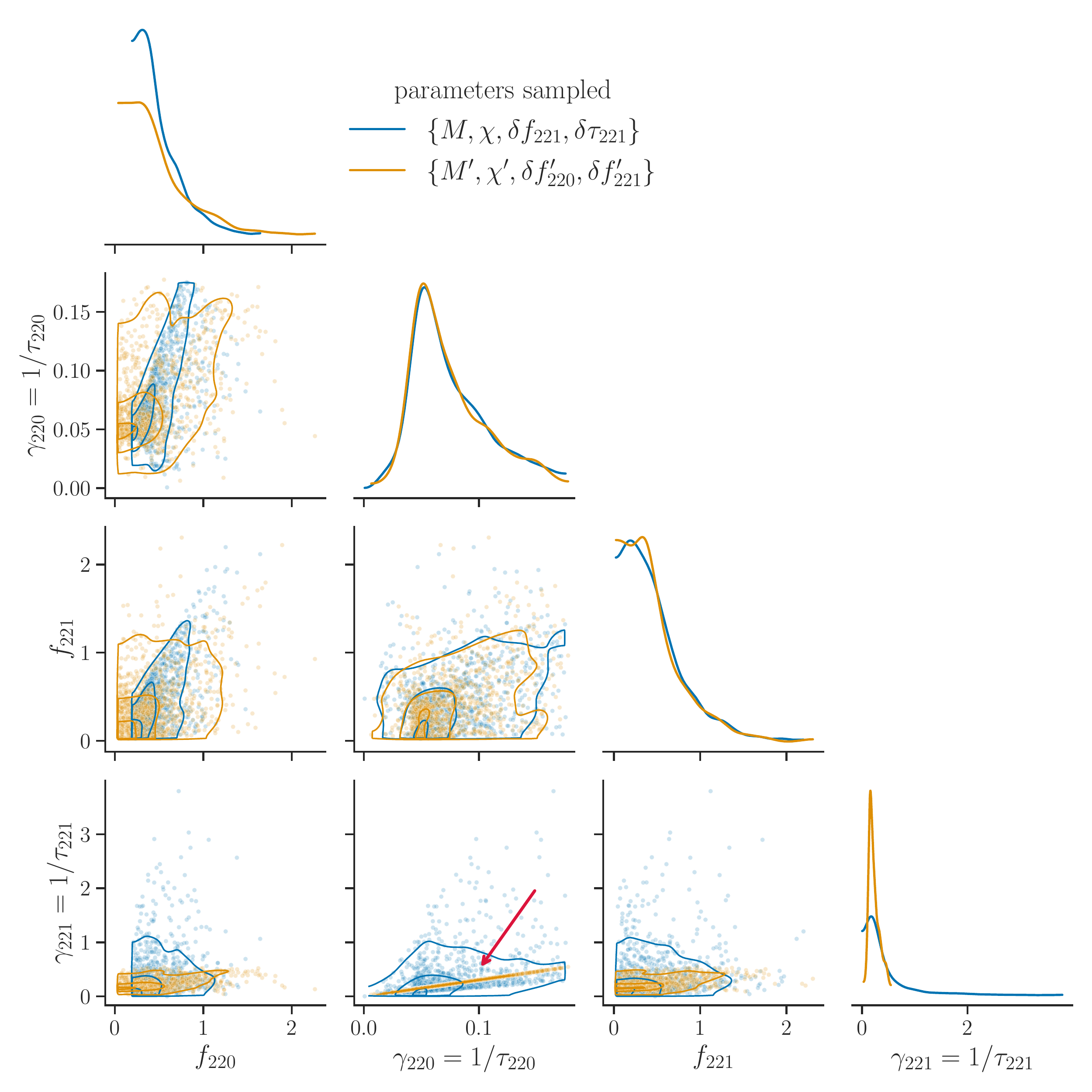}
  \caption{\label{fig:four-parameter-posterior} Comparison between the
  distributions of $f_{220}$, $\gamma_{220} = 1/\tau_{220}$, $f_{221}$, and
  $\gamma_{221} = 1 / \tau_{221}$ induced by flat priors on $1/2 < M, M' < 2$ (arbitrary units),
  $0 < \chi, \chi' < 1$, $-0.9 < \delta f_{220}' < 0.9$, $-0.9 < \delta f_{221},
  \delta f_{221}' < 0.9$, and $-0.9 < \delta \tau_{221} < 0.9$ in the
  \citet{Isi:2019aib} parameterization of Eqs.~\eqref{eq:nongr_df1_dtau1} (blue ) and the
  parameterization of Eqs.\ \eqref{eq:nongr_df0_df1} (orange).
  The degeneracy in the latter parameterization is
  apparent in that it produces $\gamma_{220}$--$\gamma_{221}$ pairs that lie
  (nearly) on a single plane, while the \citet{Isi:2019aib} one explores a finite volume
  in the four-dimensional frequency-damping rate space (highlighted by red arrow).}
\end{figure*}

We can proceed similarly to get an expression for $M$ as function of primed quantities by solving
\begin{equation}
  f_{220} = f_{220}^\mathrm{(GR)}\left( M, \chi \right) = f_{220}^\mathrm{(GR)}\left( M', \chi' \right) \left( 1 + \delta f_{220}' \right)
\end{equation}
for $M$.
With $M$ and $\chi$ in hand, we can find $\delta f_{221}$ from
\begin{align}
  f_{221} &= f_{221}^{\mathrm{(GR)}}\left( M', \chi' \right) \left( 1 + \delta f_{221}' \right) \\
  &= f_{221}^\mathrm{(GR)}\left( M, \chi \right) \left( 1 + \delta f_{221} \right), \nonumber
\end{align}
and, finally, $\delta \tau_{221}$ from
\begin{equation} \label{eq:dtau1_map}
  \tau_{221} = \tau_{221}^\mathrm{(GR)}\left( M', \chi' \right) = \tau_{221}^\mathrm{(GR)}\left( M, \chi \right) \left( 1 + \delta \tau_{221}\right).
\end{equation}

The above equations suggest that, excepting the $Q_{220} \geq Q_{\rm min}$ restriction, there is a one-to-one mapping between the primed and unprimed coordinates.
However, the structure of the Kerr spectrum implies that Eq.~\eqref{eq:dtau1_map} can only produce a small range $\delta \tau_{221}$ values, and so the coordinate transformation is effectively degenerate.
This becomes apparent by considering the fundamental-to-overtone damping time ratio as written in each parameterization,
\begin{equation}
  \label{eq:ratio-of-taus}
  \frac{\tau_{220}}{\tau_{221}} = \frac{\tau_{220}^\mathrm{(GR)}\left( M', \chi' \right)}{\tau_{221}^\mathrm{(GR)}\left( M', \chi' \right)} = \frac{\tau_{220}^\mathrm{(GR)}\left( M, \chi \right)}{\tau_{221}^\mathrm{(GR)}\left( M, \chi \right)} \frac{1}{1 + \delta \tau_{221}} .
\end{equation}
Figure \ref{fig:ftau-ratios-chi} shows this dimensionless ratio as a function of the spin parameter; notably, the ratio varies by only $\sim \pm 1.2\%$ across the entire allowed range of $0 \leq \chi < 1$.
Thus, irrespective of the values of $\chi$ and $\chi'$, Eq.~\eqref{eq:ratio-of-taus} requires $\left( 1 + \delta \tau_{221} \right) \simeq 1$, or $\delta \tau_{221} \simeq 0$.
Therefore, varying $M'$, $\chi'$, $\delta f_{220}'$, and $\delta f_{221}'$ is nearly identical to varying $M$, $\chi$, and $\delta f_{221}$ with $\delta \tau_{221} = 0$, and the seemingly four-parameter model of Eq.~\eqref{eq:nongr_df0_df1} all but reduces to the three-parameter model $\{M,\, \chi,\ \delta f_{221}\}$.
The degeneracy can only be broken by a $\tau_{221}$ measurement with better than ${\sim}1\%$ precision.

Figure \ref{fig:four-parameter-posterior} demonstrates this near-degeneracy empirically.
We show the joint distribution of $f_{220}$, $\gamma_{220}=1/\tau_{220}$, $f_{221}$, and $\gamma_{221} = 1/\tau_{221}$ induced by a uniform prior on $1/2 < M, M' < 2$ (arbitrary units), $0 < \chi, \chi' < 1$, $-0.9 < \delta f_{220}' < 0.9$, $-0.9 < \delta f_{221}, \delta f_{221}' < 0.9$, and $-0.9 < \delta \tau_{221} < 0.9$ in the parameterization of Eq.~\eqref{eq:nongr_df1_dtau1} (un-primed) and the parameterization of Eqs.\ \eqref{eq:nongr_df0_df1} (primed).
Using this latter parameterization, the prior only permits ratios of $\gamma$ values that lie on a (nearly) degenerate plane of (nearly) constant $\gamma_{220}/\gamma_{221}$; using the former parameterization, the prior allows frequencies and damping rates that explore a finite volume in the four-dimensional space.
(See also Sec.~\ref{sec:analysis:pert:alt}.)

As for the damping times, Fig.~\ref{fig:ftau-ratios-chi} also shows the ratio of Kerr ringdown frequencies is a small $\sim \pm 3.5\%$ over $0 \leq \chi < 1$; this suggests that a parameterization with $\delta \tau_{220}$ and $\delta \tau_{221}$ would suffer from a near-degeneracy analogous to Eq.~\eqref{eq:nongr_df0_df1}.

\subsubsection{Three-parameter models}
\label{sec:3D-parameterizations}

The first overtone's damping time, $\tau_{221}$, is often quite poorly-measured \cite{Isi:2019aib}.
Under the assumption that it is effectively unconstrained by the data, the two-mode problem reduces to a three-dimensional parameter space.
This is equivalent to $\delta \tau_{221}$ in Eq.\ \eqref{eq:tau221-default} being unconstrained.
Therefore, the fully-general (up to the $Q_{220} > Q_\mathrm{min}$ constraint) three-parameter likelihood  can be parameterized by Eqs.\ \eqref{eq:nongr_df1_dtau1} with $\delta \tau_{221} =0$.

\begin{figure}
  \includegraphics[width=\columnwidth]{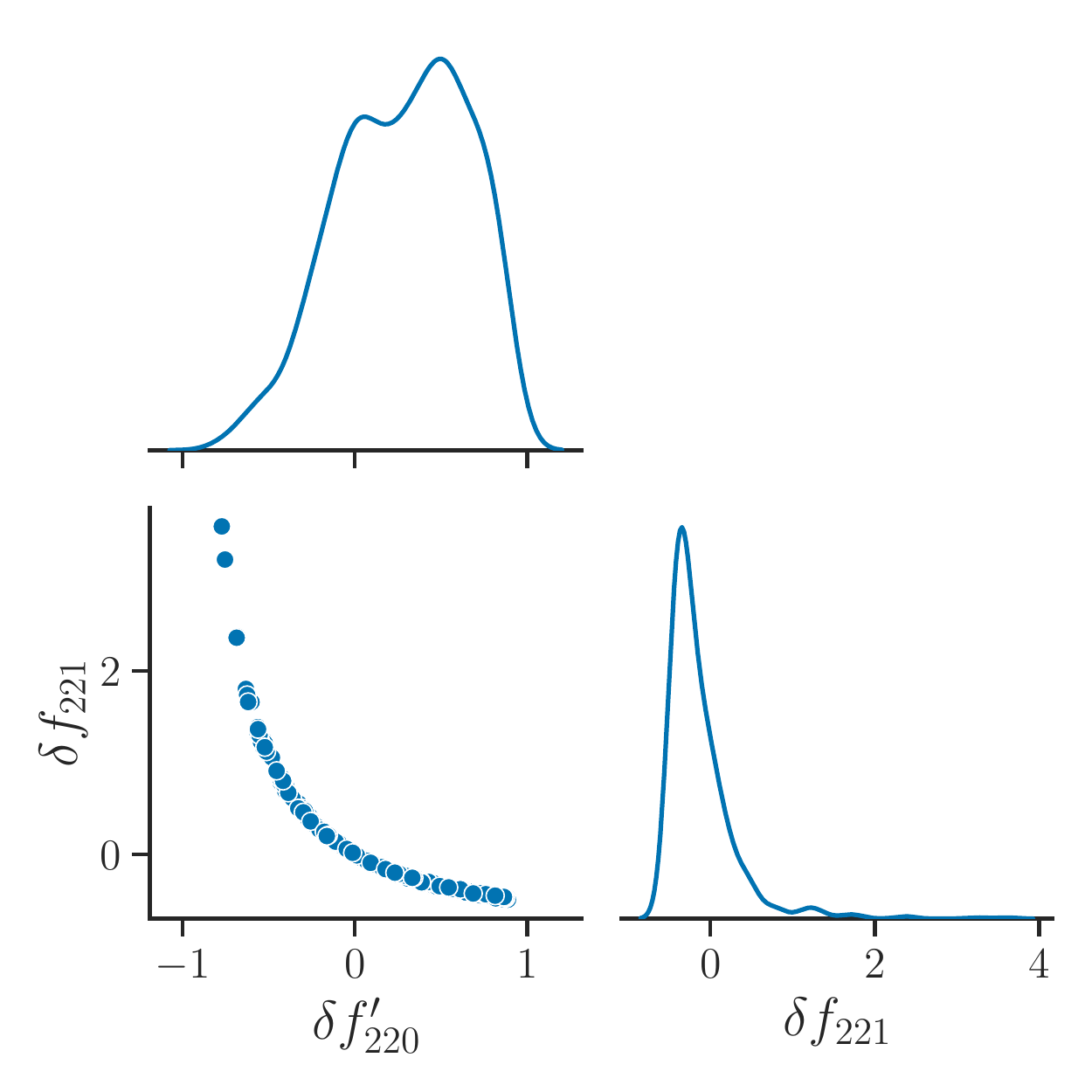}
  \caption{\label{fig:three-parameter-prior} Joint distribution of $\delta
  f_{220}'$ and $\delta f_{221}$ under a uniform prior on $0.5 < M' < 2$, $0 <
  \chi' < 1$, and $-0.9 < \delta f_{220}' < 0.9$ under the one-to-one mapping
  between the reduced three-dimensional parameterizations in the text.  In spite
  of the range of $M'$ and $\chi'$ values permitted at fixed $\delta f_{220}'$,
  the relation between $\delta f_{220}'$ and $\delta f_{221}$ is exceedingly
  tight.}
\end{figure}

As above, it is useful to consider an alternative parameterization like
\begin{subequations}
\begin{align}
  f_{220} &= f_{220}^\mathrm{(GR)}\left( M', \chi' \right) \left( 1 + \delta f_{220}' \right)\label{eq:f220-df0} \\
  \tau_{220} &= \tau_{220}^\mathrm{(GR)}\left( M', \chi' \right) \label{eq:tau220-df0} \\
  f_{221} &= f_{221}^\mathrm{(GR)}\left( M', \chi' \right) , \label{eq:f221-df0}
\end{align}
\end{subequations}
where again primes denote quantities in the alternative parameterization.
By the same logic we applied to Eq.~\eqref{eq:Q220-transf}, it is clear that we can transform between primed and unprimed coordinates in this case also, as long as $\delta f_{220}'$ does not push $Q_{220}$ below the Kerr constraint.
Figure \ref{fig:three-parameter-prior} shows the joint distribution of $\delta
f_{220}'$ and $\delta f_{221}$ under this one-to-one mapping when $1/2 < M' <
2$ (arbitrary units), $0 < \chi' < 1$ and $-0.9 < \delta f_{220}' < 0.9$ are
drawn uniformly within their respective ranges.  In spite of the wide
distribution of $M'$ and $\chi'$ values, there is a very narrow one-to-one
correspondence between $\delta f_{220}'$ and $\delta f_{221}$ (insensitive to
the arbitrary $M$ range chosen).

The points in the $\delta f_{220}'-\delta f_{221}$ plane closely follow a
hyperbola. We can see that this is the case by considering the ratio of frequencies
\begin{align}
\frac{f_{220}}{f_{221}} &= \frac{f_{\rm 220,Kerr}(M', \chi')}{f_{\rm 221,Kerr}(M', \chi')} \left(1+\delta f_{220}\right)  \nonumber \\
&= \frac{f_{\rm 220,Kerr}(M, \chi)}{f_{\rm 221,Kerr}(M, \chi)} \frac{1}{\left(1+\delta f_{221}\right)} \, .
\end{align}
The ratios of Kerr frequencies are independent of mass and only weakly dependent
on the spin (Fig.~\ref{fig:ftau-ratios-chi}), so they are approximately equal no
matter the values of $\chi$ and $\chi'$. To good accuracy, this implies
\begin{equation}
  \label{eq:f-hyperbola}
  \left(1 + \delta f_{220}'\right)\left(1
+ \delta f_{221}\right) \approx \text{constant}\, ,
\end{equation}
which describes a hyperbola that asymptotes to $\delta f_{220}' = -1$ as $\delta
f_{221}\rightarrow \infty$, and vice versa. Therefore, if we restrict $-1
<\delta f_{220}'< 1$, we will necessarily have $-0.5 < \delta f_{221} < \infty$,
with the upper (lower) limit on $\delta f_{220}'$ setting the lower (upper)
limit on $\delta f_{221}$. This explains the range of the $\delta f_{221}$
marginal distribution in Fig.~\ref{fig:three-parameter-prior} (lower right). The
full shape of that curve can be well approximated analytically by applying the
Jacobian implied by Eq.~\eqref{eq:f-hyperbola} to the $\delta f_{220}'$
distribution.

Since there is a tight, nearly one-to-one mapping between $\delta f_{220}'$ and
$\delta f_{221}$ even over a broad range of $M'$ and $\chi'$, using both parameterizations to fit real ringdown data would not return independent measurements unless the overtone damping time can be constrained to sub-percent precision.

\section{Inference framework}
\label{sec:inference}

A signal template, such as discussed above, is only one of the many moving pieces that make up the Bayesian infrastructure required to extract information from ringdown data.
In this section, we motivate and describe our time domain domain framework for Bayesian inference with ringdown signals (Sec.~\ref{sec:td}), and show it is mathematically equivalent to a recently proposed alternative (Sec.~\ref{sec:fd-inpainting}); we also provide implementation details (Sec.~\ref{sec:imp}).

The reason for spending time carefully examining data analysis techniques is that the framework standard to LIGO-Virgo transient analyses is not readily suitable for ringdown-only studies.
The overwhelming majority of such analyses are constructed in the Fourier domain, the fundamental reason for this being that nominal instrumental noise is well described as a stationary Gaussian process.
Assuming periodic boundary conditions, this means that the noise covariance matrix diagonalizes in the frequency domain.  The noise Fourier amplitudes become independent random variables with variance described by some one-sided power spectral density (PSD) as a function of frequency, $S(f)$ \cite{Allen:2001ay,UNSER1984231}.
This vastly simplifies Bayesian computations because, in that case, the log-likelihood for a signal $\tilde{s}(f)$ is given by a simple noise-weighted inner product,
\begin{equation} \label{eq:lnlike_fd}
\ln P(\tilde{d} \mid \tilde{s}) = - 2 {\Delta f}\sum_{i=0}^{N-1} \frac{|\tilde{d}_i - \tilde{s}_i|^2}{S(f_i)} + \mathrm{const.},
\end{equation}
for Fourier-domain data $\tilde{d}$ sampled at $N$ frequencies $f_i \geq 0$, with $\Delta f = f_{i+1} - f_i$.
Evaluating Eq.~\eqref{eq:lnlike_fd} takes $O(N)$ computations---an improvement over the $O(N^3)$ required for an arbitrary covariance matrix.
The properties of the signal are secondary, as long as it is smooth, contained within the frequency band, and sufficiently short for the stationarity and periodic-boundary assumptions to apply.

\begin{figure}
\includegraphics[width=0.9\columnwidth]{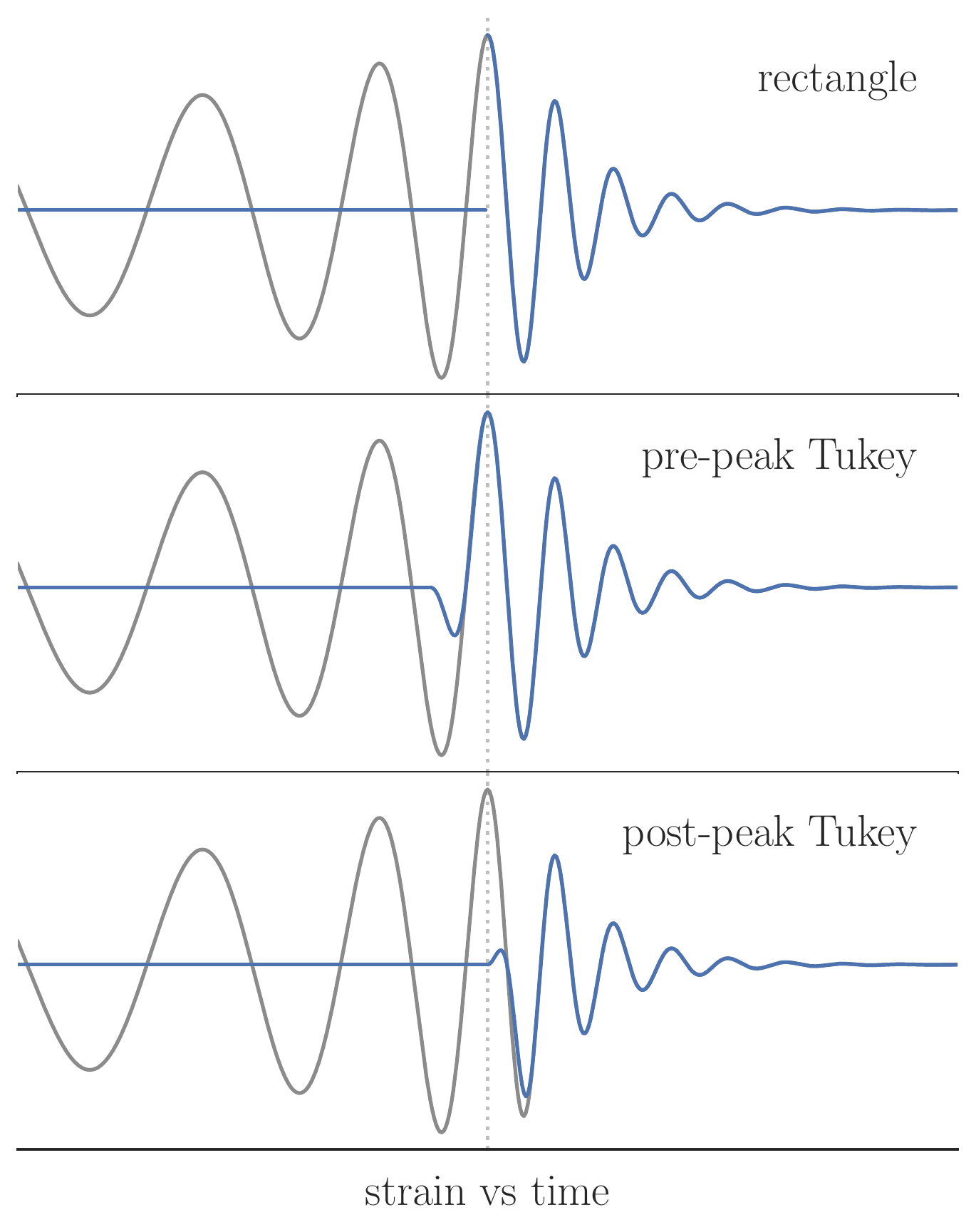}
\caption{The challenge of ringdown-only analyses. Targeting the ringdown portion of an IMR signal requires discarding the data before some truncation time $t_0$, here the peak of the signal (dotted gray line).
A regular Fourier domain analysis requires smoothly windowing the data segment---not doing so is equivalent to applying a sharp rectangular window (top) that will lead to spectral leakage. Avoiding spectral leakage requires a smooth roll-on (e.g., a Tukey window), which is not compatible with a signal that has significant content at the edge of the segment: a window that offers support for $t < t_0$ will allow contamination from non-ringdown data (middle), while one that does not will necessarily corrupt the start of the ringdown signal (bottom).
The issue can be circumvented by remaining in the time domain (Sec.~\ref{sec:td}), as we propose, or by applying nontrivial modifications to the frequency-domain likelihood (Sec.~\ref{sec:fd-inpainting}), as was proposed in \cite{Capano2021,Zackay2019}.
}
\label{fig:window}
\end{figure}

Unfortunately, analyses aimed at isolating the ringdown fail to meet these basic requirements:
it is not possible to enforce a cyclic boundary, and avoid spectral leakage, without corrupting part of the signal.
Regular CBC analyses based on Eq.~\eqref{eq:lnlike_fd} rely on stretches of time-domain detector data chosen such that the signal lies a safe distance away from the segment edges, which are tapered smoothly to prevent leakage when Fourier transforming \cite{LIGOScientific:2019hgc}.
However, that is not an option when targeting the ringdown in isolation: the signal in which we are interested is necessarily located at the edge of the data segment, and would be corrupted by any windowing procedure (Fig.~\ref{fig:window}).
Retaining data before the ringdown is not viable because the whitening filter for colored noise will necessarily couple points across the cutoff, contaminating the targeted signal \cite{Cabero:2017avf}.
Simply Fourier transforming an unwindowed data segment starting at the beginning of the ringdown will both cause spectral leakage and correlate the beginning of the segment to its end.

In spite of such fundamental obstacles, there have been several attempts to formulate ringdown analyses fully or partially in the traditional Fourier domain framework \cite{TheLIGOScientific:2016src,Prix:T1500618,Cabero:2017avf,CalderonBustillo:2020tjf}.
The original LIGO-Virgo study in \cite{TheLIGOScientific:2016src} followed a mixed strategy, which consisted of inverse-Fourier transforming overwhitened detector data to compute the likelihood in the time domain \cite{Prix:T1500618}.
Although this method avoids explicitly Fourier-transforming a discontinuous template, it is mathematically equivalent to Eq.~\eqref{eq:lnlike_fd}: the overwhitening filter couples data before and after the truncation point, contaminating the ringdown measurement as illustrated in Fig.~\ref{fig:lvc_performance}.
The approach was successful for GW150914 in \cite{TheLIGOScientific:2016src} because the chosen truncation times lay far away from the signal peak: with low SNR immediately before the truncation point, the systematics caused by overwhitening were likely smaller than the statistical noise; however, this would fail for times closer to the peak or for louder signals.
Other proposed strategies, like zeroing out data before the cutoff \cite{Cabero:2017avf}, or replacing it with detector noise \cite{CalderonBustillo:2020tjf}, fail to prevent the whitening filter from coupling data across the boundary---which occurs irrespective of the nature of the pre-cutoff data (whether the data before the cutoff are zeros or noise is immaterial)---and can also be subject to spectral leakage induced by sharp features in the data.
None of these problems arise if we formulate our analysis in the time domain without enforcing cyclic boundary conditions.

\subsection{Time domain likelihood}
\label{sec:td}

\begin{figure}
\centering
\begin{minipage}{0.5\columnwidth}
\centering
\includegraphics[width=0.9\columnwidth]{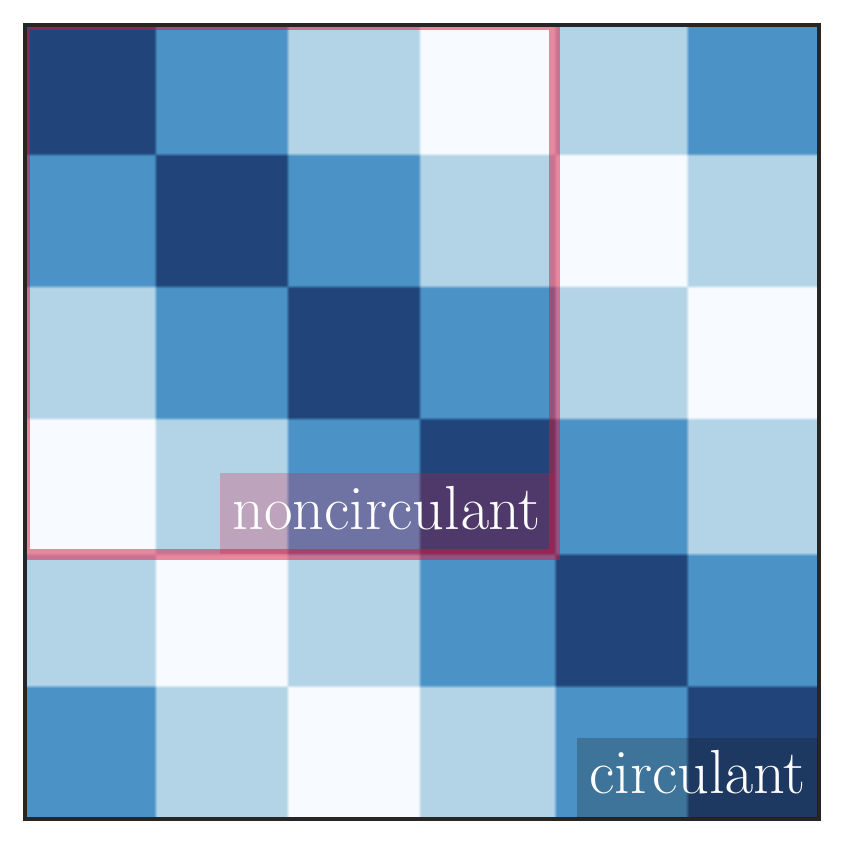}%
\end{minipage}%
\begin{minipage}{0.5\columnwidth}
\centering
\textsc{symmetric\\noncirculant \\Toeplitz matrix}
\begin{equation}
\begin{pmatrix}
\rho_0 & \rho_1 & \dots & \rho_{N-1}\\
\rho_1 & \rho_0 & \dots & \rho_{N-2}\\
\vdots &  & \ddots & \vdots \\
\rho_{N-1} & \dots & \rho_1 & \rho_{0}\\
\end{pmatrix} \nonumber
\end{equation}
\end{minipage}
\caption{Symmetric Toeplitz matrices. \emph{Left:} illustration of the symmetric-Toeplitz matrix structure of Eq.~\eqref{eq:cov_td} in both circulant (main matrix) and noncirculant (highlighted submatrix) forms; the shade of each square represents the value of a matrix entry $C_{ij}$: the circulant case satisfies $C_{ij} = C_{(N-i)j}$ for any entry $0 \leq i,j < N$, while the noncirculant case does not.
\emph{Right:} schematic form of a noncirculant Toeplitz matrix, i.e., a matrix that satisfies $C_{ij}=C_{i+1,j+1}$ and $C_{ij}=C_{ji}$, or equivalently $C_{ij} = \rho(|i-j|)$ as in Eq.~\eqref{eq:cov_td}, but not $C_{ij} = C_{(N-i)j}$, just as in the highlighted submatrix on the left.
}
\label{fig:matrix}
\end{figure}

We model instrumental noise as a discrete-time random process, represented by a set of random variables $n_i \equiv n(t_i)$ when sampled at arbitrary times $t_i$.
Assuming Gaussianity, the process will be fully characterized by its mean $\mathrm{E}\left[n_k \right]$ and covariance matrix
$
\cov_{ij} \equiv \mathrm{E}\left[n_i n_j - \mathrm{E}\left[n_k \right]\right] ,
$
where $\mathrm{E}$ denotes expectation values.
In general, the mean can be trivially enforced to be zero $\mathrm{E}\left[n_k \right]=0$, while the only restriction on $\cov$ is that it be positive semidefinite.
For an $N$-vector $n= \{n_0,\dots,n_{N-1}\}$ drawn from such a process, $\cov$ will be an $N \times N$ matrix,%
\footnote{In this section, and only in this section, $N$ stands for the number of samples analyzed, not number of overtones in a template.}
and the log-probability of the draw will be given by
\begin{equation} \label{eq:noise_pdf_td}
\ln P(n) = - \frac{1}{2} \sum_{i,j} n_i\, \cov_{ij}^{-1}\, n_j + \mathrm{const.},
\end{equation}
where $\cov^{-1}$ is the inverse covariance matrix.
Equation \eqref{eq:noise_pdf_td} requires $O(N^2)$ computations if the covariance matrix is pre-processed into a convenient form.  Such preprocessing costs $O(N^3)$.

To reduce the computing cost per likelihood evaluation, we may for example Cholesky-decompose the covariance matrix into a lower-triangular factor $L$ and its transpose, such that
\begin{equation} \label{eq:cholesky}
\cov_{ij} = \sum_{k} L_{ik} L_{kj} \, .
\end{equation}
The inverse of $L$ can be computed by backsubstitution efficiently, and can act on an arbitrary time series $x$ to produce a new, uncorrelated (``whitened'') time series,
\begin{equation} \label{eq:whiten}
\white{x}_i = \sum_jL_{ij}^{-1} x_j \, .
\end{equation}
If the covariance of $n_i$ is given by Eq.~\eqref{eq:cholesky}, then each component of $\white{n}$ will be independently drawn from a unit normal, i.e.,~$\white{n}_i \sim \mathcal{N}(0,1)$.
With that definition, the likelihood in Eq.~\eqref{eq:noise_pdf_td} simplifies to
$\ln P(n \mid \cov) = - \frac{1}{2} \sum_{i,j} \white{n}_i\,  \white{n}_j + \mathrm{const.}$

\begin{figure*}
\includegraphics[width=0.329\textwidth]{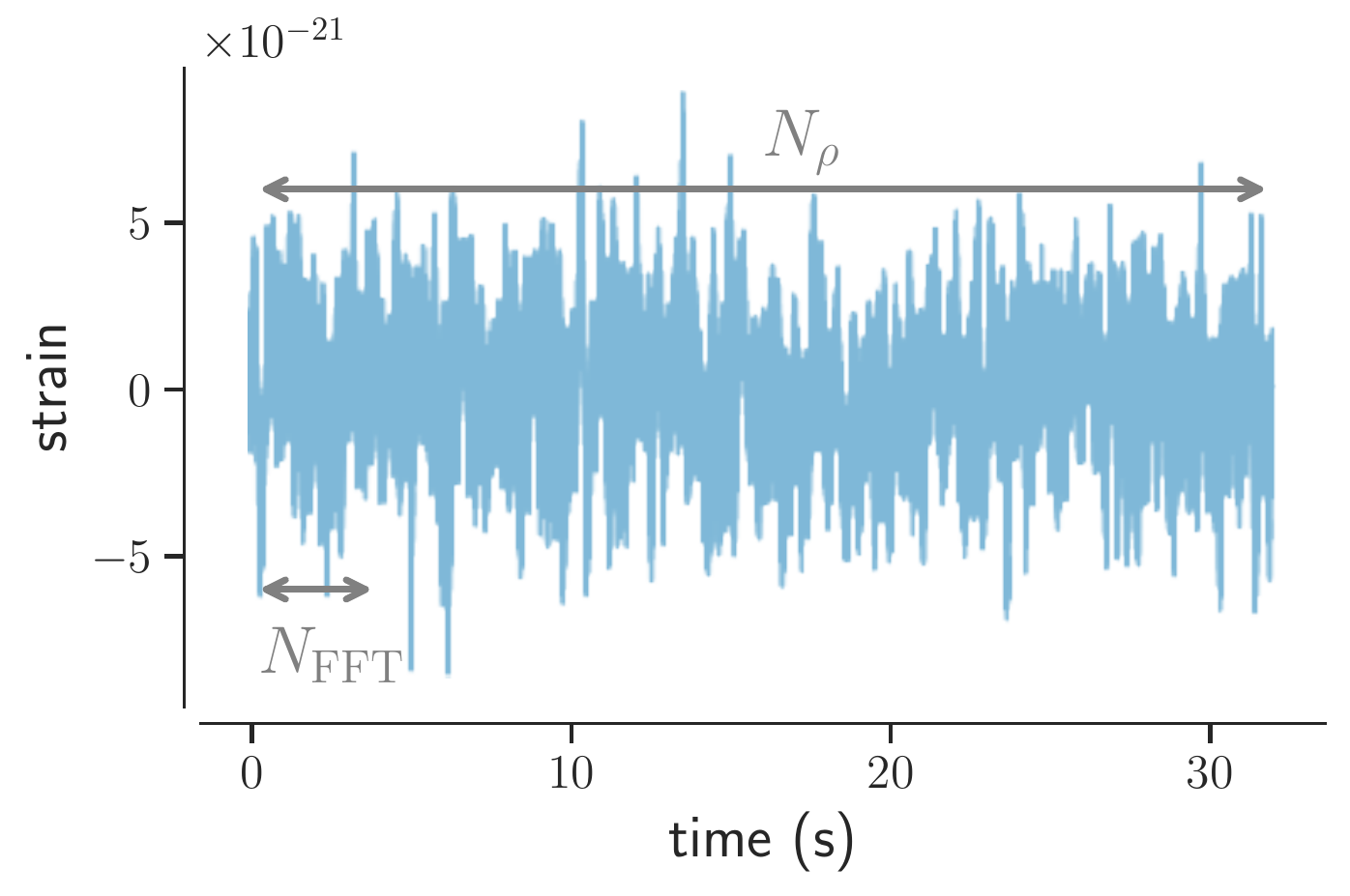}
\includegraphics[width=0.329\textwidth]{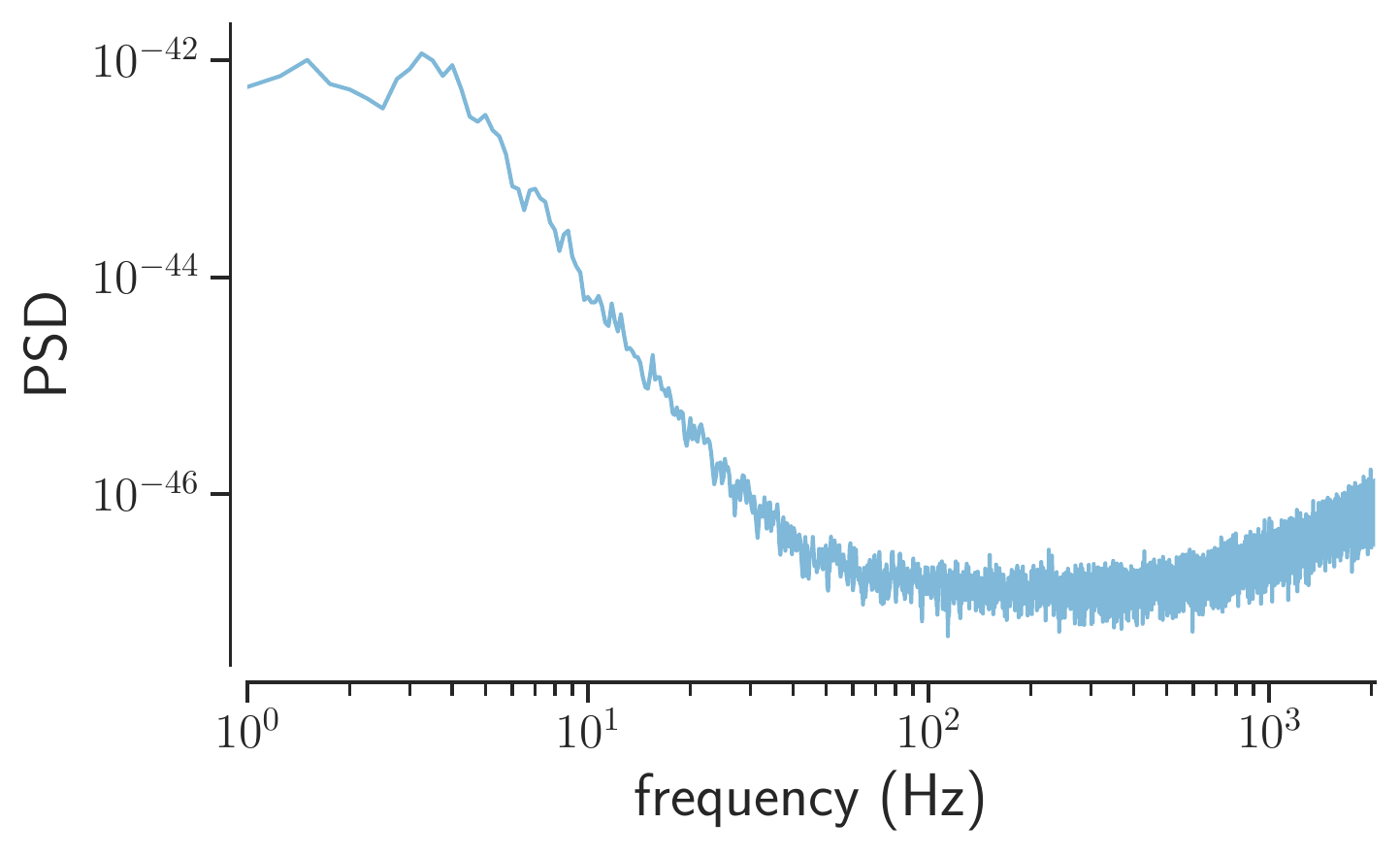}
\includegraphics[width=0.329\textwidth]{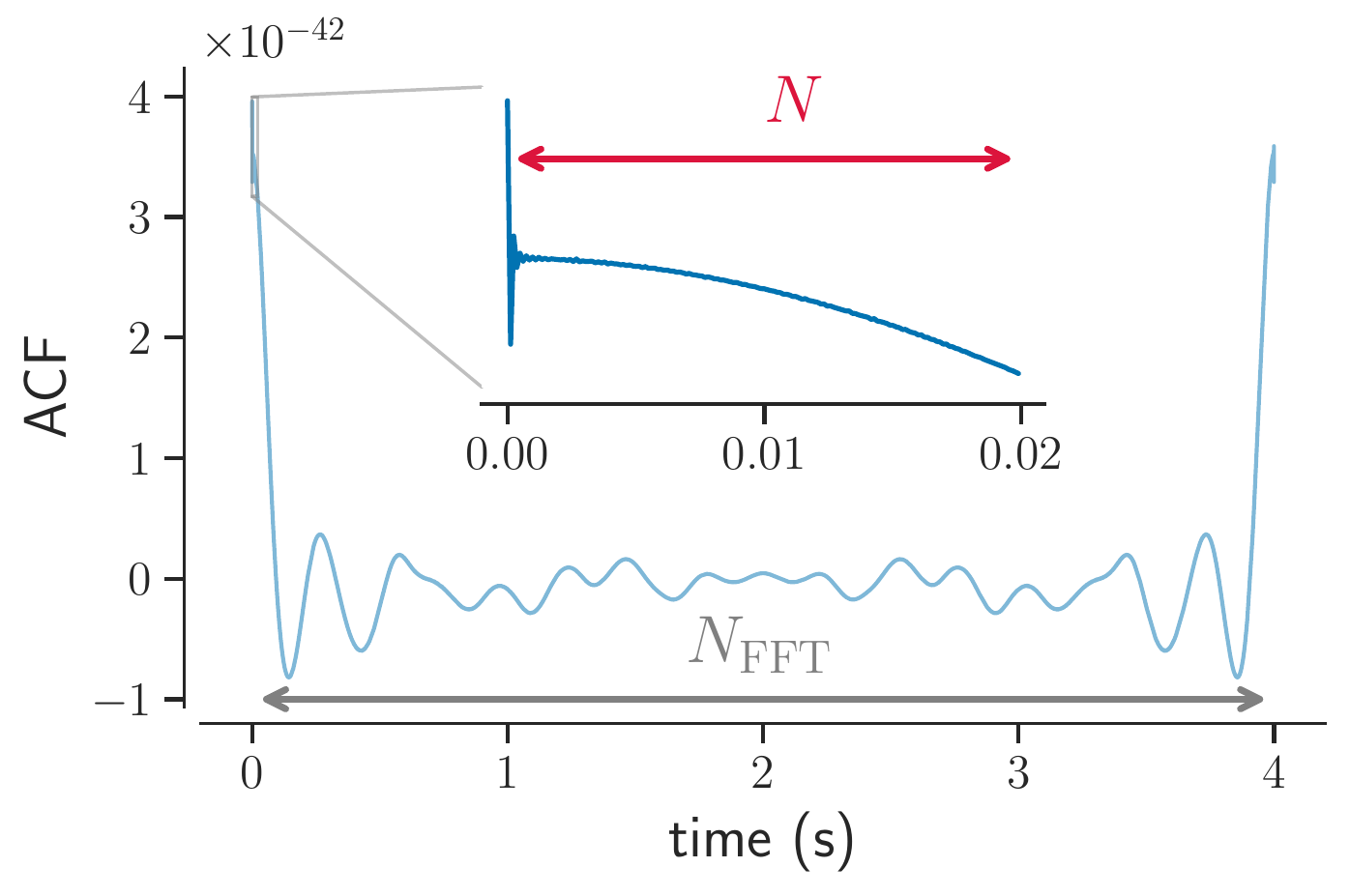}
\caption{Obtaining an aperiodic ACF: starting from a long stretch of time-domain noise with duration $N_\rho \Delta t$ (left), we can obtain a PSD through usual techniques like Welch estimation with a coherent integration length of $N_{\rm FFT}$ (middle); we can then inverse-Fourier transform the PSD to obtain a long, periodic estimate of the noise ACF (right, main panel), which we can finally truncate to the desired analysis length $N$ (right, inset) to form the covariance matrix through Eq.~\eqref{eq:cov_td}. Alternatively, we could estimate the ACF directly by autocorrelating the time domain data, or we could start the process from a PSD obtained by a different method \cite{Littenberg2015,Zackay2019}.
}
\label{fig:psd_acf}
\end{figure*}

For a stationary random processes, the covariance takes a particularly simple (symmetric Toeplitz) form,
\begin{equation} \label{eq:cov_td}
\cov_{ij} = \acf\left( \left| i - j \right| \right) ,
\end{equation}
where $\acf(k)$ is the autocovariance function (ACF).
This can be estimated empirically by autocorrelating a long stretch of noise-only data (i.e., a segment of length $N_\acf \gg N$),
\begin{equation} \label{eq:rho}
\hat{\acf}(k) = \frac{1}{N_\acf} \sum_{i=0}^{N_\acf - 1} n_i n_{i+k}\, ,
\end{equation}
for $|k| < N_\acf$, and zero otherwise\footnote{This is a biased estimator of the true ACF; the bias at lag $k$ is a factor of $N_\acf/(N_\acf - |k|)$.  We favor Eq.~\eqref{eq:rho}, however, because it reduces the influence of large lag terms where the variance in the ACF estimate is larger; it is thus more stable than the unbiased estimator.}.

If, in addition to stationarity, we impose periodic boundary conditions, then $\acf(k) = \acf(N-k)$ and $\cov$ will be circulant (Fig.~\ref{fig:matrix}).
Circulant matrices are diagonalized by the discrete Fourier transform \cite{UNSER1984231}, meaning the noise Fourier amplitudes $\tilde{n}_i$ will be drawn from a Gaussian process with covariance matrix
\begin{equation} \label{eq:cov_fd}
\tilde{\cov}_{ij} \equiv E[\tilde{n}_i\tilde{n}_j] =  \frac{1}{2}\, T\, S (|f_i|)\, \delta_{ij} \, ,
\end{equation}
where $T=N \Delta t$, $S(|f|)$ is the one-sided PSD, and $\delta_{ij}$ is the Kronecker delta.
The PSD is derived from the cyclic ACF through a discrete Fourier transform,
\beq \label{eq:psd_rho}
S (|f_j|) = 2 \Delta t \sum_{k=0}^{N-1} \acf(k)\, e^{-2 \pi i j k / N}\, ,
\eeq
for $f_j = j \Delta f = j / T$, and $0 \leq j < N$.
Naturally, because $\acf(k)$ is real-valued, $S(f_j) = S(f_{N - j})$.

In the limit of infinite observation time ($N\to \infty$), Eq.~\eqref{eq:cov_fd} is a consequence of the Wiener-Khinchin theorem and the diagonalization it implies is exact, no matter the specific structure of $\acf(k)$; for large but finite $N$, the diagonalization is only approximate, unless $\cov$ is exactly circulant \cite{UNSER1984231,Rover2011}.
In either case, we can always choose to work with Fourier-domain quantities without loss of generality.
However, $\tilde{\cov}$ will only be diagonal if and only if $\cov$ is circulant.
If that is not the case, Eq.~\eqref{eq:lnlike_fd} fails to apply and there is little incentive to switch to the Fourier domain.%
\footnote{There will always exist a coordinate transformation (known as Karhunen-Lo\'eve transformation) to diagonalize any covariance matrix \cite{UNSER1984231}, but there will generally not exist an efficient algorithm (like the fast Fourier transform) to effect it.}

Conversely, the Fourier domain likelihood of Eq.~\eqref{eq:lnlike_fd} is predicated on Eq.~\eqref{eq:cov_fd} and, therefore, assumes that the time-domain data are periodic.
Starting from a given PSD estimate $\hat{S}(|f|)$ sampled at $N$ frequencies $f_j$, we may invert Eq.~\eqref{eq:psd_rho} to obtain the corresponding ACF estimate,
\begin{equation} \label{eq:rho_psd}
\hat{\acf}_{S}(k) = \frac{1}{2 T} \sum_{j=0}^{N-1} \hat{S}(|f_j|)\, e^{2 \pi i j k /N}\, .
\end{equation}
This estimator is normalized in the same way as Eq.~\eqref{eq:rho}, but will be cyclic by construction.
This additional symmetry is not implied by Eq.~\eqref{eq:cov_td}, or Eq.~\eqref{eq:rho}, but is rather induced by Eq.~\eqref{eq:cov_fd}.
With $\cov_{ij}$ derived from $\hat{\acf}_S(k)$, the time-domain expression in Eq.~\eqref{eq:noise_pdf_td} is formally equivalent to Eq.~\eqref{eq:lnlike_fd}, with $n_i = d_i - s_i$.

The key to our approach is to do without the covariance structure imposed by Eq.~\eqref{eq:cov_fd}, enabling us to analyze a short segment of ringdown data without corrupting it.
Without tapering, our analysis is also free from spurious Fourier bin covariances \cite{Talbot:2021igi}.
We can achieve this simply by using the time-domain likelihood of Eq.~\eqref{eq:noise_pdf_td} with a covariance matrix constructed from an acyclic estimate of the ACF.
This can be obtained directly from noise samples in the time domain via Eq.~\eqref{eq:rho}.
It can also be derived from a preexisting PSD estimate using Eq.~\eqref{eq:rho_psd}, as long as the PSD was estimated from data segments of length $N_{\rm FFT}$ much longer than the analysis segment length $N$, with $N_{\rm FFT}$ replacing $N$ in Eq.~\eqref{eq:rho_psd}.
A long segment allows us to truncate the resulting $N_{\rm FFT}$-long $\hat{\acf}_S(k)$ to length $N$ before constructing the covariance matrix (Fig.~\ref{fig:psd_acf}).
After truncation, $\hat{\acf}_S(k) \neq \hat{\acf}_S (N-k)$, thus breaking circularity.

Once we have constructed an acyclic covariance matrix, we can analyze a noisy data stream $d(t)$ with a model defined only after some truncation time $t_{\rm start}$.
Assuming the data were presampled at some set of times $t_i$, we will index them such that $i=0$ corresponds to the first sample at or after the specified truncation time,
\beq \label{eq:tstart}
t_0 \equiv \min(\left\{t_i \text{ such that } t_{\rm start} \leq t_i \}\right) .
\eeq
With this convention, the log-likelihood for data after $t_{\rm start}$ containing a signal $s$ is nothing but
\begin{equation} \label{eq:lnlike_td}
\ln P\left(d \mid s\right) = -\frac{1}{2}\sum_{i,j=0}^{N-1} \left(d_i - s_i\right) \cov^{-1}_{ij} \left(d_j - s_j\right) ,
\end{equation}
up to a constant, as implied by Eq.~\eqref{eq:noise_pdf_td} and $d_i = s_i + n_i$ for $i \geq 0$.
This likelihood is completely agnostic about times before $t_0$, and does not impose a periodic boundary.

Note that it is essential to discard the data preceding the truncation point before evaluating Eq.~\eqref{eq:lnlike_td}: it is not sufficient to set $s(t) = 0$ for $t < t_0$, as was done in \cite{Carullo:2019flw}.
Such a model predicts no signal until $t_0$, at which point the template turns on sharply---this is decidedly not agnostic about data for $t < t_0$.
If one adopts such a model, whether the covariance is cyclic or not becomes irrelevant: the beginning of the ringdown will be unavoidably corrupted by the zeroes that preceed it, resulting in shortcomings similar to those of the Fourier-domain strategy in \cite{Cabero:2017avf}.

The likelihood of Eq.~\eqref{eq:lnlike_td} is the key element required to obtain a Bayesian posterior on the parameters $\theta$ of a potential signal in the data, $p(\theta\mid d)$.
By Bayes' theorem,
\begin{equation}
p(\theta \mid d) \propto p(d \mid s_\theta)\, p(\theta)\, ,
\end{equation}
where $s_\theta \equiv s(t;\theta)$ is the signal corresponding to parameters $\theta$, $p(d\mid s_\theta)$ is the likelihood given by Eq.~\eqref{eq:lnlike_td}, and $p(\theta)$ is the prior.
If multiple, independent data streams are available (e.g., from different GW detectors), the posterior and likelihood generalize trivially, such that
\begin{equation} \label{eq:lnlike_multi}
p(\theta \mid \{ d_I \}) \propto p(\theta) \prod p(d_I \mid s_\theta^I)\, ,
\end{equation}
where $d_I$ represents each independent data stream, $s^I_\theta$ is the signal expected at the $I$\ts{th} stream given parameters $\theta$, and the product is over all values of $I$.

Just as with the likelihood, it is convenient to have a notion of SNR that is able to account only for data from some truncation point onward.
Such a notion is useful to quantify the amount of signal power in the ringdown for a given signal in the data, and to make projections about ringdown detectability.

Equation \eqref{eq:noise_pdf_td} provides a natural definition for an inner product incorporating data only after $t_{\rm start}$, for any two time series $x_i = x(t_i)$ and $y_i = y(t_i)$:
\begin{equation} \label{eq:ip1}
\langle x \mid y \rangle_{t_0} \equiv \sum_{i,j=0}^{N-1} x_i\, \cov^{-1}_{ij}\, y_j = \sum_{i,j=0}^{N-1} \white{x}_i \white{y}_j\, ,
\end{equation}
where we extend the sum $N-1$ steps up to our last data point at time $t_{N-1} = t_0 + (N-1) \Delta t = t_0 + T - \Delta t$.
We may use the notion of distance induced by Eq.~\eqref{eq:ip1} to define the \emph{time-bounded optimal SNR}, $\snropt[t_0]$, of an arbitrary timeseries, $x_i$, as the norm of the timeseries, namely
\begin{equation} \label{eq:snr}
\snropt[t_0] \equiv ||x||_{t_0} \equiv \left\langle x \mid x \right\rangle_{t_0}^{1/2}\, .
\end{equation}
By the same token, the \emph{time-bounded matched-filter SNR}, $\snrmf[t_0]$, of a signal, $h_i$, in some noisy data, $d_i=s_i+n_i$, is
\begin{equation} \label{eq:snrmf}
\snrmf[t_0] \equiv {\left\langle s \mid d \right\rangle_{t_0}}{\left\langle s \mid s \right\rangle^{-1/2}_{t_0}}\, .
\end{equation}
We will usually drop simplify notation by dropping the ``$[t_0]$'' specifier when the chosen truncation time is clear from context.
With these definitions in place, Eq.~\eqref{eq:lnlike_td} can be written succinctly as:
$
\ln P(d \mid s) = -\frac{1}{2}  || d-s ||^2_{t_0} + \text{const.}
$

For multiple detectors, Eq.~\eqref{eq:lnlike_multi} implies that the expressions for the SNR generalize as usual to the network SNR,
\begin{equation} \label{eq:snr_net}
\mathrm{SNR}_{\rm net}[t_0] \equiv \sqrt{\sum \left(\mathrm{SNR}_I[t_0 + \delta t_I]\right)^2}\, ,
\end{equation}
where the sum is over each instrument $I$, and $\mathrm{SNR}_I[t_0 + \delta t_I]$ is the SNR at the $I$\ts{th} instrument (matched filter or optimal), evaluated at respective truncation times shifted by $\delta t_I$ with respect to $t_0$.
This global reference time can be chosen to be the truncation time at one of the detectors (for which $\delta t_I$ would vanish by definition), or some arbitrary reference like the geocenter.

If $\cov$ is derived from a circular ACF as in Eq.~\eqref{eq:psd_rho}, then Eq.~\eqref{eq:ip1} is equivalent to the usual Fourier-domain inner product implied by Eq.~\eqref{eq:lnlike_fd}, which weights data by the PSD (see, e.g., \cite{Creighton2012}).
Accordingly, Eqs.~\eqref{eq:snr} and \eqref{eq:snrmf} reduce to the usual quantities in that case, assuming the analysis segment is extended to encompass the full signal.

\subsection{Modified frequency-domain likelihood}
\label{sec:fd-inpainting}

An alternative to our approach above is to circumvent truncation issues while remaining in the frequency domain via nontrivial modifications to the likelihood as proposed in \citep{Capano2021,Zackay2019}.
This approach destroys the diagonality of the Fourier covariance matrix, thus reducing the incentive for switching to the frequency domain in the first place, and generally increases the computational cost.
However, it is a valid strategy that is formally equivalent to ours, so we review it here for completeness.

Consider a long stretch of data with enough samples before $t_0$ to permit windowing so that the complete stretch can be treated as periodic, with a cyclic, stationary noise covariance matrix $C$; let $M$ be the total length of this segment, with $N$ samples corresponding to $t \geq t_0$ preceded by $I \equiv M - N$ samples with $t < t_0$.
The frequency domain noise covariance for the $M$-long segment, $\tilde{C}$, is diagonal as in Eq.~\eqref{eq:cov_fd}, but the usual likelihood function of Eq.~\eqref{eq:lnlike_fd} is not appropriate because it depends on data and signal values for $t < t_0$, as detailed in the introduction to this section.
However, we can eliminate this dependence by modifying the time-domain noise covariance matrix, adding large terms along the diagonal for entries $i < I$ corresponding to times before the desired analysis start time:
\begin{equation}
  C \to C + a \sum_{i < I} e_i e_i^T,
\end{equation}
where $e_i$ is the $i$\ts{th} time-domain basis vector, with zeros in every entry except $i$ (i.e., with the $k$\ts{th} component given by $\delta_{ik}$), and we will take $a \to \infty$ at the end of our calculation to render the modified likelihood completely insensitive to data and signal entries $i < I$.
The effect of this manipulation is to artificially impose infinite uncertainty about the noise contribution to the data before $t_0$.

With the modified covariance matrix, the likelihood becomes
\begin{align}
  \ln P\left( d \mid s \right) &= -\frac{1}{2} \left( d - s \right)^T \left( C + a \sum_{i < I} e_i e_i^T\right)^{-1} \left( d - s \right) \nonumber\\
  &= -\frac{1}{2} \left( \tilde{d} - \tilde{s} \right)^\dagger \left( \tilde{C} + a \sum_{i<I} \tilde{e}_i \tilde{e}_i^\dagger \right)^{-1} \left( \tilde{d} - \tilde{s} \right),
\end{align}
where $\tilde{e}_i$ is the Fourier transform of $e_i$ (not the $i$th basis vector in the Fourier domain), $T$ denotes a transpose and $\dagger$ a Hermitian conjugate.
The additional terms in the frequency domain covariance are not diagonal, so the
modified likelihood cannot be computed in $O(M)$ time in the frequency domain.

Consider the singular value decomposition (SVD) of the extra terms:
\begin{equation}
a \sum_{i<I} \tilde{e}_i \tilde{e}_i^\dagger = U \Sigma V^\dagger
\end{equation}
where $\Sigma$ is diagonal $I \times I$ and $U$ and $V$ are unitary $M \times
I$.  We have $\Sigma = O(a)$ so $\Sigma^{-1} \to 0$ as $a \to \infty$.  The
Woodbury matrix identity states that
\begin{multline}
  \left( \tilde{C} + a \sum_{i<I} \tilde{e}_i \tilde{e}_i^\dagger \right)^{-1} \\ = \tilde{C}^{-1} - \tilde{C}^{-1} U \left( \Sigma^{-1} + V^\dagger \tilde{C}^{-1} U \right)^{-1} V^\dagger \tilde{C}^{-1}
\end{multline}
Taking $a \to \infty$ introduces infinite uncertainty about the noise
contribution to times $t_i$ with $i < I$, eliminating any dependence in the
likelihood on the data or signal at these times.  The effect is to eliminate the
$\Sigma^{-1}$ term:
\begin{multline}
  \lim_{a \to \infty} \left( \tilde{C} + a \sum_{i<I} \tilde{e}_i \tilde{e}_i^\dagger \right)^{-1} \\ = \tilde{C}^{-1} - \tilde{C}^{-1} U \left( V^\dagger \tilde{C}^{-1} U \right)^{-1} V^\dagger \tilde{C}^{-1}.
\end{multline}
The modified likelihood thus becomes
\begin{equation}
  \label{eq:fd-likelihood}
  \ln P\left( d \mid s \right) = -\frac{1}{2} \left( \tilde{d} - \tilde{s} \right)^\dagger \tilde{C}^{-1} F \left( \tilde{d} - \tilde{s} \right)
\end{equation}
where $F$ is a projection matrix ($F^2 = F$) defined by \citep{Zackay2019}
\begin{equation}
  F = 1 - U\left( V^\dagger \tilde{C}^{-1} U \right)^{-1} V^\dagger \tilde{C}^{-1}.
\end{equation}
The effect of $F$ is to project out the parts of the data and signal that would
render the likelihood sensitive to samples before the desired start time.  If
used with a covariance matrix $C_{ij}$ that agrees with the matrix used in Eq.\
\eqref{eq:lnlike_td} for $i, j \geq I$, the likelihood functions will be
identical.

The likelihood function in Eq.\ \eqref{eq:fd-likelihood} uses the noise
covariance in the frequency domain $\tilde{C}$, much like standard parameter
estimation analyses \citep{Veitch2015}, except the matrix is no longer diagonal.
However, the frequency-domain method
requires that the data be circular; in practice, this means that there must be
enough padding before the start of the ``analysis segment'' and after the
effective end of the ringdown to implement a tapering of the data to enforce the
circularity assumption that renders $\tilde{C}$ diagonal.  This means that $I$
must be fairly large (perhaps even comparable to the total length of the initial data
segment, $M$, if the ringdown portion is short compared to the necessary smooth
taper).

The computational cost of the method at startup is $I^3$ to compute the
SVD and $\left( V^\dagger \tilde{C}^{-1} U \right)^{-1}$, $IM$ to compute the
matrix multiplications, and $M \log M$ to compute the PSD that determines
$\tilde{C}^{-1}$; at each iteration the cost is $I M + I^2$ to implement the
projection $F$.  This is compared to $N^3$ at startup and $N^2$ for the
pure time-domain likelihood in Eq.\ \eqref{eq:lnlike_td}.  For $I \ll
M$ (short taper, long signal) the frequency-domain likelihood will be more
efficient; for $I \sim M$ (long
taper, short signal), usually the case for ringdown signals, the time-domain likelihood is more efficient.

\subsection{Implementation}
\label{sec:imp}

We will use the time domain likelihood of Eq.~\eqref{eq:lnlike_td}, as we did in \cite{Isi:2019aib}.
Besides this function, we also need to specify priors for the signal parameters.
In the following, we apply uniform priors almost exclusively, but the quantities for which the prior is uniform depends on the specific parameterization used, as detailed in each respective section.
For instance, when using the elliptical QNM template of Eq.~\eqref{eq:h_j}, we will apply flat priors on the parameters in Eq.~\eqref{eq:params_gr} or, if deviations from Kerr are allowed, Eq.~\eqref{eq:params_nongr}.
The elliptical-mode parameterization has the advantage of allowing us to directly define priors for the overall amplitude of each QNM, while remaining agnostic about polarization content; this would not be straightforward had we chosen a different parameterization, e.g., in terms of the right- and left-handed polarization amplitudes $|C_{\ell\pm m n}'|$ in Eq.~\eqref{eq:ellip}, as in \cite{Abbott:2020tfl,Abbott:2020mjq,Abbott:2020jks}: a uniform prior on those quantities would induce a triangular prior on the overall QNM amplitude $A_{\ell |m| n}$, disfavoring $A_{\ell |m| n}=0$.

In order to estimate posterior densities, it is straightforward to implement the time-domain likelihood of Eq.~\eqref{eq:lnlike_td} to work with any standard Markov chain Monte Carlo (MCMC) sampler.
We choose to use the \textsc{Stan} platform \cite{Stan,Carpenter:2017} in order to take advantage of its efficient no-U-turn Hamiltonian Monte Carlo (HMC) algorithm \cite{Neal:2011,Hoffman:2014no}.
Whatever the sampler used, it is often the case that parameterizations of the likelihood that are most conceptually meaningful are ill suited for sampling purposes.
This is certainly true for our elliptical-mode template, Eq.~\eqref{eq:h_j}: although the $\{A_j, \epsilon_j, \theta_j, \phi_j\}$ quantities are a natural choice due to their self-evident meaning, they are cumbersome to sample directly.
For example, the $\theta_j$ parameter becomes meaningless when $\epsilon = \pm 1$, resulting in an ambiguity that carries over also to $\phi_j$ (cf., Fig.~\ref{fig:ellipse}).
This can be circumvented by reparameterizing the likelihood in terms of more convenient variables, and applying the requisite Jacobian.
When sampling with the elliptical-mode template, we have found that efficiency is optimized with the reparameterization offered in App.~\ref{app:jacobian}.

To study measurements in colored noise, we synthesize data from the Advanced LIGO (aLIGO) design PSD \cite{aLIGOpsd}; when studying multiple-detector measurements, we use that PSD for all instruments to avoid confounding factors arising from a heterogeneous detector network.
Unless otherwise stated, we simulate the treatment of real data by first generating 4096 s of time-domain noise sampled at 8192 Hz, then bandpassing between $[20,\, 2048]\, \mathrm{Hz}$ (an appropriate range for the signals we study), downsampling to a sample rate 4096 Hz and, finally, truncating at $t_0$; we always inject signals \emph{before} conditioning.
As we would with real data, we estimate the ACF from a long stretch of signal-free noise samples.
We speed up the likelihood computation through the Cholesky decomposition of $\cov$, Eq.~\eqref{eq:cholesky}, but one could also take advantage of algorithms tailored to Toeplitz systems, like Levinson recursion \cite{Levinson1946,Durbin1960}.

\section{Analyzing damped sinusoids}
\label{sec:ds}

We first demonstrate some basic features of ringdown analyses under our framework, starting from simple superpositions of damped sinusoids without the complexities of the template introduced in Sec.~\ref{sec:template}.
For a single data stream at times $t \geq t_0$, the simplest template that captures the features in which we are interested is a generic superposition of some $\nmode$ damped sinusoids,
\begin{equation} \label{eq:ds}
s(t) = \sum_{n=0}^{\nmode-1} A_n e^{-(t-t_{\rm ref})/\tau_n}\cos[2\pi f_n (t - t_{\rm ref}) + \phi_n] \, ,
\end{equation}
with frequencies $f_n$ and damping times $\tau_n$, as well as amplitudes $A_n$ and phases $\phi_n$ defined at some arbitrary reference time $t_{\rm ref}$ (unless otherwise stated, we will set $t_{\rm ref} = t_0$ by default).
To prevent label switching degeneracies, it is helpful to impose an ordering on $f_n$ or $\tau_n$; since BH overtones will be the main focus of our analysis (Sec.~\ref{sec:analysis}), in this paper we define $\tau_n > \tau_{n+1}$.
(This can be implemented through a coordinate transformation to an un-constrained space \cite{Buscicchio:2019rir,Stan}.)
Even for enhanced models that account for multiple polarizations and detector sensitivities as in Eq.~\eqref{eq:s}, the goal will always be to infer some set of $\{f_n, \tau_n, A_n, \phi_n\}$ from the data using the acyclic time-domain likelihood of Eq.~\eqref{eq:lnlike_td}.

\begin{figure}
\includegraphics[width=\columnwidth]{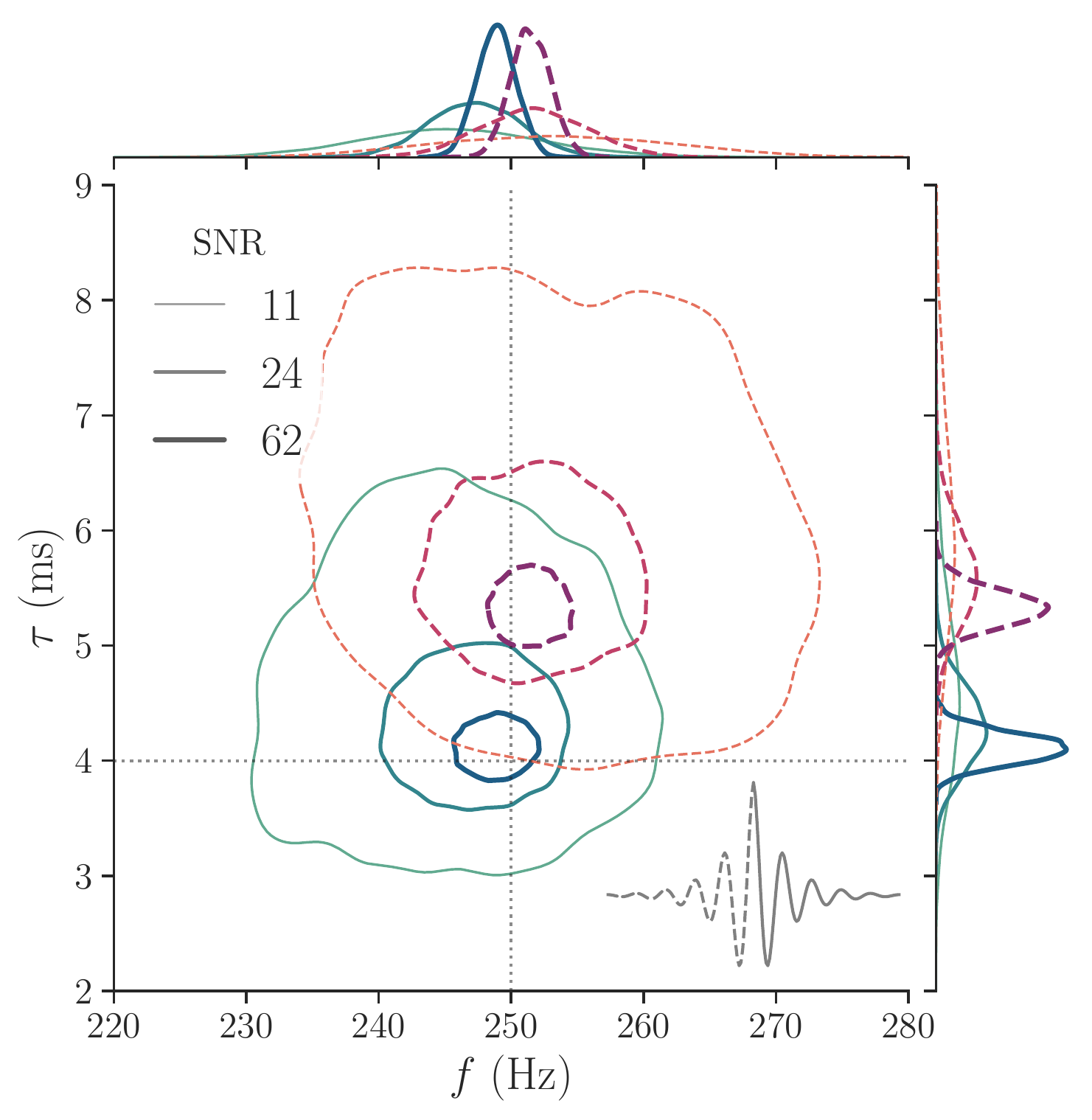}
\caption{Parameter recovery of our method (solid green) versus the our implementation of the original LIGO-Virgo method in \cite{TheLIGOScientific:2016src} (dashed red), on a ringup-ringdown injection (bottom right diagram).
From the peak onward (solid trace in diagram), the injection is a simple damped sinusoid with frequency (abscissa) and damping time (ordinate) indicated by the dotted lines.
Contours represent the 90\%-credible levels obtained from each method for varying injected postpeak SNR (line thickness and tone), Eq.~\eqref{eq:snrmf}; the top and right panels show the corresponding marginals.
The red distributions are biased by the prepeak data, which is coupled into the analysis by the frequency-domain analysis method as discussed in the introduction to Sec.~\ref{sec:inference}.)
}
\label{fig:lvc_performance}
\end{figure}

\subsection{Method efficacy}
\label{sec:ds:efficacy}

By construction, our template is left undefined for times before $t_0$.
This means that we can apply it even if we do not expect data at those times to be at all described by Eq.~\eqref{eq:ds}.
We demonstrate this in Fig.~\ref{fig:lvc_performance}, where we synthesize noise from the aLIGO design PSD \cite{aLIGOpsd}, and add it to a signal conforming to Eq.~\eqref{eq:ds} only after some $t_0$.
We make use of a ``ringup-ringdown'' injection,
\begin{equation} \label{eq:rurd}
s_{\rm inj}(t) =  A\, e^{-|t-t_0|/\tau} \cos[2\pi f (t - t_0)] \, ,
\end{equation}
for all times before and after some arbitrary $t_0$, with $f=250\, {\rm Hz}$ and $\tau=4\, {\rm ms}$ (consistent with the fundamental mode of the GW150914 remnant \cite{TheLIGOScientific:2016src,Isi:2019aib}), and $A$ chosen to yield the post-peak SNRs indicated by the legend.
We condition the data as we normally would (including filtering and downsampling), and discard times before $t_0$ to carry out a Bayesian analysis based on Eq.~\eqref{eq:lnlike_td}, with a single damped sinusoid as a template.
The green distributions in Fig.~\ref{fig:lvc_performance} show that we are able to correctly infer the properties of the injected signal, even though the injection departs from Eq.~\eqref{eq:ds} before $t_0$.
On the other hand, a different method based on Eq.~\eqref{eq:lnlike_fd} shows a bias induced by the pre-$t_0$ data (see App.~\ref{app:lvc} for details).

The example in Fig.~\ref{fig:lvc_performance} is designed to concretely illustrate the need for our time-domain formalism, with the ringup-ringdown injection as a proxy for the full inspiral-merger-ringdown signal.
When other analysis strategies are adopted, the nature and magnitude of the bias depends on the specific properties of the noise and signal, as well as the details of the method implementation (including conditioning).
In general, we may expect the bias to be more pronounced when targeting short-lived modes that are only visible near the truncation time, where contamination from pre-$t_0$ data will tend to be greatest.
This is why we found the time-domain formalism to be indispensable when searching for overtones in the GW150914 ringdown \cite{Isi:2019aib}.

To demonstrate the robustness of our method, we must go beyond the concrete example in Fig.~\ref{fig:lvc_performance} to show that we can recover arbitrary signals without systematic bias, as long as they conform to Eq.~\eqref{eq:ds} after $t_0$ and no matter what came before it.
We do this through a probability-probability (PP) test \cite{Cook2006,Veitch2015,Talts2018,Romero-Shaw2020}, also called ``simulation based calibration.''  We simulate the measurement of a large number of signals drawn from our prior, with random instantiations of noise drawn from our likelihood.
If our method is working properly, we expect that the resulting posteriors will be such that the true values are recovered with X\% credibility for X\% of the simulations.
Importantly, we do not only simulate times $t \geq t_0$, but instead make use of ringup-ringdown templates similar to Eq.~\eqref{eq:rurd}, and put them through the same conditioning process that we would apply to real data (Sec.~\ref{sec:imp}).
This ensures that we are sensitive to any potential degradation induced by the bandpassing and downsampling filters.
Therefore, the PP test allows us to validate the entirety of our setup: conditioning, truncation, and likelihood treatment.

\begin{figure}
\includegraphics[width=\columnwidth]{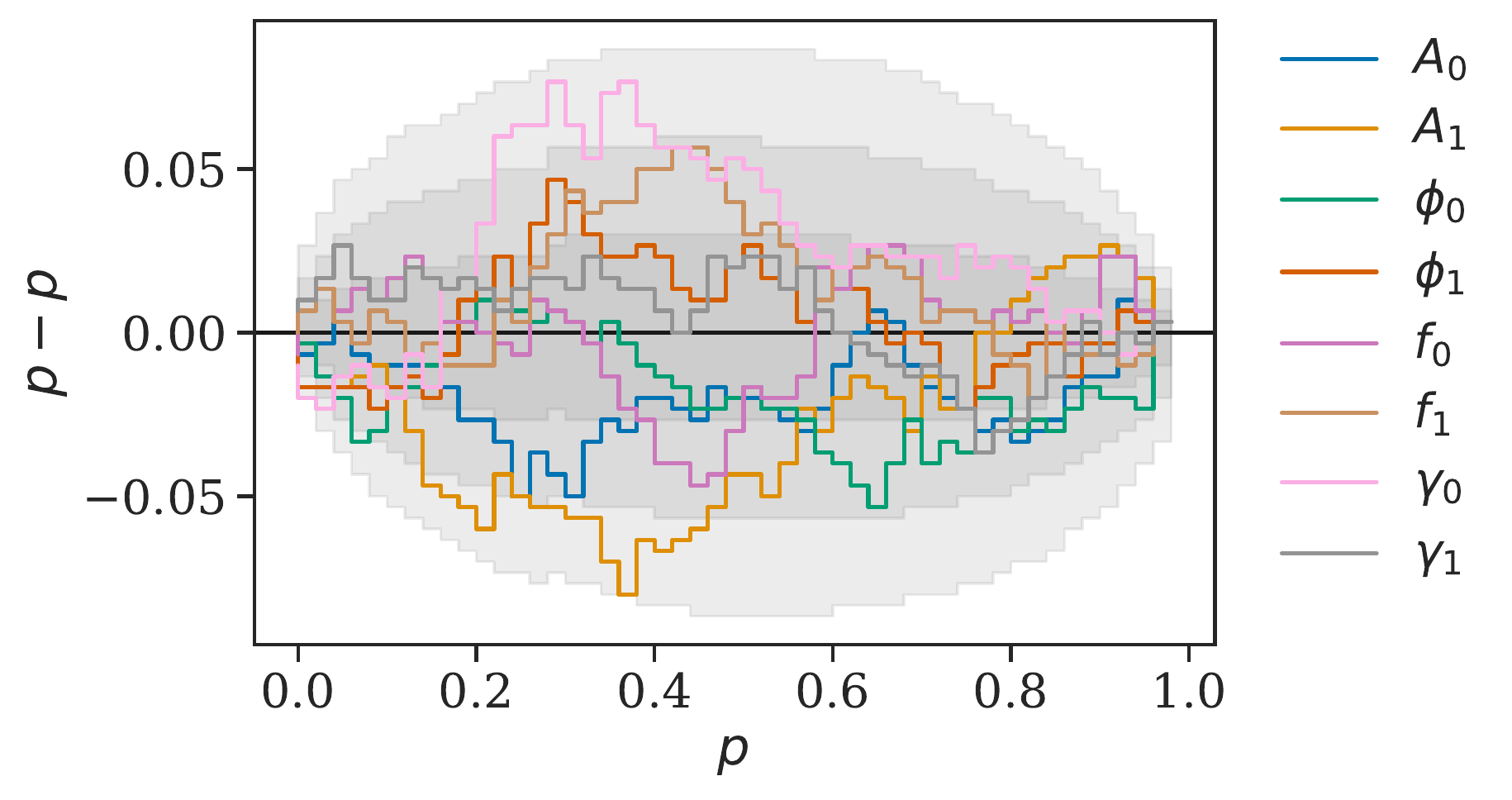}
\caption{PP plot. Deviation from uniformity in the recovered quantiles for {300} injections drawn from the prior, as described in the main text. Gray bands mark the expected $1\sigma$, $2\sigma$ and $3\sigma$ variations under the null hypothesis.}
\label{fig:ds_pp}
\end{figure}

Since we will later focus mostly on two-mode models, we pick $D=2$ for this test and generalize Eq.~\eqref{eq:rurd} so that our injections are now described by
\begin{equation}
s_{\rm inj}(t) =  \sum_n A_n\, e^{-|t-t_0|/\tilde{\tau}_n(t)} \cos[2\pi f_n (t-t_0) + \phi_n] \, ,
\end{equation}
where we sum over two modes ($n = 0,\, 1$), and allow the pre-$t_0$ $e$-folding time $\tau'_n$ to differ from the regular post-$t_0$ value $\tau_n$ by defining the piecewise auxiliary function $\tilde{\tau}_n(t)\equiv \tau'_n \Theta(t_0-t) + \tau_n\Theta(t-t_0)$, in terms of Heaviside steps $\Theta$.
This choice is designed to provide a variety of pre-$t_0$ morphologies, while preserving smoothness across the $t_0$ boundary.
For each simulation, we synthesize time-domain noise based on the aLIGO design PSD, and draw injection parameters $A_n \sim \mathcal{U}(0,3\times10^{-21})$, $\phi_n \sim \mathcal{U}(0,2\pi)$, $f_n/\mathrm{Hz} \sim \mathcal{U}(635,1285)$, $\ln(\gamma_n/\mathrm{Hz}) \sim \mathcal{U}(\ln 635, \ln 1285)$ and same for $\ln(\gamma_n'/\mathrm{Hz})$, for both $n=0,1$, and enforce $\tau_0 > \tau_1$ with no imposed ordering on $\tau_{0/1}'$; these distributions match our prior---except for the $\gamma_n'\equiv1/\tau_n'$ parameters, which only control the pre-$t_0$ morphology of the injection and, thus, are not part of our recovery template, Eq.~\eqref{eq:ds}.
The resulting posteriors conform to our statistical expectations, as shown by the agreement between the measurements (colored lines) with our statistical expectation (gray bands) in Fig.~\ref{fig:ds_pp}.

\subsection{Detecting modes}
\label{sec:ds:det}

\begin{figure}
\includegraphics[width=\columnwidth]{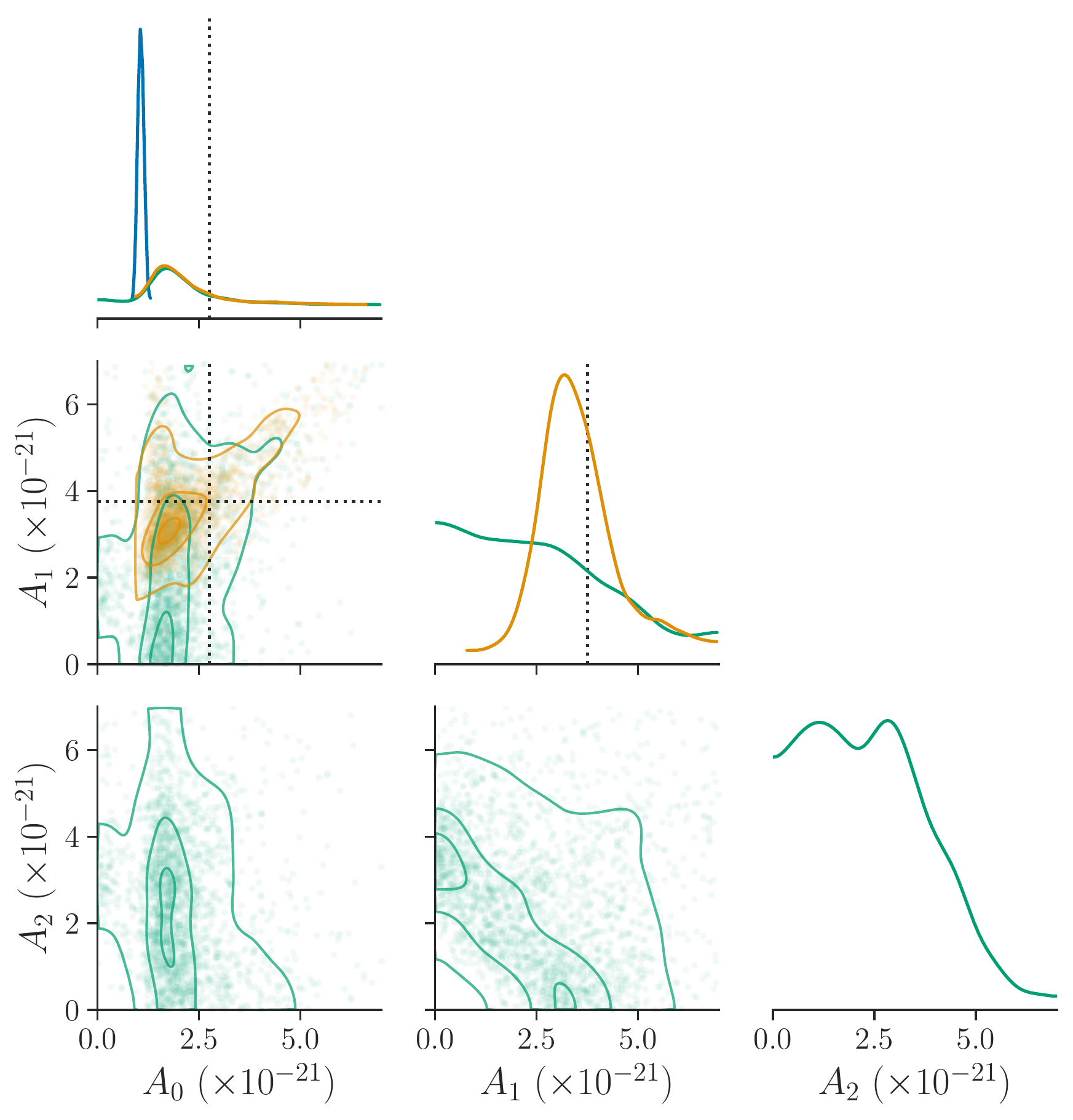}
\caption{Amplitude posteriors from the analysis of two damped sinusoids ($\nmode_{\rm true} = 2$) in noisy data, using templates with a varying number of modes in Eq.~\eqref{eq:ds}:
$\nmode=1$ ({blue}), $\nmode=2$ ({orange}), $\nmode=3$ ({green}).
The diagonal displays amplitude marginals, while the upper and lower corners respectively show credible levels ({90\%, 50\%, 10\%}) and individual samples; dotted lines mark the true values (Table~\ref{tab:ds_detection}).
Noise follows the aLIGO design PSD, resulting in an injected matched filter SNR of {20}, as defined in Eq.~\eqref{eq:snrmf}.
The posterior for $\nmode=3$ allows some of the amplitudes to vanish, so we conclude $\nmode_{\rm best}=2$.
}
\label{fig:ds_detection}
\end{figure}

A template like Eq.~\eqref{eq:ds} can accommodate an arbitrary number of damped sinusoids $\nmode$.  Actual BBH merger ringdowns in GR excite a (countably) infinite number of modes.
Determining how many modes are \emph{necessary} to adequately describe a given set of data is a model selection problem: to identify the number of modes that are clearly detected in a fully Bayesian setting, we could compare the relative evidences of models made up of different combinations of modes.
Formally, this procedure is encapsulated by the Bayesian odds between models, or just their Bayes factors if we have no \emph{a priori} preference between the different alternatives.
However, Bayes factors are only meaningful insofar as our priors actually represent our belief about the expected distribution of parameter values.
Assigning compelling priors to the ringdown parameters can be quite challenging, especially when it comes to the damped sinusoid amplitudes and relative phases.  Therefore we do not advocate for using relative evidences between models to choose the ``correct'' number of modes in a signal.  The procedure we outline below is considerably less sensitive to prior choices, depending only on how those choices change the \emph{shape} of the inferred posterior, unlike the model evidences that depend on the prior in parameter regions without any posterior support.

We propose to determine the number of modes $\nmode_{\rm best}$ that the data require by demanding that:
\begin{enumerate}
\item if $\nmode = \nmode_{\rm best}$, the posterior indicate $A_n > 0$ with a reasonable degree of certainty (we use 90\% credibility in this work), for \emph{all} $0 \leq n < \nmode$; while,
\item if $\nmode = \nmode_{\rm best}+1$, this is no longer be the case.
\end{enumerate}
In other words, we stop adding modes to the template once the only effect of doing so is to introduce degeneracies.
This procedure relies on the existence of an intrinsic hierarchy of modes, so that there is a natural order in which to progressively add them to the model.
For example, modes with shorter durations are generally harder to measure, so we may add modes in order of decreasing damping time (suitable for a series of overtones).
Alternatively, we can add modes to the model in order of expected amplitude (suitable for a series of angular harmonics).

We illustrate this in Fig.~\ref{fig:ds_detection}, where we inject two damped sinusoids ($\nmode_{\rm true}=2$) into Gaussian noise and carry out measurements with $\nmode = 1, 2, 3$ in Eq.~\eqref{eq:ds}.
The posterior for a single-mode template ({blue} distribution, top left) confidently indicates that the data contain a signal, inconsistent with zero amplitude.
Including a second damped sinusoid in the template results in a posterior ({orange} distribution) that is consistent with the injection, and that offers support for neither $A_0 = 0$ nor $A_1 = 0$.
This stops being true with the addition of a third mode, in which case the posterior ({green} distribution) allows any \emph{one} of the amplitudes to vanish, as long as the other two do not.
The fact that no two amplitudes may vanish simultaneously is reflected in the characteristic ``arching'' of the 2D joint marginals for each amplitude pair, excluding $A_n = A_m =0$ for any choice of $n \neq m$.
At this point, it is no longer helpful to continue adding modes to the template, so we conclude $\nmode_{\rm best} = 2$ for this example.

\begin{figure}
\includegraphics[width=\columnwidth]{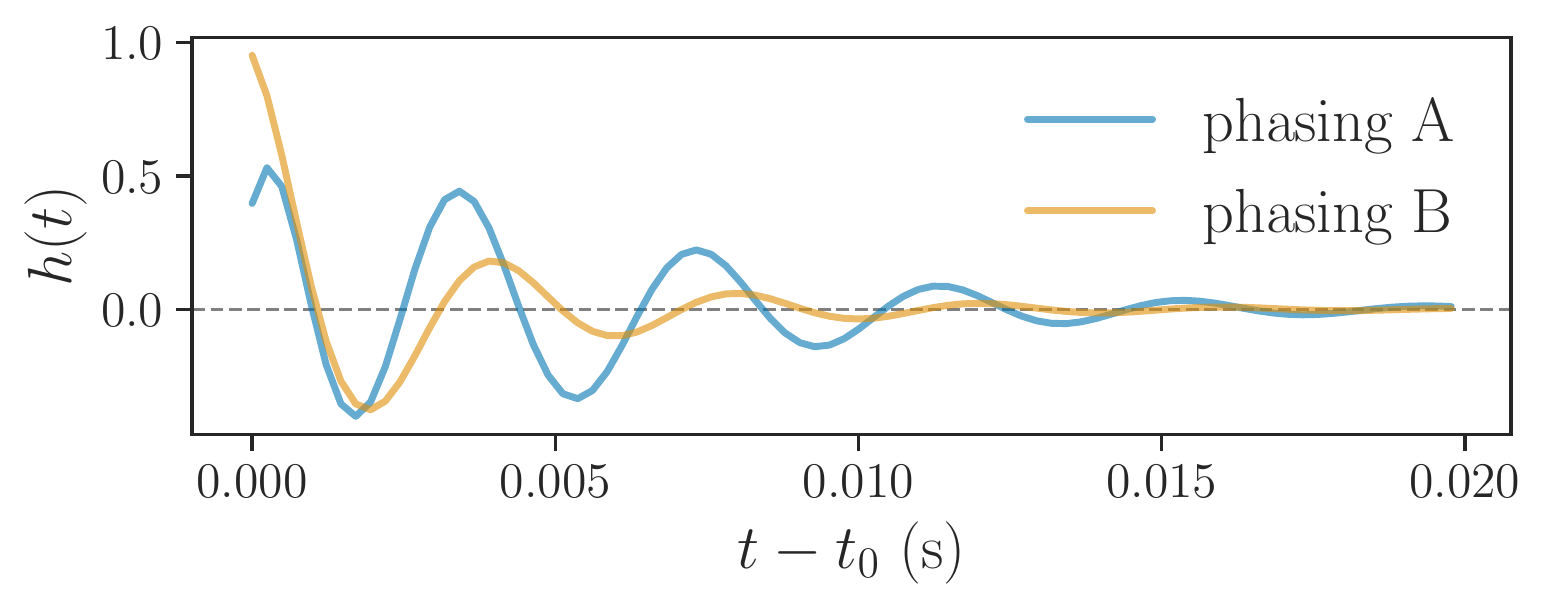}
\caption{Effect of phasing on mode detectability, illustrated through a pair of $D=2$ signals sharing a spectrum $f_n, \tau_n$, and mode-amplitude ratio $A_1/A_0$, but with different mode phases $\phi_n$. Although both signals integrate to the same SNR, the $n=1$ mode is down-weighted in example A ({blue}) relative to example B ({orange}), potentially making it harder to detect.}
\label{fig:phasing_example}
\end{figure}

The number of modes required by the data $\nmode_{\rm best}$ need not correspond to the true number $\nmode_{\rm true}$, and it certainly will not in the analysis of true (as opposed to synthetic) BH ringdowns, for which $\nmode_{\rm true} = \infty$ always.
Since $\nmode_{\rm best}$ is the number of \emph{discernible} modes, it will depend on the overall SNR.
It will also be contingent on the specific characteristics of the signal (particularly the relative mode phases in the true signal) and the specific noise realization.
Any of those factors may impact the features of the $\nmode$-dimensional amplitude posterior, and therefore our determination of $\nmode_{\rm best}$.
For instance, identifying a critically damped ($Q=1/2$) or overdamped ($Q < 1/2$) mode would require exceptionally high SNR, since such modes do not exhibit any cycles.
Moreover, even for underdamped signals, the specific phasing determines the distribution of SNR across modes, and can thus have a big impact on $\nmode_{\rm best}$ (Fig.~\ref{fig:phasing_example}).

\begin{figure}
  \includegraphics[width=\columnwidth]{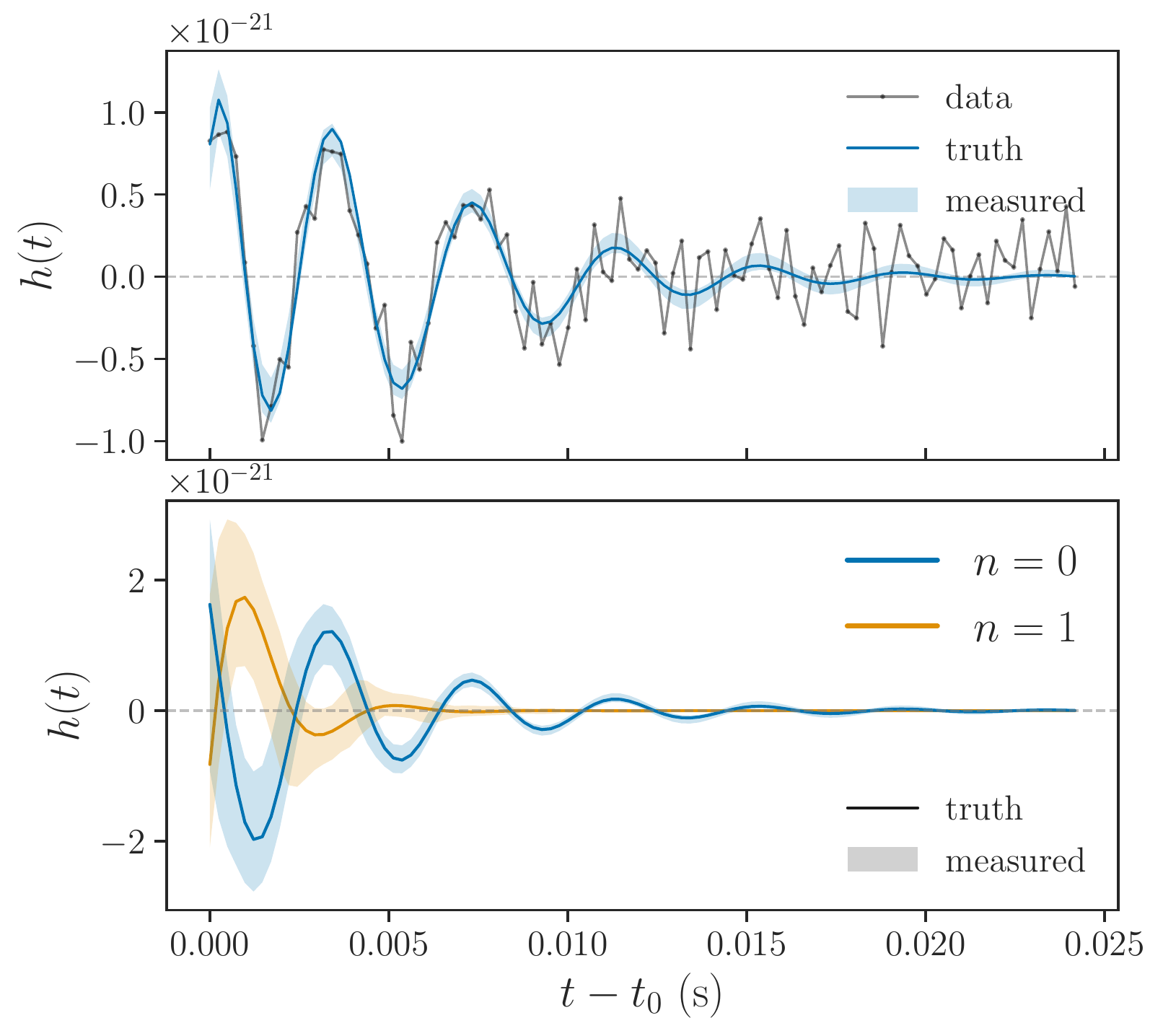}
  \caption{Reconstruction of two damped sinusoids injected into simulated aLIGO noise, as recovered using a model with $\nmode=2$ modes ({orange} distribution in Fig.~\ref{fig:ds_detection}).
The top panel shows the injected signal ({blue line}) and the 90\%-credible reconstruction ({blue shading}), laid over the noisy data ({black}).
The bottom panel shows the two individual modes making up the injection (lines), and their respective reconstructions (shading).
The true parameters are based on a GW150914-like remnant BH (Table~\ref{tab:ds_detection});
the shortest-lived mode ($n=1$) exhibits little over one full cycle.}
  \label{fig:analytic_strain_M2}
\end{figure}

Having identified certain number of damped sinusoids in the data, we may reconstruct them (Fig.~\ref{fig:analytic_strain_M2}) and quantify their properties.
Our definition of $\nmode_{\rm best}$ is designed such that this a meaningful thing to do: if the data favored $A_n=0$ for some $n$, then the posterior on the corresponding $f_n$ and $\tau_n$ would necessarily span the range of the prior, making constraints harder to interpret (e.g., the 90\%-credible interval would necessarily vary with the prior bounds, which are usually arbitrary).

\begin{table}[tb]
\caption{GW150914-like parameters injected in Fig.~\ref{fig:ds_detection}.}
\label{tab:ds_detection}
\begin{tabular}{l@{\qquad}l@{\qquad}l@{\qquad}l@{\qquad}l@{\quad}}
\toprule
         {\quad}$n$ &   $f_n$ & $\tau_n$ & $A_n$ & $\phi_n$ \\ \midrule
         {\quad}0 &   250 Hz & 4.0 ms & $3\times 10^{-21}$ & $5.34$ rad \\
         {\quad}1 &   245 Hz & 1.4 ms & $4\times 10^{-21}$ & $1.79$ rad \\ \bottomrule
\end{tabular}
\end{table}

\subsection{Characterizing modes}
\label{sec:ds:char}

\begin{figure}
\includegraphics[width=\columnwidth]{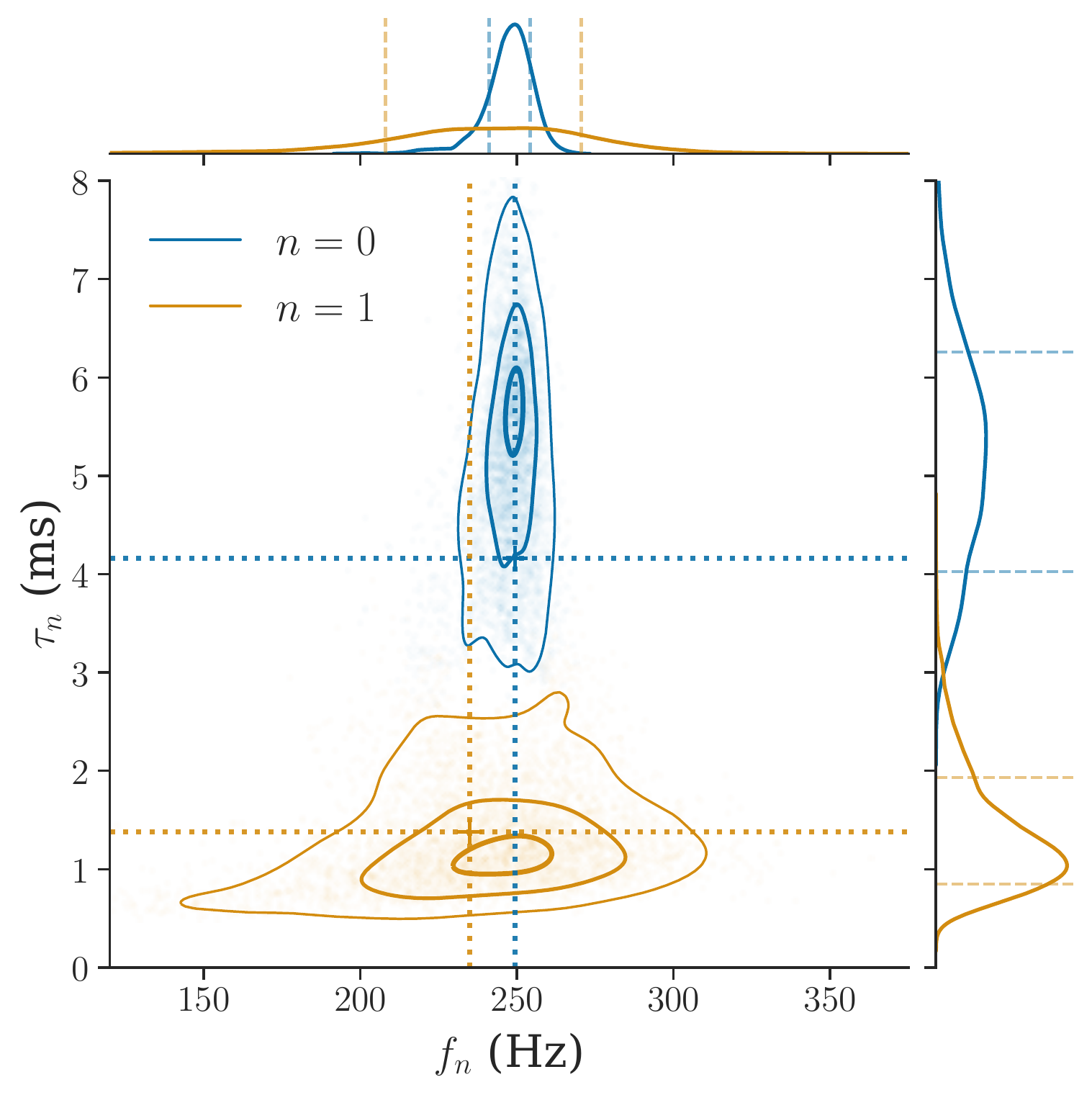}
\caption{Joint frequency (abscissa) and damping rate (ordinate) posterior for the longest- ({blue}) and shortest- ({orange}) lived damped sinusoids in the $\nmode=2$ analysis from Fig.~\ref{fig:ds_detection}.
Contours enclose {90\%, 50\%, and 10\%} of the probability mass;
crosses mark the injected values (Table~\ref{tab:ds_detection}).
For comparison to Eqs.~\eqref{eq:rayleigh_f} and \eqref{eq:rayleigh_tau}, we show the symmetric 68\%-credible ($1\sigma$) intervals for each marginal (dashed lines, top and right).
Note that this figure hides some of the information yielded by the Bayesian analysis, since it does not show the correlations between the $n=0$ and $n=1$ parameters.
}
\label{fig:ds_ftau}
\end{figure}

\subsubsection{Resolving modes}

Having a preference for $A_n > 0$ is also a {sufficient} condition for a meaningful measurement of $f_n$ and $\tau_n$ under a specific model.
This is in spite of claims to the contrary in the ringdown literature.
For instance, Refs.~\cite{Bhagwat:2019dtm,Forteza:2020hbw} presuppose that a frequency measurement in an $\nmode=2$ model is only possible if the $f_n$ are ``resolved,'' by which they mean that the frequencies satisfy
\begin{equation} \label{eq:rayleigh_f}
|f_0 - f_1| > \max(\sigma_{f_0}, \sigma_{f_1}) \, ,
\end{equation}
where $\sigma_{f_n}$ is the standard deviation of the $f_n$ posterior samples.
By the same token, constraining the damping times would only be possible if
\begin{equation} \label{eq:rayleigh_tau}
|\tau_0 - \tau_1| > \max(\sigma_{\tau_0}, \sigma_{\tau_1}) \, .
\end{equation}
An insistence that modes be spectrally separated in this way led Refs.~\cite{Bhagwat:2019dtm,Forteza:2020hbw} to conclude that a constraint of the overtone frequencies in a BH ringdown would require $\mathrm{SNR} \sim 100$, disagreeing with the results in \cite{Isi:2019aib,Abbott:2020jks}.

\begin{figure*}
  \includegraphics[width=\columnwidth]{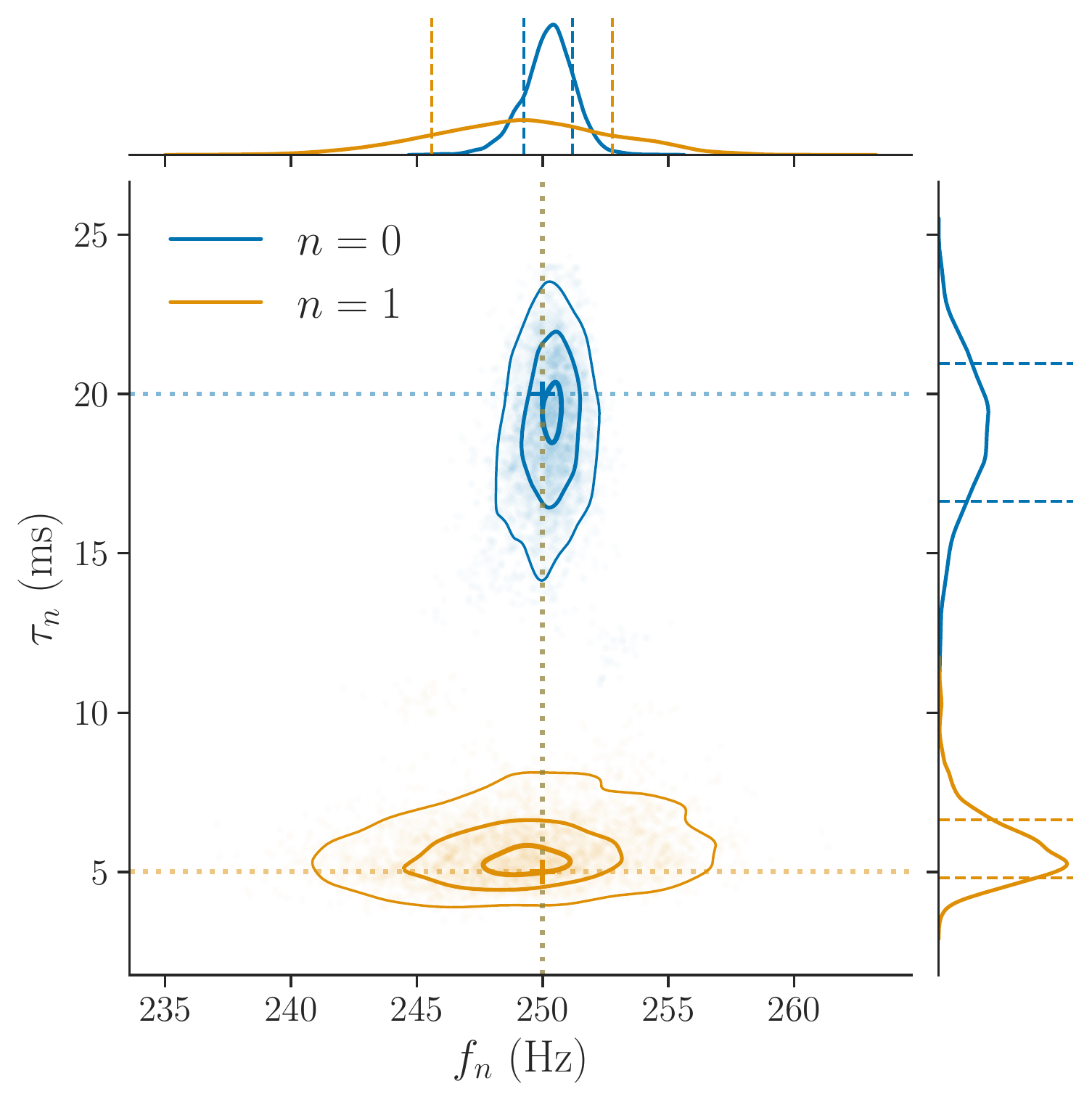}
  \includegraphics[width=\columnwidth]{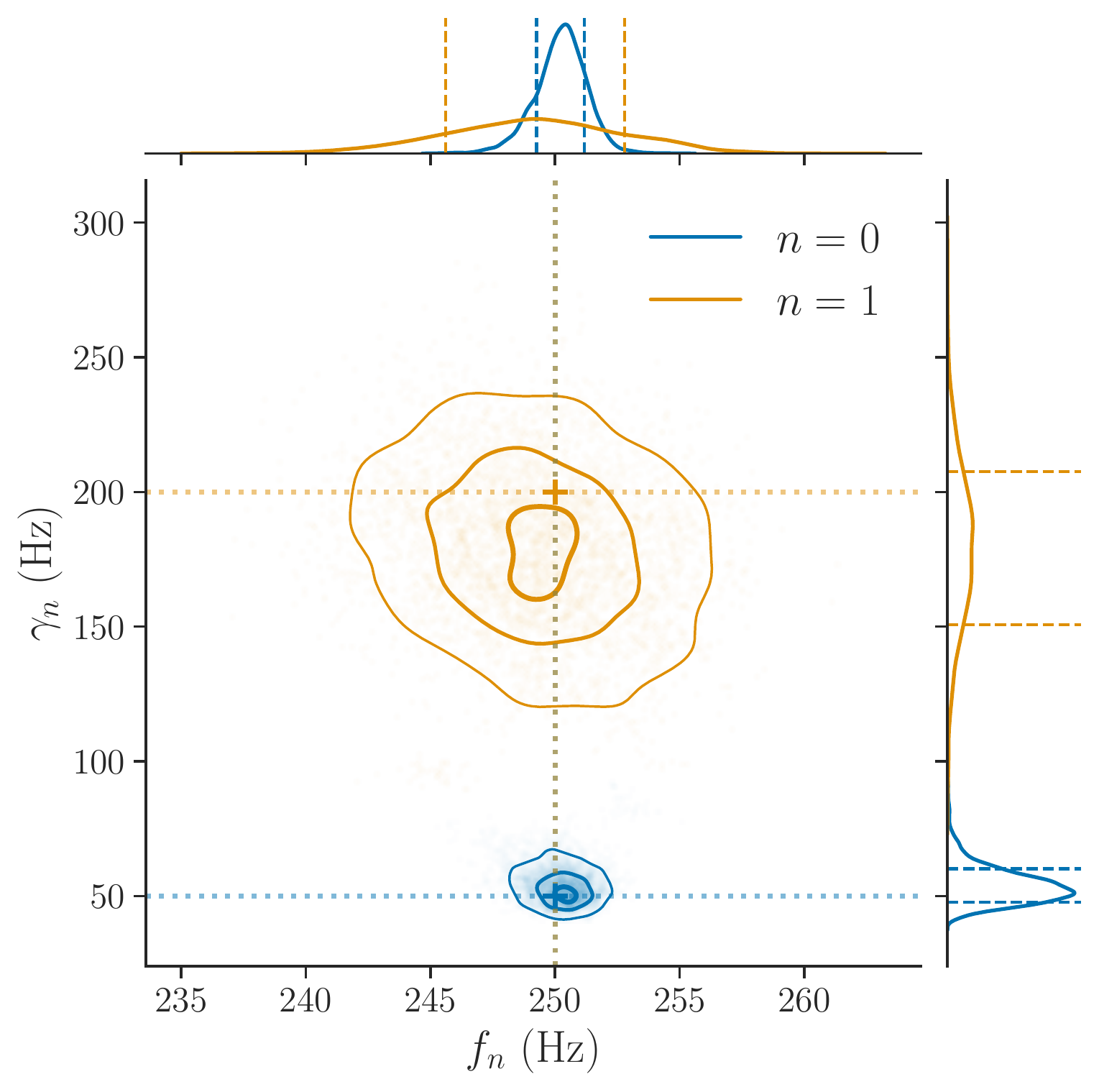}
  \caption{Reconstruction of signal made up of two damped sinusoids (color) with equal frequency $f_0=f_1 = 250\, \mathrm{Hz}$ (abscissa) but different damping times $\tau_0 = 20\, \mathrm{ms}$ and $\tau_1 = 50\, \mathrm{ms}$ (ordinate, left panel), that is, damping rates $\gamma_0 = 50\, \mathrm{Hz}$ and $\gamma_1 = 200\, \mathrm{Hz}$ (ordinate, right panel). Contours enclose {90\%, 50\% and 10\%} of the probability mass, while vertical lines over the marginals mark the {68\%} credible interval. From the left panel, it is clear that $\tau_1$ is more sharply constrained than $\tau_0$, which is what we expect if $\gamma_1$ is less constrained than $\gamma_0$ (right).}
  \label{fig:ds_ftau_extra}
\end{figure*}

However, the possibility of measuring $f_n$ or $\tau_n$ does not hinge on any such condition.
In a Bayesian analysis, all damped sinusoids in the template are modeled simultaneously and self consistently.
A $(2\times\nmode)$-dimensional posterior captures all potential degeneracies between the different $f_n$ and $\tau_n$.  Two modes will be distinguishable whenever the multidimensional posterior inferences about their frequencies and damping times do not overlap.  Non-overlapping frequency and damping time posteriors do not necessarily produce non-overlapping one-dimensional marginal posteriors on the frequency or damping time.  The correct condition for resolvability is that the separation between the frequency and damping time of two modes be large compared to the posterior \emph{covariance} of these quantities.  When the posteriors do not have significant correlations between $f$ and $\tau$ the distinguishability condition looks like
\begin{equation}
  \label{eq:correctRayleigh}
  r^2 \equiv \frac{\left(f_0 - f_1\right)^2}{\sigma_{f_0}^2 + \sigma_{f_1}^2} + \frac{\left( \tau_0 - \tau_1 \right)^2}{\sigma_{\tau_0}^2 + \sigma_{\tau_1}^2} \gtrsim 1.
\end{equation}
Whenever the posterior distributions for the two modes' parameters are distinguishable, each mode's frequency and damping rate can be constrained.
In other words, we can meaningfully constrain the values of the $f_n$'s from the data even if Eq.~\eqref{eq:rayleigh_f} is not satisfied, and same for $\tau_n$ and Eq.~\eqref{eq:rayleigh_tau}.

As an example, consider the $\nmode=2$ measurement in Fig.~\ref{fig:ds_detection} ({orange}).
We summarize the associated frequency and damping time posteriors in Fig.~\ref{fig:ds_ftau}, by overlaying the joint distributions for $\{f_0,\tau_0\}$, and for $\{f_1,\tau_1\}$ ({blue} and {orange}, respectively).
The injected frequencies are much closer than the characteristic width of the corresponding marginals (top panel), and Eq.~\eqref{eq:rayleigh_f} is flagrantly violated.
Yet we constrain the frequency to be $f_0 = 252^{+8}_{-15}\, \mathrm{Hz}$ for the longest-lived mode, and $f_1 = 229^{+36}_{-38}\, \mathrm{Hz}$ for the shortest-lived one (at 68\% credibility, or $1\sigma$, to match the resolvability criteria), and the two modes making up the signal can be reconstructed reasonably well (Fig.~\ref{fig:analytic_strain_M2}).

We chose the parameters injected in Fig.~\ref{fig:ds_ftau} to be consistent with the ringdown of a GW150914-like remnant (Table~\ref{tab:ds_detection}).
However, the observation that frequency measurements are not predicated on Eq.~\eqref{eq:rayleigh_f} is general and applies to more extreme cases.
This includes hypothetical signals in which the true mode frequencies are exactly identical.
For example, in Fig.~\ref{fig:ds_ftau_extra} we constructed an injection with $f_0 = f_1$; in spite of this, the two modes are correctly identified thanks to their different decay rates, and we are able to measure the two frequencies and damping times successfully.
Of course, the point of this example is not to argue that modes with similar parameters are \emph{ideal} for this kind of measurement, but simply that they are not \emph{a priori} unsuitable.

Correlations aside, the Rayleigh criteria can indeed be a useful guide in forecasting $\nmode_{\rm best}$ for hypothesized measurements, especially if high SNRs can be expected \cite{Berti:2005ys,Berti:2007zu}.
That is because, in the absence of cross-mode correlations, failing \emph{both} Eq.~\eqref{eq:rayleigh_f} and Eq.~\eqref{eq:rayleigh_tau} would imply that the data are consistent with a shared frequency and damping time for the two modes.
Whatever their amplitudes and phases, the sum of two damped sinusoids with identical $f$ and $\tau$ is just another damped sinusoid.
Therefore, if the data prefer $f_0 = f_1$ and, simultaneously, $\tau_0 = \tau_1$, then one of the modes in the $\nmode=2$ template can be zeroed out, and we would not have found $\nmode_{\rm best} = 2$ in the first place, i.e., we would have stopped at $\nmode=1$ per the procedure outlined above (Sec.~\ref{sec:ds:det}).

\subsubsection{Relative versus absolute precision}

\begin{figure}
  \includegraphics[width=\columnwidth]{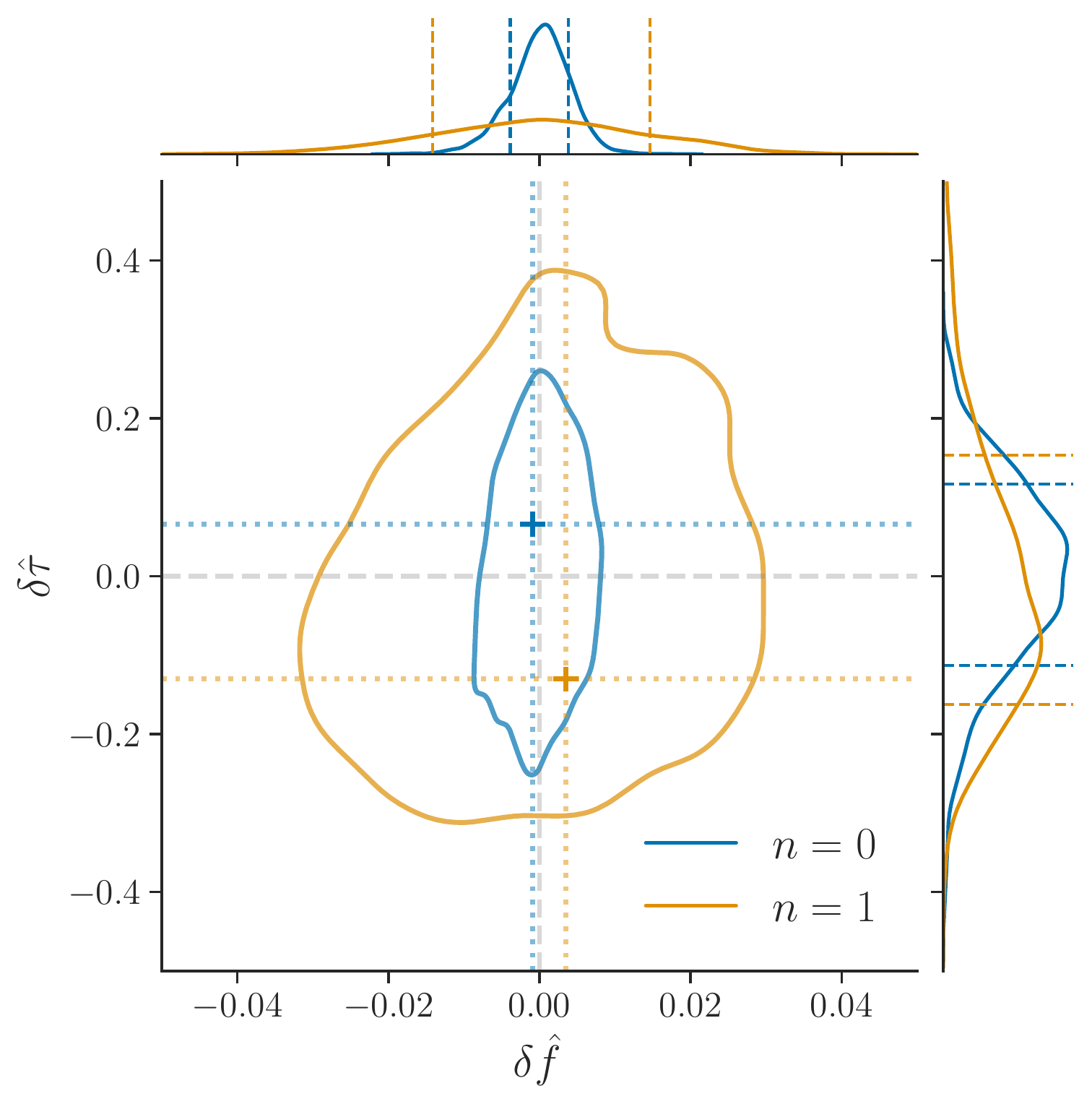}
  \caption{Fractional uncertainties corresponding to the measurement in Fig.~\ref{fig:ds_ftau_extra}, as proxied by the posterior on the quantites in Eq.~\eqref{eq:reluncert}. The origin represents the location of the mean of the distributions in the left panel of Fig.~\ref{fig:ds_ftau_extra}, $\hat{f}_n$ and $\hat{\tau}_n$, and crosses mark the relative locations of the truth values of $f_n$ and $\tau_n$; contours enclose 90\% of the probability, while vertical lines over the marginals indicate the 68\% creidble interval. The fractional uncertainties are similar for $\tau_0$ and $\tau_1$, but the fractional uncertainty in $f_1$ is larger than $f_0$.}
  \label{fig:ds_ftau_extra_relative}
\end{figure}

Figure \ref{fig:ds_ftau_extra} serves to illustrate another important point: being able to separate two modes in the $f, \tau$ plane does not imply we can measure those parameters with high \emph{relative} precision.
From the left panel of Fig.~\ref{fig:ds_ftau_extra}, it is clear that the analysis can very successfully distinguish the two modes in the signal through their damping times, and that the $\tau_n$ measurement is more accurate for $n=1$ than $n=0$.
This is a statement about the absolute precision with which we can determine $\tau_0$ or $\tau_1$.
Instead, we may ask about the relative precision of these measurements by looking at the quantities
\begin{equation} \label{eq:reluncert}
\delta\hat{\tau}_n \equiv \frac{\tau_n - \hat{\tau}_n}{\hat{\tau}_n}~~,~~
\delta\hat{f}_n \equiv \frac{f_n - \hat{f}_n}{\hat{f}_n}\,
\end{equation}
where $\hat{\tau}_n$ is the $\tau_n$ posterior mean as estimated from our samples, and same for $f_n$.
This definition gives us a proxy for fractional deviations of the parameters around the true mean.

Figure \ref{fig:ds_ftau_extra_relative} shows the posterior on these quantities as derived from Fig.~\ref{fig:ds_ftau_extra}.
Even though we were able to separate $\tau_0$ and $\tau_1$ cleanly in Fig.~\ref{fig:ds_ftau_extra}, both the $\delta\hat{\tau}_0$ and $\delta\hat{\tau}_1$ measurements are quite broad percentage-wise: $\delta \hat{\tau}_0 \approx \pm 20\%$ and $\delta \hat{\tau}_1 \approx \pm 30\%$, at 90\% credibility.
On the other hand, the $\delta\hat{f}_n$ frequency deviations are much more tightly constrained: $\delta \hat{f}_0 \approx \pm 1\%$ and $\delta \hat{f}_1 \approx \pm 2\%$.
So, even though the two modes have quite different damping times, it is the $\delta \hat{f}$'s which are determined more precisely.
Therefore, it should come as no surprise that, when we attempt to test the Kerr hypothesis using the first overtone of the $\ell = |m|=2$ mode, we can better constrain fractional deviations in its frequency than its damping time (as was found in \cite{Isi:2019aib,Abbott:2020jks}, and as we will show in Sec.~\ref{sec:analysis:pert}).

Incidentally, although $\tau_n$ is more sharply constrained for $n=1$ than $n=0$ (Fig.~\ref{fig:ds_ftau_extra}, left), casting the measurement in terms of $\gamma_n = 1/\tau_n$ inverts the relationship between the modes: $\gamma_0$ is better measured than $\gamma_1$ (Fig.~\ref{fig:ds_ftau_extra}, right).
This straightforward observation reveals the fact that $\omega_n$ and $\gamma_n$ are the natural basis that diagonalizes this measurement.
In any case, this is of little consequence when it comes to the precision of fractional deviation measurements.

This concludes our exploration of generic damped sinusoids based on the simplified template of Eq.~\eqref{eq:ds}.
In the following section, we turn to more realistic examples of BH ringdowns, for which we will apply the full template of Eq.~\eqref{eq:s}.
Our examples will focus on tones of the $\ell = |m| =2$ harmonic.

\section{Analyzing black-hole ringdowns}
\label{sec:analysis}

\begin{figure*}
  \includegraphics[width=\columnwidth]{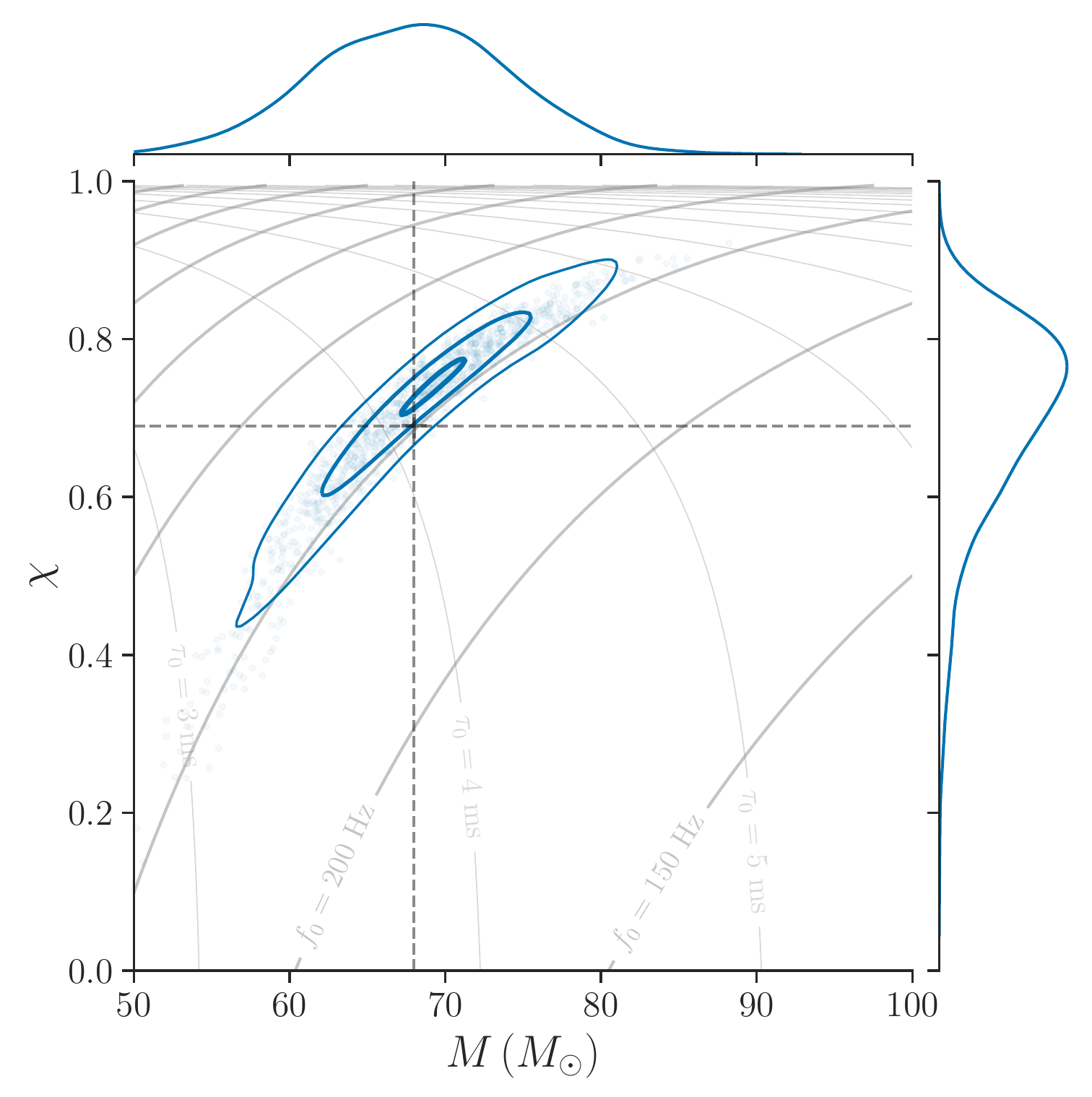}
  \includegraphics[width=\columnwidth]{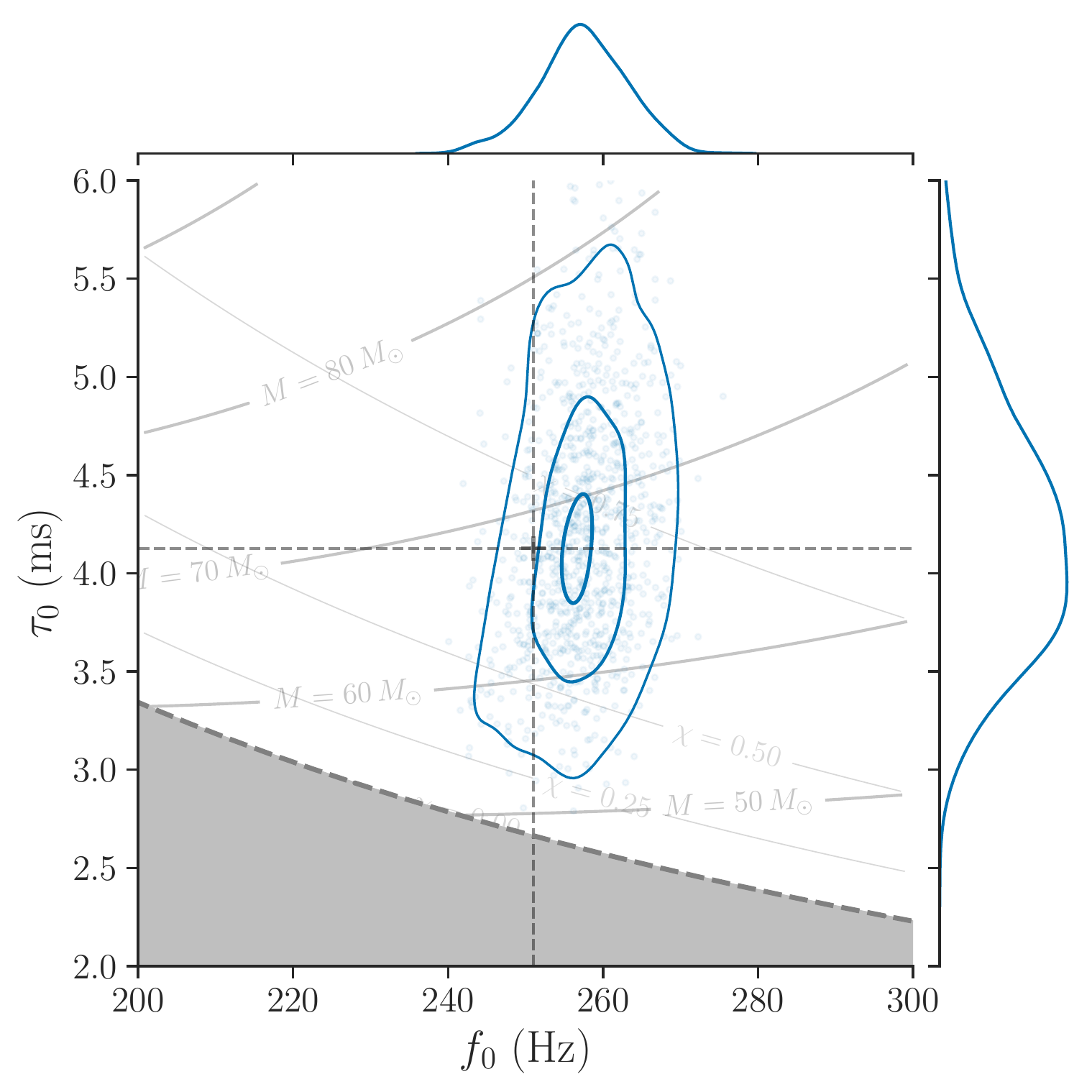}
  \caption{Simulated measurement of the least-damped $\ell=|m|=2$ mode for a GW150914-like remnant BH.
  The left panel shows the joint posterior for $M$ and $\chi$, quantities in which the measurement parameterized, with blue contours enclosing 90\%, 50\% and 10\% of the probability mass; the right panel shows the implied $f_0$ and $\tau_0$ distribution. On the left (right) gray lines mark contours of equal $f_0$ and $\tau_0$ ($M$ and $\chi$).
  A uniform prior on $(M,\chi)$ induces a nonuniform prior on $(f_0,\tau_0)$, and  $(f_0,\tau_0)$ values below the $\chi=0$ contour are disallowed (gray region on the right).
  Crosshairs mark the true values; parameters not shown are $\phi_0 = 5.34\, \mathrm{rad}$, $\epsilon=-1$ and amplitude chosen to yield ringdown $\mathrm{SNR} = 14$ over the network.
  }
  \label{fig:analytic_gr_n0}
\end{figure*}

The study of GW signals is not fundamentally different from the simplified damped-sinusoid examples in the previous section.
This is true even though the existence of multiple polarizations and detectors complicates the implementation in practice, as is reflected in the increased complexity of both the template, Eq.~\eqref{eq:s}, and the likelihood, Eq.~\eqref{eq:lnlike_multi}.

In the general case, having two GW polarizations introduces an additional pair of degrees of freedom per QNM, represented by $\epsilon_j$ and $\theta_j$ in Eq.~\eqref{eq:ellip}.
It also affects the structure of the template in Eq.~\eqref{eq:s}, making it always a linear combination of the two polarizations, with the antenna patterns $F^I_{+/\times}$ serving as respective weights for each detector $I$.
However, since we can neglect the polarization angle $\psi$ (Sec.~\ref{sec:template}), the only material effect of the $F^I_{+/\times}$ factors consists of regulating the expected amplitude \emph{ratios} of the signal as measured by different detectors.
Such ratios, as well as the relative time delays $\delta t_I$, implicitly make the template depend on the source sky location.
If this was not known a priori, we would have to include the right ascension $\alpha$ and declination $\delta$ as additional degrees of freedom in our analysis.
However, for signals seen by multiple detectors, we always have an accurate measurement of both the $\delta t_I$'s and the amplitude ratios from the full IMR analysis; for signals seen by a single detector, we effectively only measure a single polarization, so the template reduces to Eq.~\eqref{eq:ds} and we can ignore the sky location altogether.

The IMR analysis also provides information about the overall time of arrival of the signal, which is important in choosing a start time for the ringdown analysis.
Specifically, we can use the IMR results to reconstruct the time, $t_{\rm peak}$, at which the peak of the complex strain envelope, $|h|^2 = h_+^2 + h_\times^2$, arrived at the geocenter, with $t_{\rm peak}^I \equiv t_{\rm peak} + \delta t_I$ the corresponding times at each detector.
We can use this reference point to define the truncation times, $t_0^I$, at which to start the ringdown analysis [cf.~Eq.~\eqref{eq:tstart}].
Generally, the peak time can be reconstructed with sufficient accuracy.
Otherwise, we could incorporate uncertainty in the arrival time and sky location by sampling over $\alpha$ and $\delta$ without varying the truncation time, as long as we ensured that the truncation time always lies on or after whatever waveform feature we use to define the start of the ringdown, but we do not implement such a model in this work.

\begin{table}
\caption{GW150914-like extrinsic parameters.}
\label{tab:extrinsic}
\begin{tabular}{l@{\quad}l@{\quad}l@{\quad}l@{\quad}}
\toprule
$\alpha$ & $\delta$ & $\psi$    & $t_0$ (GPS) \\ \midrule
1.95 rad & $-1.27$ rad & 0.82 rad & 1126259462.423 s\\
\end{tabular}
\end{table}

In the remainder of this paper, we will simulate GW signals as originating from a sky location consistent with GW150914, as specified in Table \ref{tab:extrinsic}.
Although $\psi$ plays no role in the ringdown analysis, an arbitrary choice must be made in the injection in order for the amplitudes and phases at each detector to be uniquely specified by $\{A_j, \epsilon_j, \theta_j, \phi_j\}$; our choice, shown in Table \ref{tab:extrinsic}, is the same as in \cite{Isi:2019aib}.
We also specify a geocenter-based reference GPS time $t_0$, whose only relevance is in the computation of the antenna patterns and inter-detector delays.

After properly accounting for all this additional complexity, the lessons from Sec.~\ref{sec:ds} carry over directly to the GW case, including the discussions of mode detectability and resolvability.
We can use our infrastructure to measure the properties of Kerr BHs, and to look for deviations away from the Kerr spectrum.
With recent observational results in mind \cite{Isi:2019aib,Abbott:2020jks}, we focus on modes with $\ell = |m|=2$ but many of the observations are applicable to arbitrary modes including other harmonics.
Unless otherwise stated, we place uniform priors on all quantities mentioned.

\subsection{Kerr ringdowns}
\label{sec:analysis:kerr}

Within GR, an interesting application of BH spectroscopy is to measure the mass and spin of a Kerr BH from ringdown data alone (e.g., \cite{Echeverria:1989hg,Finn:1992wt,Isi:2019aib,Abbott:2020tfl,Abbott:2020mjq,Abbott:2020jks,Isi:2020tac,Capano2021}).
Under our Bayesian framework, the most straightforward way of doing this consists of parameterizing the frequencies and damping times in Eq.~\eqref{eq:s} as a function of $M$ and $\chi$ to directly obtain a posterior on these quantities (Sec.~\ref{sec:model:kerr}).
The template can include as many QNMs as deemed appropriate, e.g., by following the procedure outlined in Sec.~\ref{sec:ds:det}.

\begin{figure*}
  \includegraphics[width=\columnwidth]{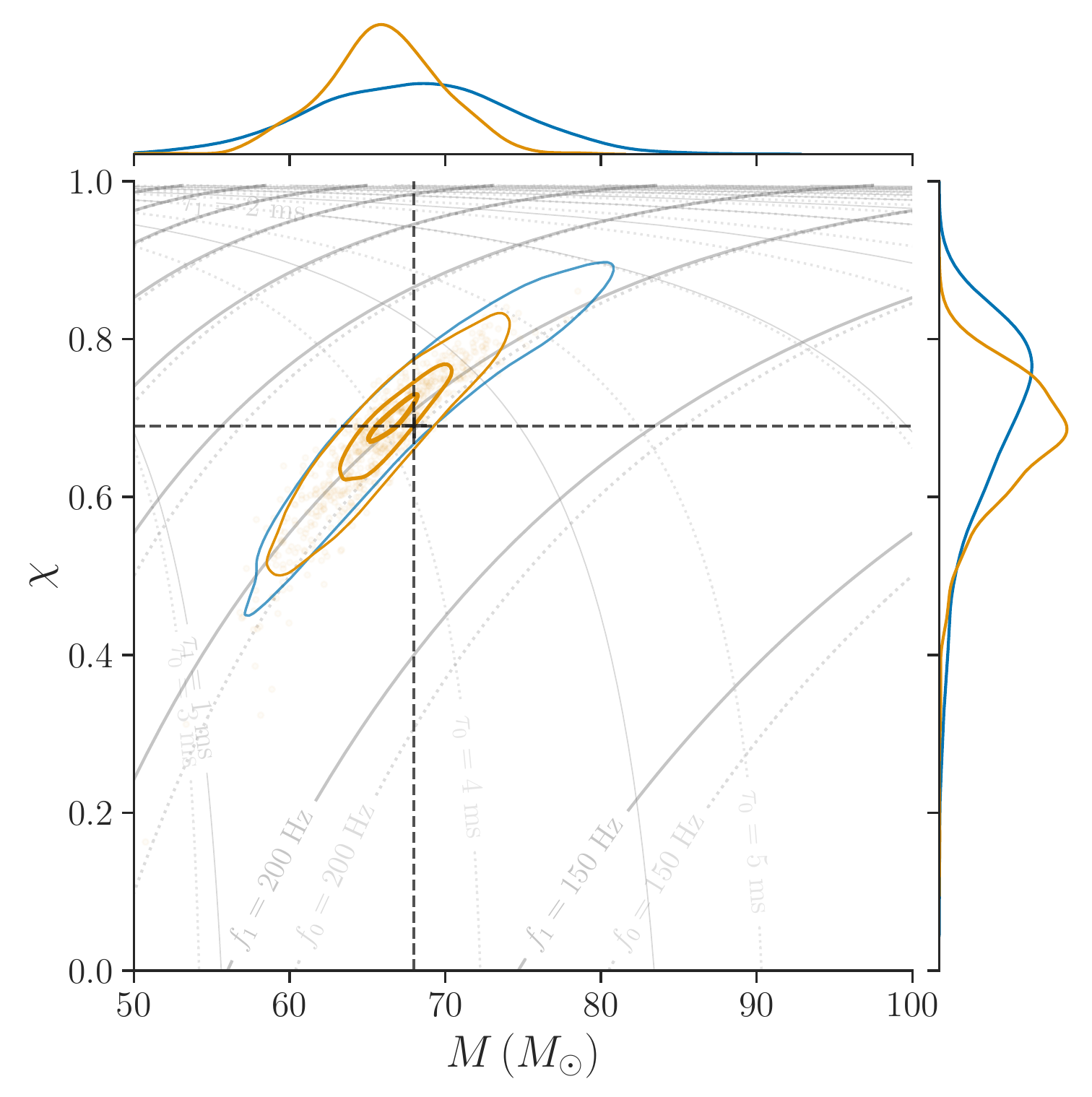}
  \includegraphics[width=\columnwidth]{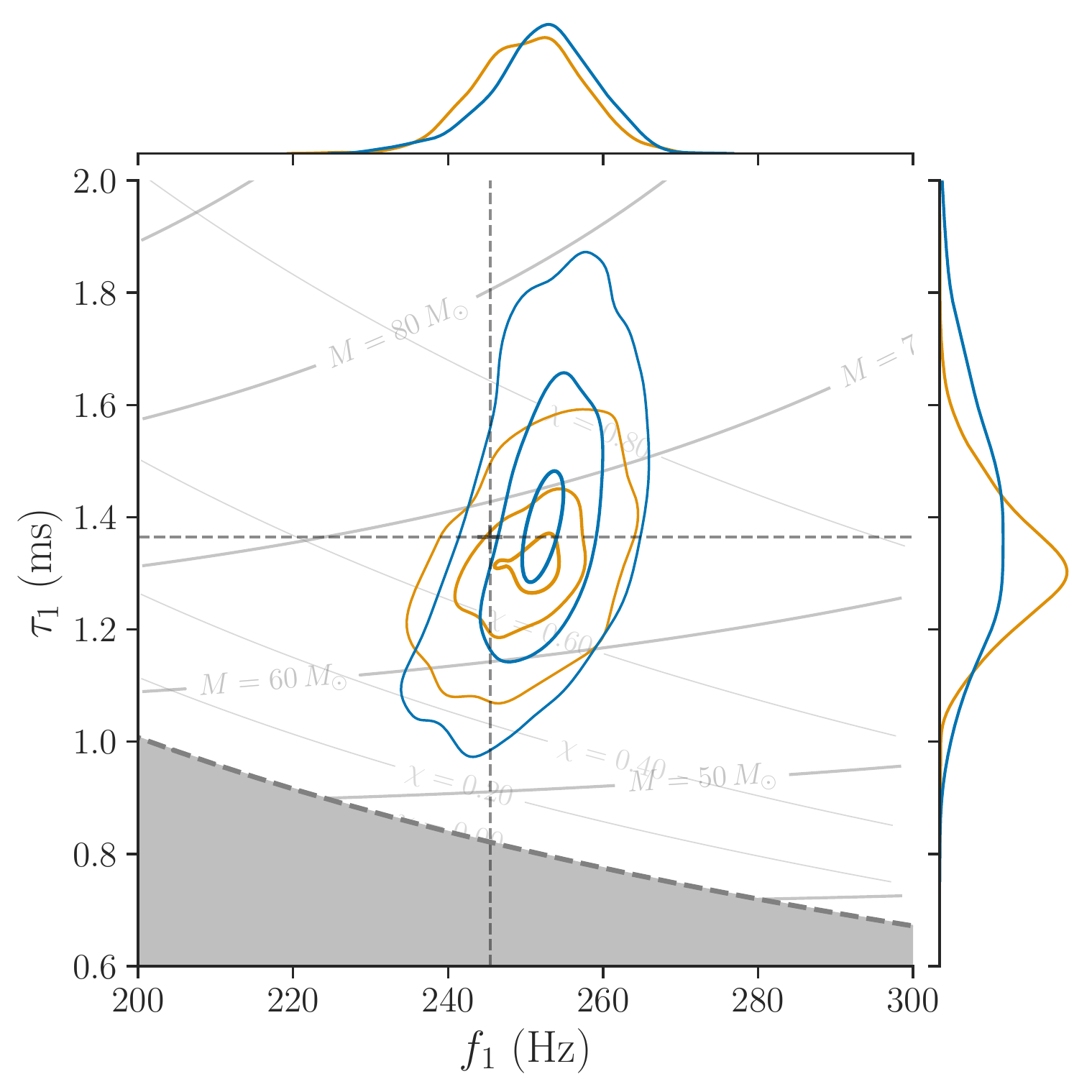}
  \caption{As in Fig.~\ref{fig:analytic_gr_n0}, but for a template and injection that incorporate the first overtone, in addition to the least-damped $\ell=|m|=2$ mode (orange).
  For comparison, we also show the 90\%-credible level from Fig.~\ref{fig:analytic_gr_n0} (blue).
  Solid gray lines mark contours of constant $(f_1,\, \tau_1)$; dotted, fainter lines mark $(f_0,\, \tau_0)$.
  Unlike Fig.~\ref{fig:analytic_gr_n0}, the right panel here shows the implied posterior on the overtone, rather than fundamental, frequency and damping rate; since the simulation in Fig.~\ref{fig:analytic_gr_n0} only included the fundamental, the blue contour on the right is fully determined by the $n=0$ parameters, which are in one-to-one correspondence thanks to the Kerr assumption.
  For this configuration, the overtone sharpens the $(M,\chi)$ recovery even with fixed SNR, but this behavior is not universal. Both here and in Fig.~\ref{fig:analytic_gr_n0}, the ringdown network SNR is 14.
  }
  \label{fig:analytic_gr_n1}
\end{figure*}

As a simple example, Fig.~\ref{fig:analytic_gr_n0} shows a simulated mass and spin measurement based on the $\ell = |m| =2,\, n=0$ mode of a GW150914-like remnant BH with $M = 68\, M_\odot$ and $\chi = 0.69$, and other injected parameters as in Table \ref{tab:intrinsic_gr}.
We simulate the signal as would be detected by the LIGO Hanford and Livingston instruments, with noise corresponding to the aLIGO design sensitivity (see Sec.~\ref{sec:imp}), and rescale the signal amplitude to yield a network SNR of 14.
The mass and spin are recovered correctly, and their joint posterior shows a characteristic, elongated shape that extends along an equal-frequency contour.
In fact, although the priors are diagonal in $M$ and $\chi$, the likelihood is effectively diagonal in $f_0$ and $\tau_0$ instead.
The uniform $(M,\, \chi)$ prior translates into a highly nonuniform $(f_0,\, \tau_0)$ prior, reflecting the coordinate geometry implied by the Kerr spectrum per Fig.~\ref{fig:ftau-chi}.
In particular, this results in the exclusion of points with $Q \lesssim 2.1$, which are unachievable for the fundamental $\ell=|m|=2$ mode with any combination of mass and spin (gray region).

\begin{table}
\caption{Kerr $\ell=|m|=2$ tones injected in Figs.~\ref{fig:analytic_gr_n0}--\ref{fig:analytic_gr_circular}. Only the amplitude ratio $A_1/A_0$ is relevant here because we rescale the signal to a given SNR; $\theta_n$ is undefined for $|\epsilon|=1$.}
\label{tab:intrinsic_gr}
\begin{tabular}{l@{\quad}l@{\quad}l@{\quad}l@{\quad}l@{\quad}l@{\quad}l@{\quad}}
\toprule
$M$ & $\chi$ & $A_1/A_0$ & $\epsilon_{0}$  & $\epsilon_{1}$  & $\phi_0$ & $\phi_1$ \\ \midrule
68 $M_\odot$ & 0.69 & 1.36 & $-1$ & $-1$ & 5.34 rad & 1.79 rad\\
\end{tabular}
\end{table}

It is possible for additional modes, if present, to reduce the uncertainty in the $(M,\, \chi)$ measurement, even for fixed SNR.
We show an example of this in Fig.~\ref{fig:analytic_gr_n1}, where we add the first overtone ($n=1$) to the signal in Fig.~\ref{fig:analytic_gr_n0}, while rescaling the amplitudes so as to keep the SNR unchanged.
Improvements of this kind are contingent on the additional modes actually appearing in the data in a way that provides new information to narrow the viable $(M,\, \chi)$ space and that overcomes the increased uncertainty due to a weakened fundamental mode (since we are keeping the SNR constant; e.g., see Fig.~\ref{fig:phasing_example}).
As underscored in Sec.~\ref{sec:ds}, whether this is the case on any specific instance will depend on the particular characteristics of the signal and noise instantiation (including nuisance parameters, like $\phi_n$, $\epsilon_n$ or $\theta_n$).
In Fig.~\ref{fig:analytic_gr_n1}, the quality of the $(M,\, \chi)$ measurement depends nontrivially on the interplay between the loss of information due to a weakened fundamental and the gain afforded by the overtone.

In realistic situations, we are interested in overtones because they can also increase the available ringdown SNR when analyzing real GW data \cite{Giesler:2019uxc,Isi:2019aib}, which is not the scenario demonstrated in Fig.~\ref{fig:analytic_gr_n1}.
Such an enhancement must generally be balanced against the growth in uncertainty associated with an increased number of degrees of freedom.
Even when additional tones do contribute, the longest-lived mode will usually be the best constrained and the structure of the $(M,\, \chi)$ posterior will not differ qualitatively from that in Fig.~\ref{fig:analytic_gr_n0}.

\begin{figure}
  \includegraphics[width=\columnwidth]{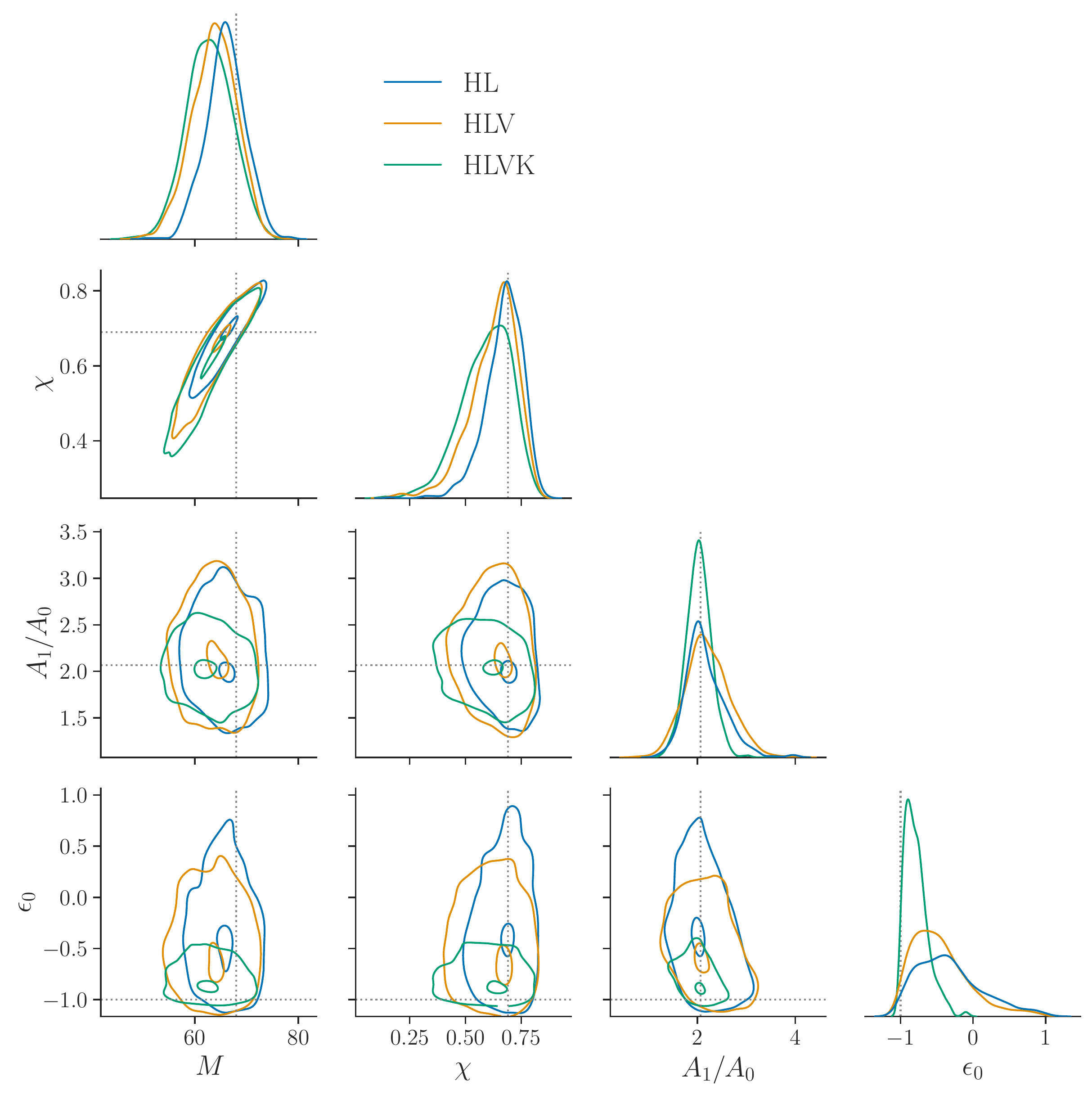}
  \caption{Effect of increasing the number of detectors on the $N=1$ analysis in Fig.~\ref{fig:analytic_gr_n1}, as manifested on the recovered mass $M$, spin $\chi$, overtone-to-fundamental amplitude ratio $A_1/A_0$, and fundamental ellipticity $\epsilon_0$, with 90\% and 10\% credibility.
  Color corresponds to networks containing interferometers at the locations of LIGO Hanford (H), LIGO Livingston (L), Virgo (V) and Kagra (K), although we take all detectors to have the design sensitivity of aLIGO.
  The blue (HL) curves correspond to the orange result in Fig.~\ref{fig:analytic_gr_n1}, and we enforce a network SNR of 14.
  Although not shown here, the result for $\epsilon_1$ is comparable to that for $\epsilon_0$;
  we have also omitted $\theta_{0/1}$ because that parameter is undefined for $|\epsilon_{0/1}|=1$, as is the case here (Table \ref{tab:intrinsic_gr}).
  }
  \label{fig:analytic_gr_ifos}
\end{figure}

For a given network SNR, having multiple detectors does not fundamentally change the above picture for $M$ and $\chi$ either.
Nevertheless, a larger network can allow for better inferences on polarization-related quantities, like $\epsilon_n$ and $\theta_n$, which can in turn lead to more accurate measurement of the spectrum by breaking degeneracies.
In Fig.~\ref{fig:analytic_gr_ifos} we show the effect of increasing the number of detectors for the $N=1$ analysis in Fig.~\ref{fig:analytic_gr_n1}, keeping the network SNR constant.
In this case, the amplitude ratio $A_1/A_0$ and the ellipticity $\epsilon_0$ are both better constrained with a larger network, and $(M,\, \chi)$ are recovered with slightly better accuracy.
Of course, in a realistic situation, having more detectors usually also brings a higher network SNR, which, all else being equal, always leads to a better measurement.
Both the SNR gain and the breaking of polarization degeneracies become more important for templates with larger numbers of modes.
Unfortunately, the two LIGO detectors are nearly coaligned, so that their ability to distinguishing GW polarizations is diminished.

\begin{figure}
  \includegraphics[width=\columnwidth]{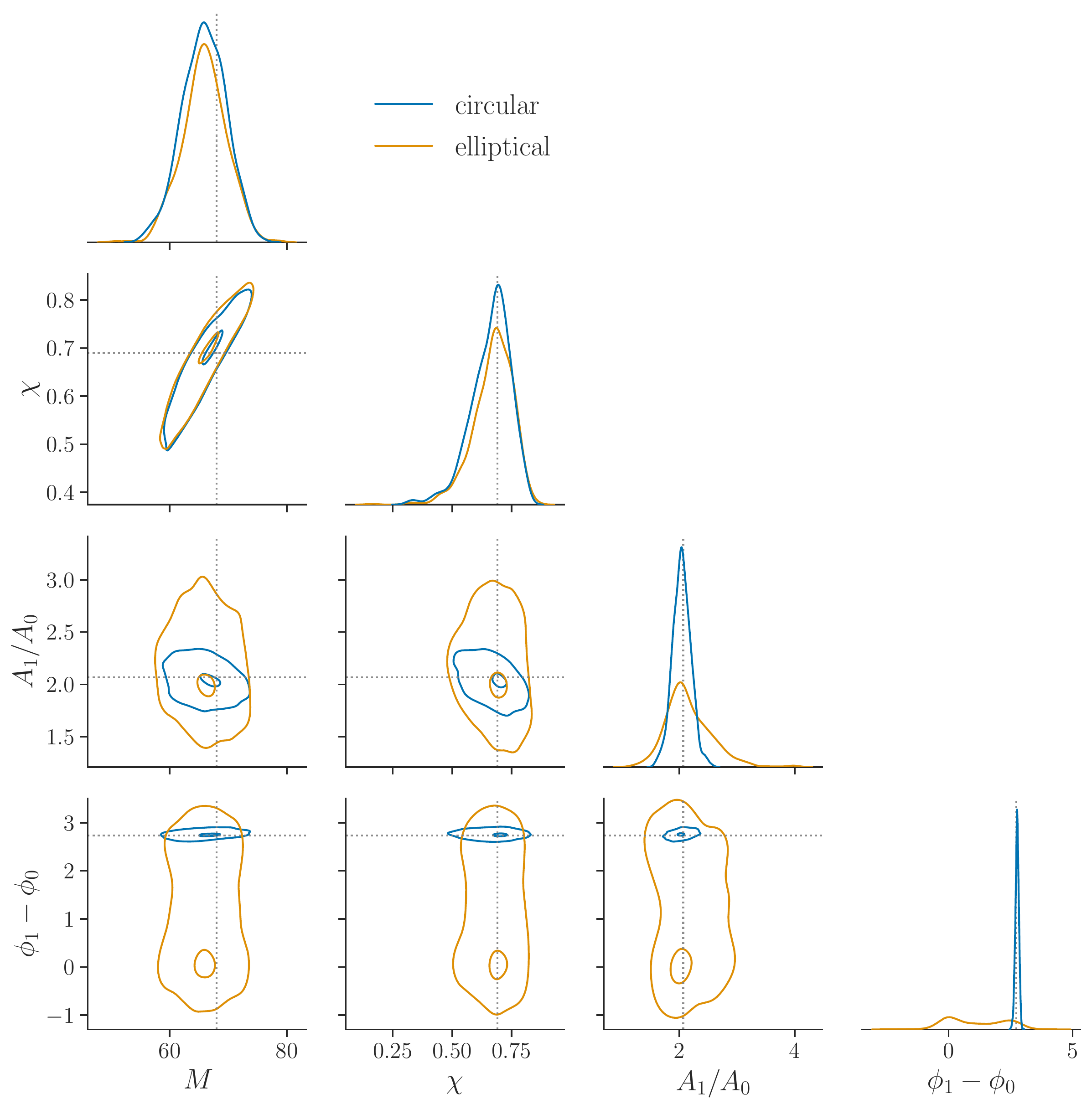}
  \caption{
  Effect of assuming a known ellipticity for all modes in the $N=1$ analysis of Fig.~\ref{fig:analytic_gr_n1}, as reflected on the BH mass $M$, spin $\chi$, mode amplitude ratio $A_1/A_0$, and mode relative phase $\phi_1 - \phi_0$, with 90\% and 10\% credibility (contours).
  Blue distributions were obtained assuming a circular-polarization model with $\epsilon_0 = \epsilon_1 = -1$, as in the injected signal; the orange distributions come from the same analysis as in Fig.~\ref{fig:analytic_gr_n1}.
  }
  \label{fig:analytic_gr_circular}
\end{figure}

Finally, there are circumstances in which it may be acceptable to use a restricted version of the elliptical model in Eq.~\eqref{eq:s}.
As mentioned above, this is the case for signals recorded by a single detector (or, to good approximation, by just the two LIGO detectors), for which the polarization parameters $\epsilon_j$ and $\theta_j$ are fully redundant.
It is also true more generally whenever the inclination is known and the system can be assumed to be (reflection) symmetric over the equator; if so, it may be reasonable to take all the modes in the signal to have a known ellipticity, e.g., $\epsilon = + 1\,(-1)$ for a face on (off) system (see App.~\ref{app:slms}).
With such a model, the recovery of amplitudes and phases is aided by the reduced number of parameters (Fig.~\ref{fig:analytic_gr_circular}).
However, this does not necessarily improve the accuracy of the $M$ and $\chi$ measurements.

\subsection{Deviations from Kerr}
\label{sec:analysis:pert}

\begin{figure*}
  \includegraphics[width=1.5\columnwidth]{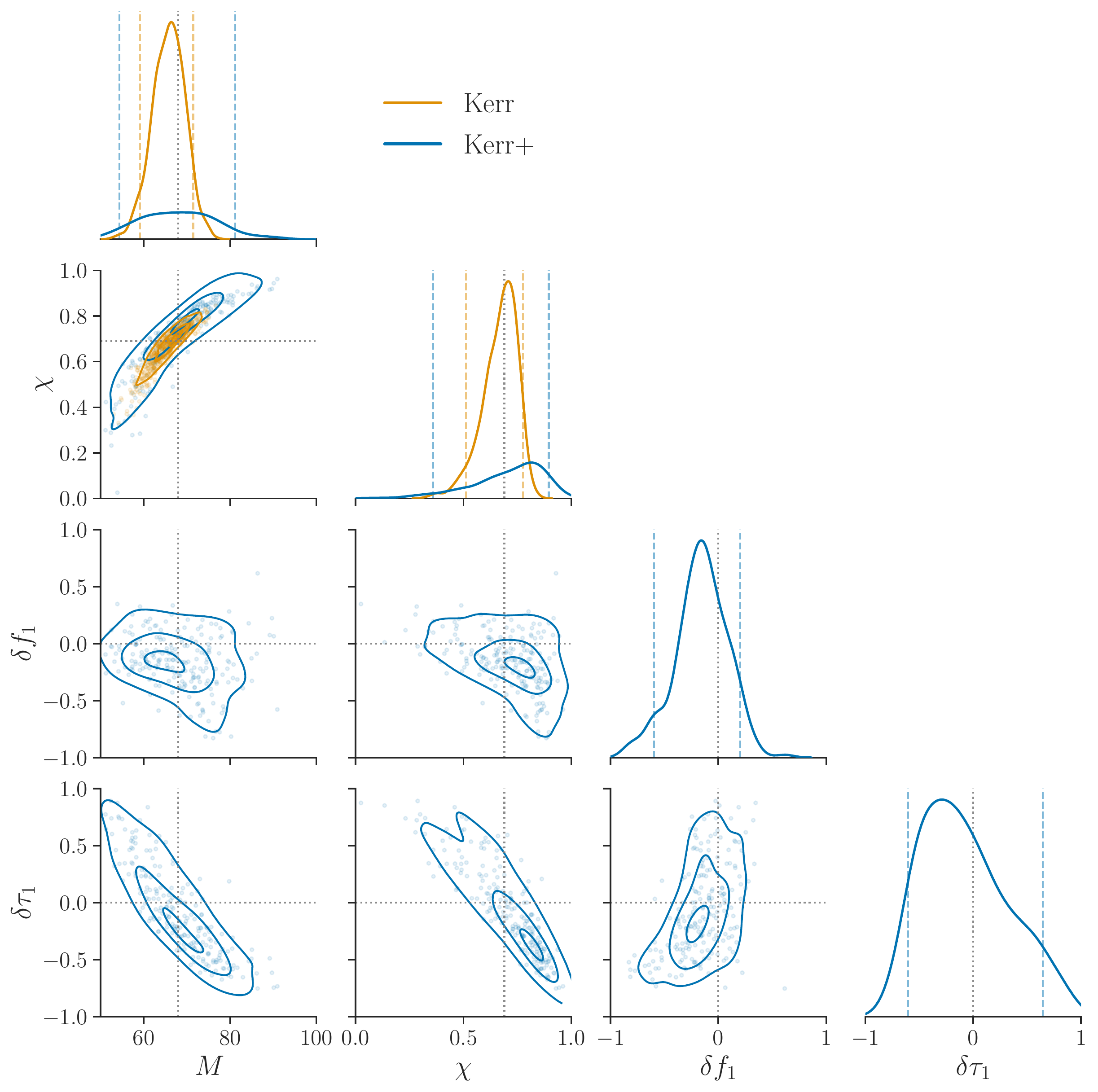}
  \caption{Simulated constraints on Kerr deviations from a model including the two longest-lived tones with $\ell=|m|=2$, using the parameterization of Eq.~\eqref{eq:nongr_df1_dtau1} on the Kerr $N=1$ injection from Fig.~\ref{fig:analytic_gr_n1}.
  The corner plot shows posterior densities for the BH mass $M$, spin $\chi$, overtone frequency deviation $\delta f_1$, and damping-time deviation $\delta \tau_1$ (blue, ``Kerr+''); for comparison, we overlay the $(M,\, \chi)$ measurement obtained assuming $\delta f_1=\delta \tau_1=0$ (orange, ``Kerr;'' same as in Fig.~\ref{fig:analytic_gr_n1}).
  In this plot, we have marginalized over nuisance parameters $\{A_n, \epsilon_n, \theta_n, \phi_n\}$ for both tones in the template ($n=0,1$), contours enclose {90\%, 50\% and 10\%} of the probability, and vertical lines over the marginals mark the 90\% credible interval; dotted straight lines indicate the truth.
  }
  \label{fig:analytic_nongr_n1_models}
\end{figure*}

Beyond mass and spin measurements within GR, we can look for signs of new physics by allowing for deviations from the Kerr prediction.
As discussed in Sec.~\ref{sec:model:pert}, the most natural way to do this is to introduce Kerr deviation parameters $\delta f$ and $\delta \tau$ (or $\delta f$ and $\delta \gamma$) for the frequency and damping time of one or more QNMs (at most $2D-2$ new parameters for a model with $D$ modes).
We can use this parameterization to establish whether any two or more damped sinusoids are consistent with a Kerr spectrum, obtaining a quantitative answer to which we can assign a definite credibility.

This kind of analysis can be carried out using however many modes are confidently detected in the data (Sec.~\ref{sec:ds:det}), whether they be tones of a given angular harmonic (same $\ell$ and $|m|$, different $n$), a series of fundamental modes with varying angular structure (different $\ell$ or $|m|$, same $n$), or any other combination.
Continuing our focus on the two-tone $\ell=|m|=2,\, N=1$ model, we will study simulated measurements of the $\delta f_{221}$ and $\delta \tau_{221}$ overtone parameters of Eqs.~\eqref{eq:nongr_df1_dtau1} (henceforth just denoted $\delta f_1$ and $\delta \tau_1$ for simplicity).
This serves the dual purpose of demonstrating our procedure and validating our infrastructure, as well as replicating some of the conclusions obtained with real data in \cite{Isi:2019aib,Abbott:2020jks}, verifying they are in agreement with expectation.%
\footnote{However, note that neither the model nor the infrastructure applied here correspond exactly to those used in \cite{Isi:2019aib} or \cite{Abbott:2020jks}. The former used a model like the one described in App.~\ref{app:slms}, while the latter applied different priors; both used different samplers.}

In Sec.~\ref{sec:analysis:pert:kerr}, we demonstrate constraints based on signals that do follow a Kerr spectrum; in Sec.~\ref{sec:analysis:pert:nonkerr}, we demonstrate measurements obtained from non-Kerr signals.

\subsubsection{Constraints from Kerr signals}
\label{sec:analysis:pert:kerr}

We begin by reanalyzing the $\ell=|m|=2$, $N=1$ Kerr signal from the previous section (Fig.~\ref{fig:analytic_gr_n1}, {orange}), this time extending our recovery model through the two Kerr-deviation parameters $\delta f_1$ and $\delta \tau_1$.
We show the resulting posterior on these parameters, as well as $M$ and $\chi$, in Fig.~\ref{fig:analytic_nongr_n1_models} ({blue}).
For this particular configuration, we obtain a reasonable constraint on the overtone frequency deviation, measuring {$\delta f_1 = -0.12^{+0.44}_{-0.47}$} with 90\% credibility, or {$\delta f_1 = -0.12^{+0.34}_{-0.28}$} with 68\% credibility; on the other hand, the damping rate is only poorly constrained to {$\delta \tau_1 > -0.34$} with 90\% credibility, with support extending to the upper edge of our prior, $\delta\tau_1=1$.
Unsurprisingly, the presence of these additional parameters results in a broader range of allowed $M$ and $\chi$ values, as seen in the comparison to the Kerr result ({orange}) in the upper left corner of Fig.~\ref{fig:analytic_nongr_n1_models}.

\begin{figure}
  \includegraphics[width=\columnwidth]{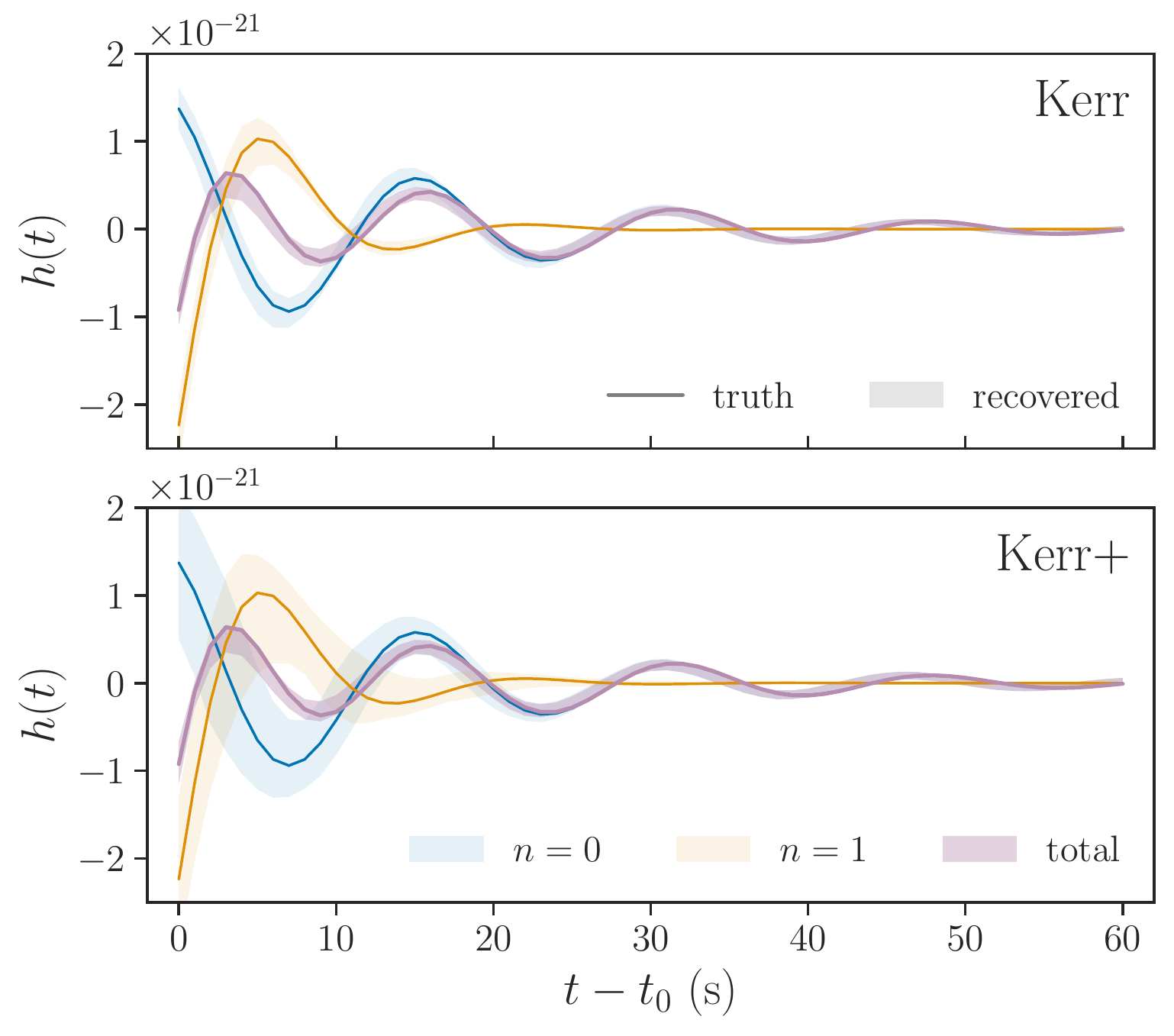}
  \caption{Waveform reconstructions from the two analyses in Fig.~\ref{fig:analytic_nongr_n1_models}, assuming a Kerr spectrum (top), or allowing deviations in the overtone (bottom).
  Solid lines show the true waveforms for the injected fundamental mode (blue), overtone (orange) and total signal ({magenta}), with colored envelopes representing the reconstruction at the 90\% credible level.
  We only show the result for the Hanford (H1) observatory, since the Livingston reconstruction is similar.}
  \label{fig:analytic_n1_wf}
\end{figure}

Figure \ref{fig:analytic_n1_wf} shows the corresponding waveform reconstructions under both the Kerr and non-Kerr models: both are consistent with the true signal with comparable uncertainty on the overall reconstruction, but the latter shows greater uncertainty on the individual modes.
This is a direct manifestation of the flexibility introduced by $\delta f_1$ and $\delta \tau_1$, which opens up more combinations of mode morphology that can add up to the observed signal.
For this configuration, the effect of the overtone on the overall waveform is to lower the amplitude near $t_0$ by interfering destructively with the fundamental mode.
At late times, once the overtone has decayed, the signal is fully given by the fundamental mode to great accuracy.

The result from this simulation is consistent with what was found in \cite{Isi:2019aib}, in particular with the expectation that, at moderate SNRs, $\delta \tau_1$ should be generally harder to pin down than $\delta f_1$ (also, e.g., \cite{Gossan:2011ha}).
In this case, the reason has to do with the correlations between the overtone deviations and $M$ and $\chi$, as displayed in Fig.~\ref{fig:analytic_nongr_n1_models}: the model is able to support high values of $\delta \tau_1$, as long as both $M$ and $\chi$ are lowered with respect to their true values.
This degeneracy is not fully intrinsic to the $N=1$ model but arises from the way in which the two damped sinusoids combine to produce the true injected signal in this example, as determined by their relative phasing (here simply the difference $\phi_1 - \phi_0$, since $\epsilon = -1$ for both modes in the injection, so the $\theta_n$ are meaningless).

\begin{figure}
  \includegraphics[width=\columnwidth]{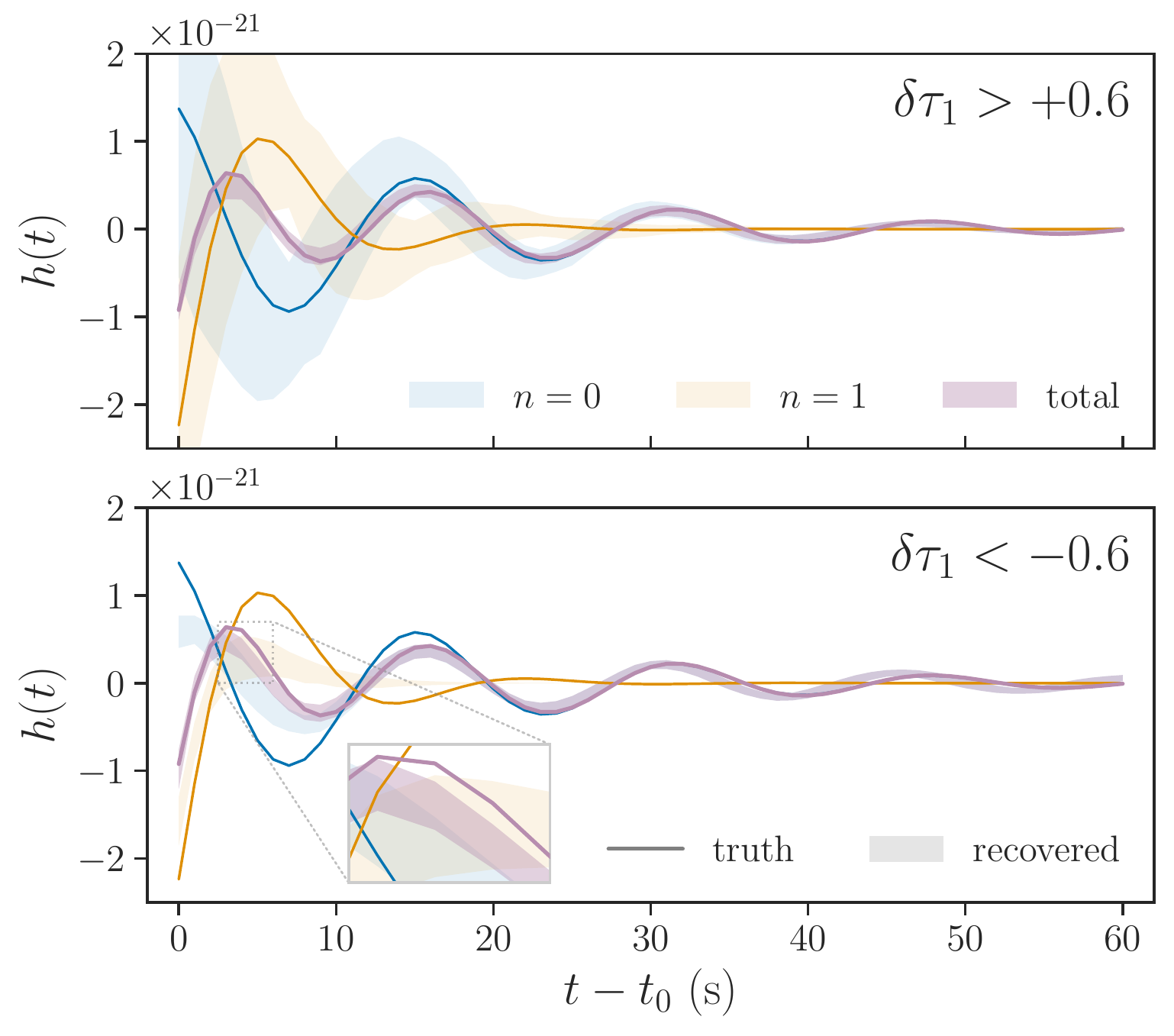}
  \caption{As in bottom panel of Fig.~\ref{fig:analytic_n1_wf}, but limiting $\delta \tau_1 > +0.6$ (top) or $\delta \tau_1 < -0.6$ (bottom).
  High values of $\delta \tau_1$ can be accommodated with small changes to the overall signal ({magenta}) by reducing $M$ and $\chi$ with respect to the true values, in such a way as to shorten the fundamental mode (blue) while keeping its frequency roughly the same, and by simultaneously tuning the amplitudes to decrease the overtone (orange) to fundamental ratio, $A_1/A_0$. No such perfect tuning is possible for overly negative values of $\delta \tau_1$, as highlighted in the inset where there is an evident disagreement between the 90\%-credible reconstruction and the true signal.
  }
  \label{fig:analytic_n1_wf_dtau1_pos_neg}
\end{figure}

Although the posterior structure will generally vary for different data, this specific configuration is representative of what we expect from binary BH numerical relativity simulations, so it is worth examining it further.
For a Kerr BH, smaller $M$ results in overall higher frequencies and lower damping times for all modes.
If the change in frequency is smaller than ${\sim}50\%$ (at this SNR), it can be counterbalanced by a decrease in $\chi$, which however will also result in lower damping times (left and center panels in Fig.~\ref{fig:ftau-chi}).
By decreasing both $M$ and $\chi$, one can thus obtain a spectrum of modes with unaltered frequencies but faster decays.
Meanwhile, $\delta \tau_1$ frees the overtone from following this pattern, so that a lower $M$ and $\chi$ can be used to decrease $\tau_0$ independently of $\tau_1$, while keeping both $f_0$ and $f_1$ approximately constant.
In fact, these correlations can be used to decrease $\tau_0$ while simultaneously \emph{increasing} $\tau_1$ through $\delta\tau_1$.
If one also alters the amplitudes so as to lower the ratio $A_1/A_0$, modifications to the waveform induced by the changes to the damping times can be made to cancel out, since one can compensate a shorter (longer) damping time with a larger (smaller) initial amplitude.
While $\delta\tau_1$ can vary over a broad range, $\delta f_1$ is more constrained, since the frequency of the overtone must remain close to the true value to properly interfere with the frequency of the fundamental mode (which is itself well determined by data at late times).
All this is encoded in Figs.~\ref{fig:analytic_n1_wf} and~\ref{fig:analytic_n1_wf_dtau1_pos_neg}.

Similar observations explain the asymmetry between positive and negative values of $\delta\tau_1$ in Fig.~\ref{fig:analytic_nongr_n1_models}.
For sufficiently negative values of $\delta\tau_1$, the overtone vanishes too quickly and the balance between the two modes cannot be restored by changes to the other parameters.
As $\delta\tau_1 \rightarrow -1$, the overtone approaches a Kronecker delta at $t_0$ and its contribution to the total signal vanishes (except at the first sample), even for high amplitudes $A_1$.
For such values of $\delta\tau_1$, then, the amplitude of the fundamental mode must be decreased to match the signal near $t_0$, but this is inconsistent with later times, during which the longest-lived mode dominates the injection.
Since the fundamental cannot be made to match both the early and late data in this regime (see inset in Fig.~\ref{fig:analytic_n1_wf_dtau1_pos_neg}), negative values of $\delta\tau_1$ are disfavored.

Note that, had the presence of the overtone been unclear (e.g., if the posterior had significant support for $A_1=0$), then letting $\delta\tau_1 \rightarrow -1$ would be a way for the sampler to get rid of this fast-decaying mode for arbitrary values of the $n=1$ parameters; this can cause an artificial preference for extremely negative values of $\delta\tau_1$, and result in sampling problems.
The LIGO-Virgo collaboration has previously encountered this issue when analyzing low-SNR ringdowns \cite{Abbott:2020jks}.  This issue could probably be eliminated by reparameterizing the modifications in terms of $\gamma = 1/\tau$ and $\delta \gamma$ because $\delta \tau \to -1$ corresponds to $\delta \gamma \to \infty$; alternately, the deviation parameters could be redefined non-linearly as $\tau \to \tau e^{\delta \tau}$, with the same result that the problematic point $\delta \tau \to -1$ is removed to infinity.  We leave explorations of these alternative parameterizations to future work.

\begin{figure}
  \includegraphics[width=\columnwidth]{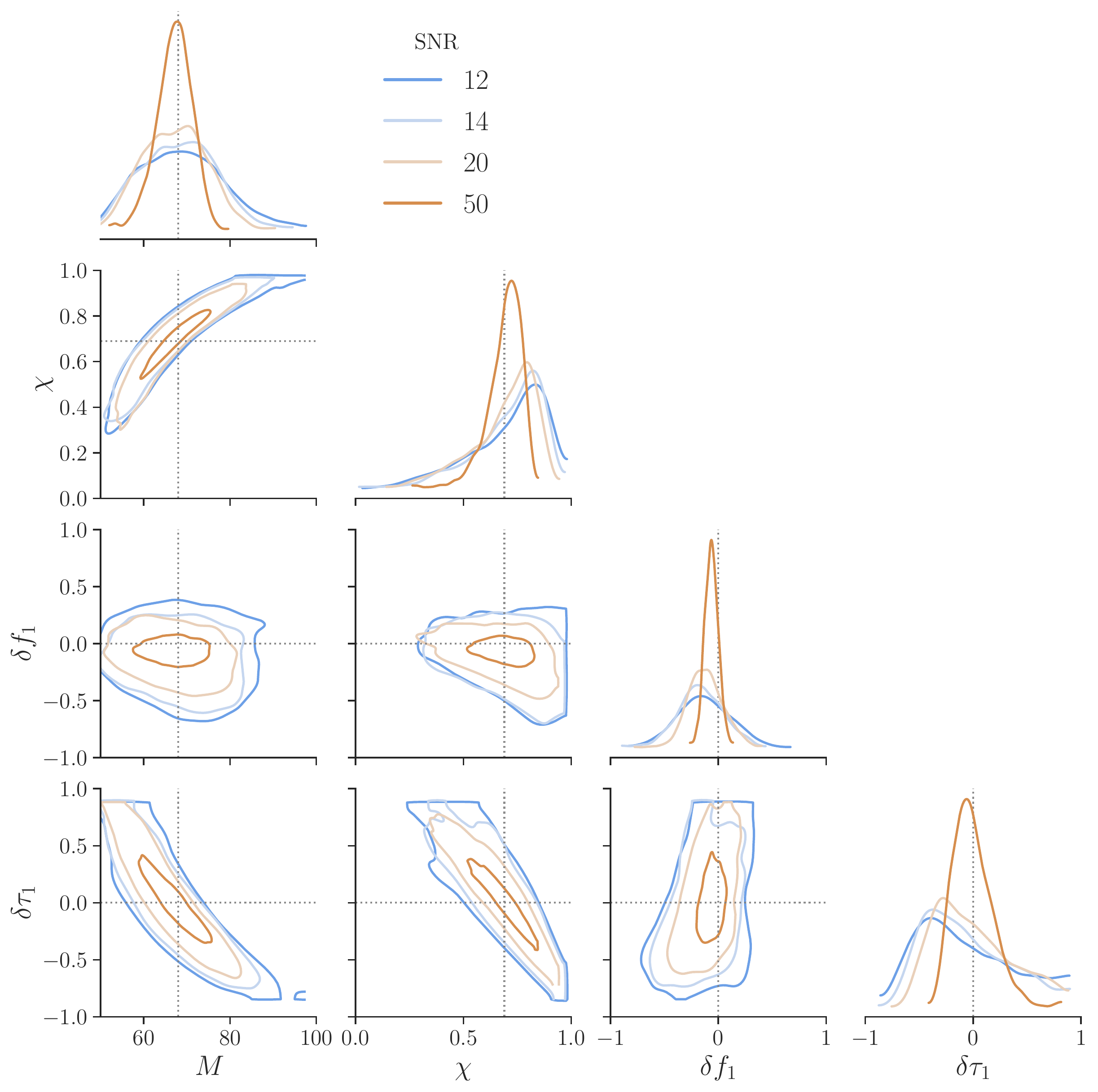}
  \caption{Constraints on Kerr deviations for increasing ringdown SNR (color). Same beyond-Kerr analysis as in Fig.~\ref{fig:analytic_nongr_n1_models}, but for varying injected network SNR. Contours enclose 90\% of the probability mass, and dotted lines mark the truth. As the SNR increases, the analysis converges on the correct values for $M$, $\chi$, $\delta f_1$ and $\delta \tau_1$ (dotted black lines), with progressively smaller uncertainties. }
  \label{fig:analytic_nongr_n1_snr}
\end{figure}

As with most other parameters, our ability to constrain $\delta f_1$ and $\delta \tau_1$ improves with SNR.
We demonstrate this in Fig.~\ref{fig:analytic_nongr_n1_snr} with the same injection as before but scaled in overall amplitude to produce progressively higher network SNRs, with the injected matched-filter SNR as defined in Eq.~\eqref{eq:snr_net}.
The $\delta\tau_1$ measurement improves slowly (bottom right panel), with support significantly restricted to $\delta\tau_1 \leq 1$ only for SNR $\geq 50$.
The tightening of $\delta f_1$ and $\delta \tau_1$ posteriors is accompanied by an improvement in the $M$ and $\chi$ measurement, for which the posteriors converge to the right values.
The individual mode amplitudes and phases are also better determined for higher SNR (not shown).

In the examples above, the model used to produce the injection matches that of the recovery.
In realistic situations, however, higher SNRs will tend to reveal additional signal features that are not explicitly captured by our QNM model; it is thus important to understand at which point the induced systematic error may mimic a deviation from Kerr.
In the context of BH overtones, for example, we may expect to see contributions from an increasing number of tones with $n > 1$ as we analyze data closer to the signal peak \cite{Giesler:2019uxc,Buonanno:2006ui,Baibhav:2017jhs,Mourier:2020mwa,Finch:2021iip}.
When fitting an $N=1$ model, such additional modes could conceivably bias the $\delta f_1$ or $\delta \tau_1$ measurements.
The magnitude of this systematic bias was quantified as a function of SNR in \cite{Giesler:2019uxc}, both through a heuristic calculation based on template mismatches, and through simulated Bayesian measurements of $M$ and $\chi$.
Here, we expand upon those results to show that, indeed, contamination due to $n > 1$ will only bias $\delta f_1$ or $\delta \tau_1$ for significantly high SNRs---loud enough for the additional modes to be detected by the procedure in Sec.~\ref{sec:ds:det}.

We simulate the same Kerr signal as above (Figs.~\ref{fig:analytic_nongr_n1_models}--\ref{fig:analytic_nongr_n1_snr}), but add to it a third mode corresponding to the second $\ell=m=2$ overtone ($n=2$).
Our choice of $M$ and $\chi$ (Table \ref{tab:intrinsic_gr}) determines the frequency and damping time of this new mode, but we have freedom to choose its amplitude and phase parameters.
In order to produce a somewhat realistic example, we set those quantities based on an $N=2$ least-squares fit of a GW150914-like numerical relativity waveform (\nrsim \cite{kidder_larry_2019_3301877,Lovelace:2016uwp,Boyle:2019kee}), modeled starting at peak strain; this serves as a proxy for features in a real signal that would not be captured by the $N=1$ ringdown model.
Based on that fit, we set $A_2 / A_0 = 1.1$, $\phi_2 = 0.83\, \mathrm{rad}$ and $\epsilon_2 = -1$ in our injection, but we do not modify the $n=1$ parameters.
Assuming only LIGO Hanford and Livingston are operational, we analyze this new $N=2$ Kerr signal with the same $N=1$ model we had been using so far, progressively increasing the injected SNR.

\begin{figure}
  \includegraphics[width=\columnwidth]{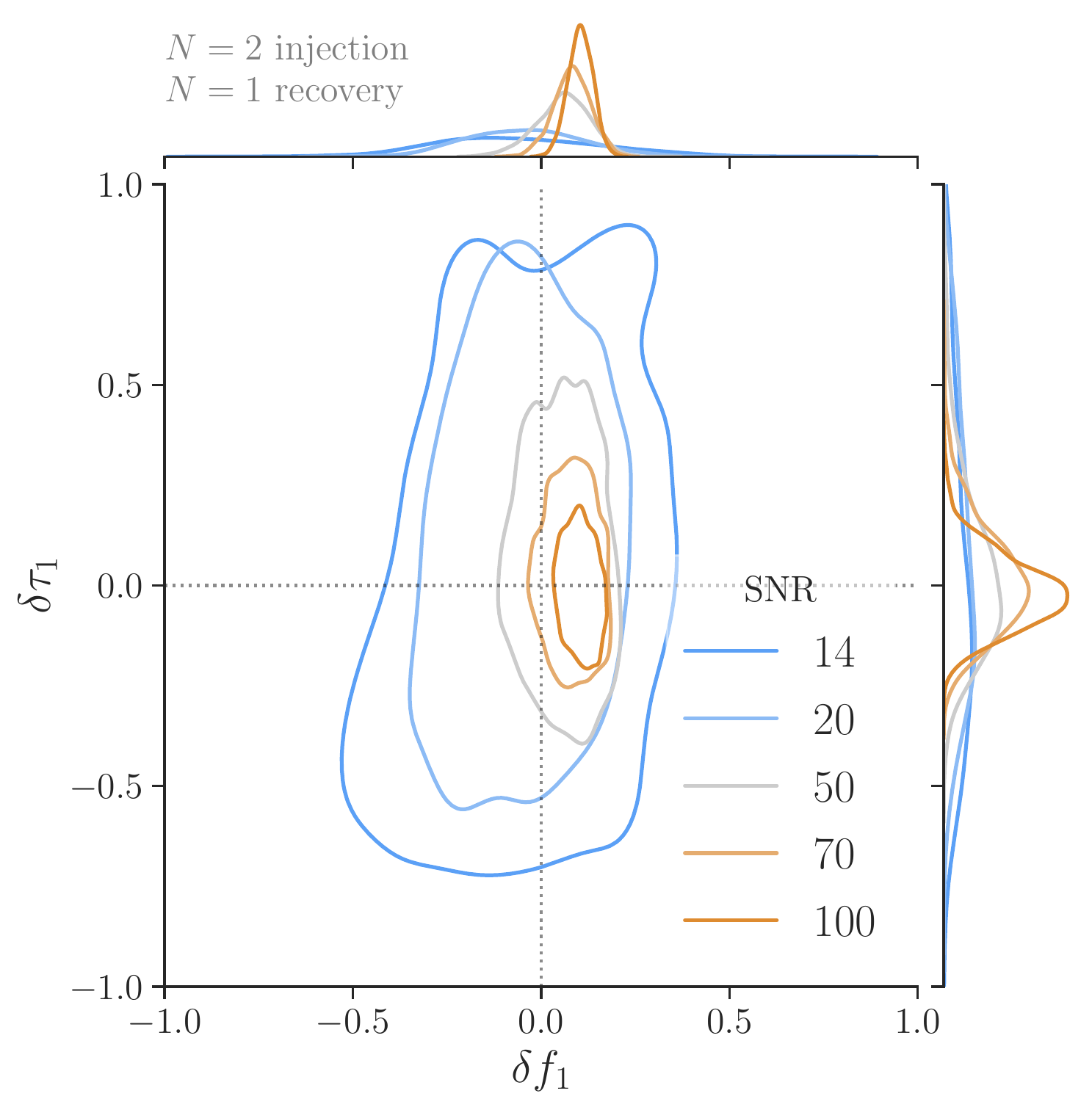}
  \caption{Effect of not accounting for the second overtone in a spectroscopic measurement assuming $N=1$ for different total SNRs (color). This is the same analysis on the Kerr signal in Fig.~\ref{fig:analytic_nongr_n1_snr}, except for the injection of the $n=2$ mode, which is not accounted in the recovery model. This causes a systematic bias in the recovered $\delta f_1$ but, for this configuration, this only becomes significant at 90\% credibility when the injected network SNR reaches 100.}
  \label{fig:analytic_df_dtau_N1_Ninj2}
\end{figure}

Figure \ref{fig:analytic_df_dtau_N1_Ninj2} shows the results in the $\delta f_1$, $\delta\tau_1$ plane.
For this configuration, the posterior density does not begin to show an obvious systematic bias until the total injected SNR reaches ${\sim} 50$.
The presence of the $n=2$ mode in the data manifests in an overall shift of the posterior towards $\delta f_1 \geq 0$, but the damping rate parameter $\delta \tau_1$ remains completely unaffected even for the highest SNR we consider.
Even though $\delta f_1$ is clearly sensitive to the $n=2$ mode, it is not until we reach $\mathrm{SNR}\gtrsim 100$ that the Kerr value ($\delta f_1 = \delta \tau_1 = 0$) falls outside the 90\%-credible region.

\begin{figure}
  \includegraphics[width=\columnwidth]{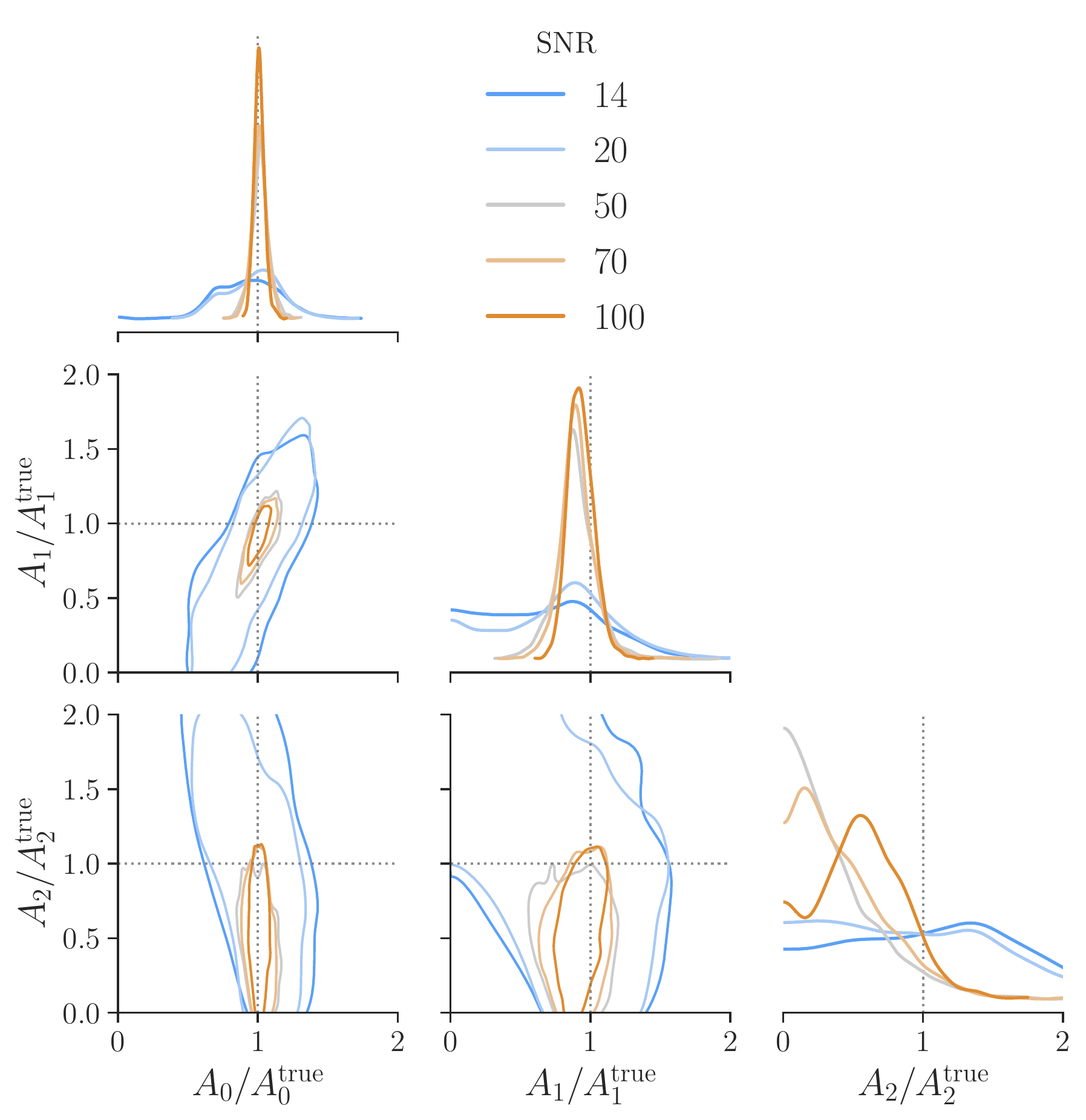}
  \caption{Amplitude posteriors from the $N=2$ analysis of the Kerr $N=2$ signal in Fig.~\ref{fig:analytic_df_dtau_N1_Ninj2}, carried out assuming a Kerr spectrum and for different total injection SNRs (color). Contours represent the 90\%-credible regions for each of the three amplitudes, measured as a fraction of the respective true injected value, so that unity represents the truth (dotted lines). For SNRs 14 and 20, the joint $A_1{-}A_2$ posterior displays the arching seen in Fig.~\ref{fig:ds_detection}; this is not the case for higher SNRs, revealing the presence of the third mode ($n=2$) and preventing us from being misled by the spurious Kerr deviation in in Fig.~\ref{fig:analytic_df_dtau_N1_Ninj2}.}
  \label{fig:analytic_amps_N2_Ninj2}
\end{figure}

However, this SNR is sufficiently high that the presence of the second overtone would have been revealed by the mode-detection procedure outlined in Sec.~\ref{sec:ds:det}.
Indeed, even before the SNR reaches ${\sim}70$, the joint $A_1{-}A_2$ posterior stops displaying the characteristic ``arching'' that we expect if only two modes are identified in the data (Fig.~\ref{fig:analytic_amps_N2_Ninj2}).
This suggests that it would be unlikely for us to be misled by the bias seen at high SNRs in Fig.~\ref{fig:analytic_df_dtau_N1_Ninj2} when looking at a signal like this.
Nevertheless, we do expect the kind of bias to compound in the batch analysis of multiple signals (as in \cite{Abbott:2020jks}), especially if extreme care is not taken to identify signals which present more visible modes than expected.

Although, the results Fig.~\ref{fig:analytic_amps_N2_Ninj2} provide some idea of the SNRs at which we expect systematics to overcome statistical uncertainty, the actual level of bias will vary with the details of the signal and noise instantiation.
{In practice, this means that potential systematics should be studied assiduously for each case, especially if there are hints of a Kerr deviation.
This task would be aided by a dedicated study of numerical relativity simulations, which we leave for future work.}

\subsubsection{Measurement of non-Kerr signals}
\label{sec:analysis:pert:nonkerr}

\begin{figure*}[p]
  \includegraphics[width=\columnwidth]{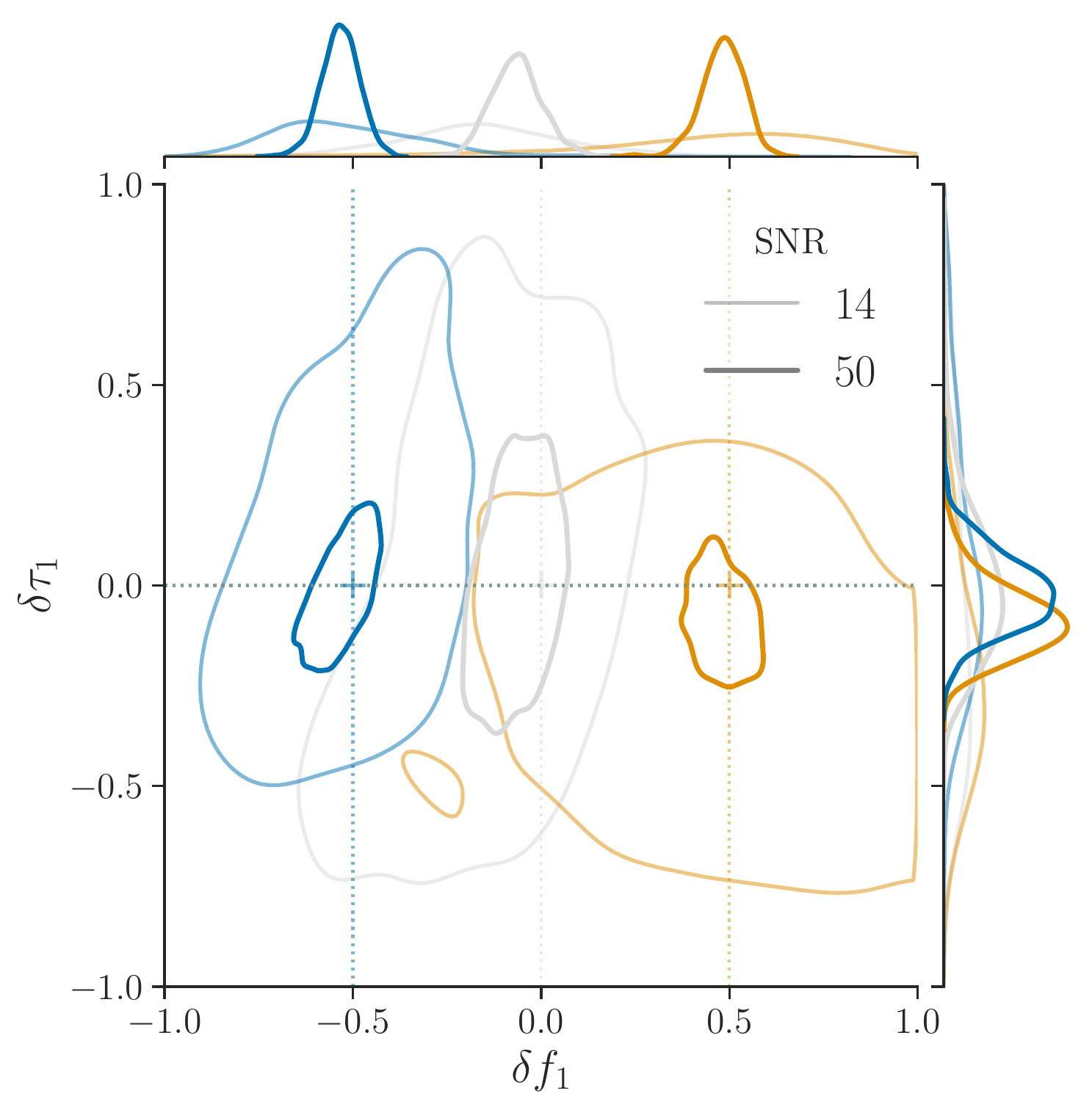}
  \includegraphics[width=\columnwidth]{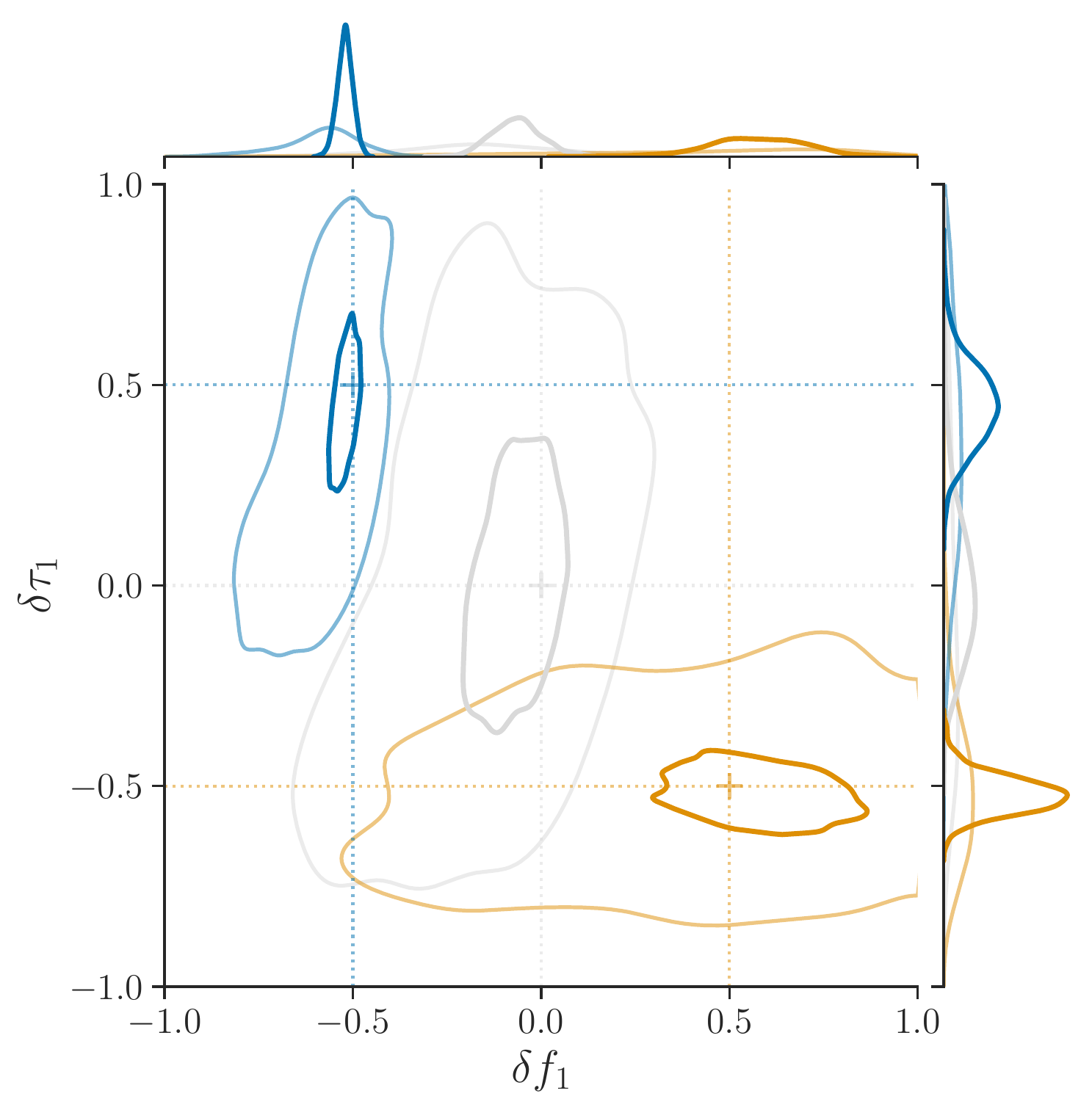}
  \includegraphics[width=\columnwidth]{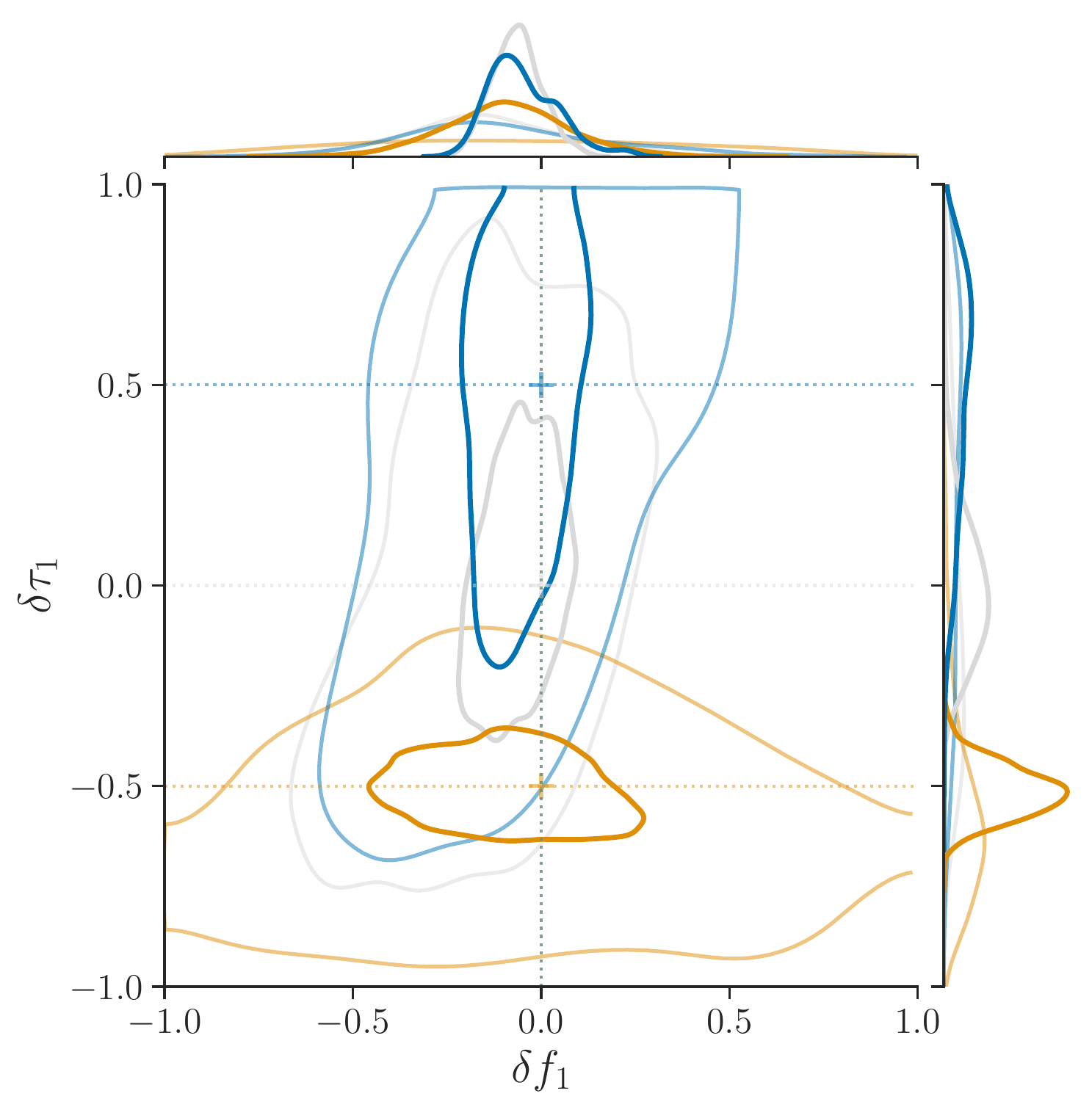}
  \includegraphics[width=\columnwidth]{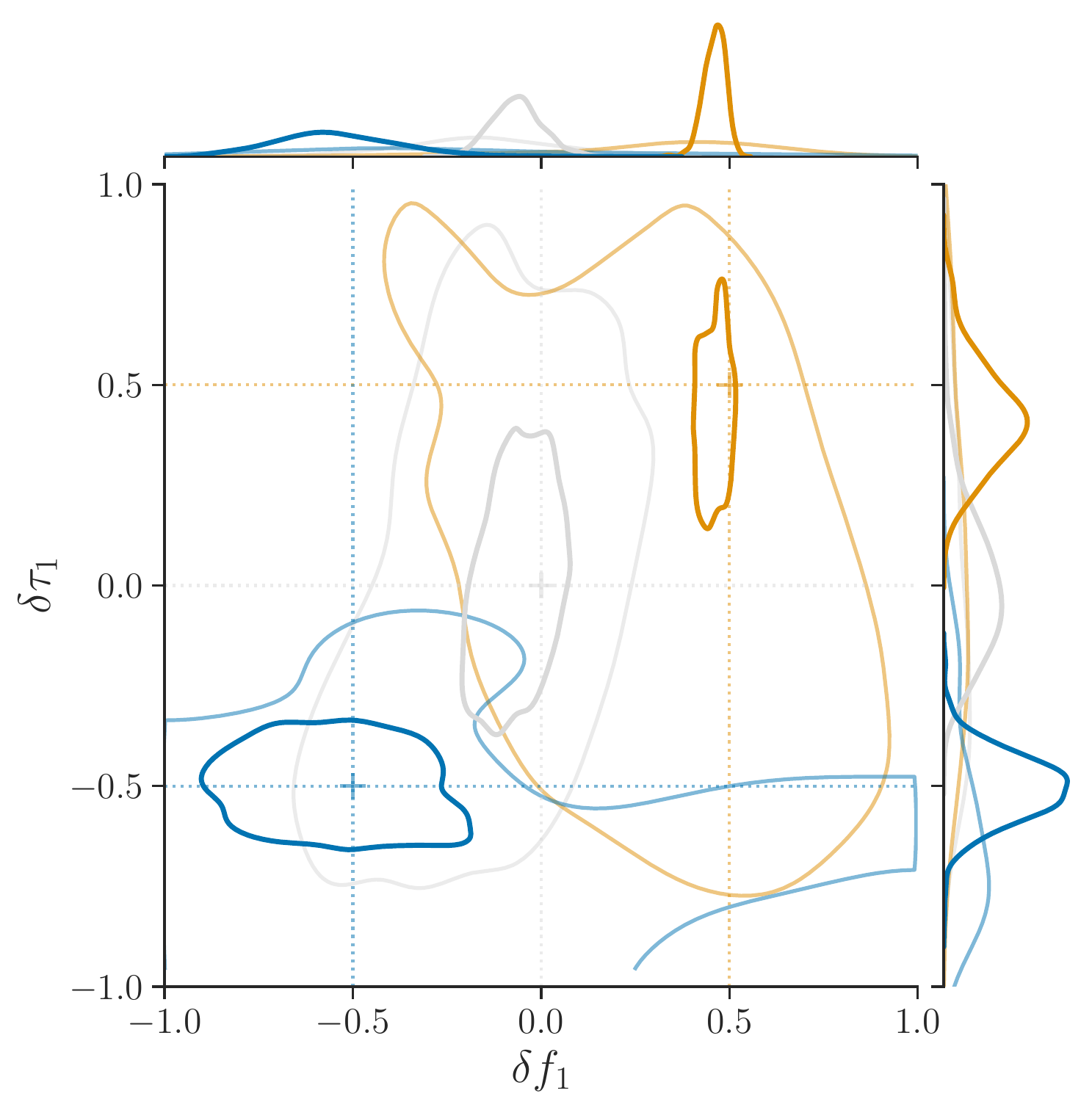}
  \caption{Simulated $\delta f_1$ and $\delta \tau_1$ measurements for different injected values (color) and  SNRs ({14}, thin trace; {50}, thick trace). The signal is as in Fig.~\ref{fig:analytic_nongr_n1_models}, except for the nonzero values of the deviation parameters (crosses, dotted lines).
  We show the case in which the true values are $\delta f_1 = \delta \tau_1 =0$ in all panels for reference (gray).
  Contours enclose 90\% of the probability mass.}
  \label{fig:analytic_df_dtau}
\end{figure*}

If a given ringdown signal does not conform to a Kerr spectrum, our model should be able to identify this.
Extending the example above, in Fig.~\ref{fig:analytic_df_dtau} we demonstrate that we can measure deviations in the spectrum of the first $\ell=m=2$ overtone of an $N=1$ injection with different true values of $\delta f_1$ and $\delta \tau_1$.
The baseline Kerr signal that we perturb is the same as in the previous section, with parameters given in Table \ref{tab:intrinsic_gr}, and we select both $\delta f_1$ and $\,\delta \tau_1$ from the set $\{-0.5,\, 0,\, +0.5\}$.
For each case, we scale the overall amplitude of the signal so that the injected network SNR is either {14} or {50}, as indicated by the label.

As before, although the posteriors in Fig.~\ref{fig:analytic_df_dtau} are specific to this configuration, some of their overarching features are worth noting.
First, as expected, the median of the distributions tracks the true value, and the precision of the measurement increases for higher SNR.
Second, $\delta f_1$ and $\delta \tau_1$ are highly interdependent, in the sense that the quality (width) of the measurement of either quantity can vary strongly with the true value of the other.
For example, from the $\delta f_1$ marginals (top sidepanels) on the right column of Fig.~\ref{fig:analytic_df_dtau}, we see that the $\delta f_1$ constraint tends to be tighter for positive values of $\delta \tau_1$, and worsens noticeably for negative values; by the same token, from the top left plot, we see that the $\delta\tau_1$ constraint tends to be narrower for nonzero values of $\delta f_1$ (on the right sidepanel, compare extents of {blue} or {orange} marginals vs the gray one).
This behavior is consistent with the intuition we built above by studying Kerr signals: for increasingly negative values of $\delta \tau_1$, the overtone approximates a Kronecker delta at $t_0$ and its frequency becomes unmeasurable; on the other hand, nonzero values of $\delta f_1$ increase the frequency beating between the two modes, making it easier to constrain $\delta \tau_1$.

Besides effects like the above, the results in Fig.~\ref{fig:analytic_df_dtau} are influenced by the fact that, as we change the true $\delta f_1$ and $\delta \tau_1$, we must also increase or decrease the overall signal amplitude to keep the SNR the same.
For example, a higher frequency overtone due to $\delta f_1 > 0$ may contribute more significantly to the total SNR, leading us to lower the overall amplitude of the injection, thus making the fundamental harder to detect.
This effect will introduce nontrivial correlations depending on the signal phasing (cf.~Fig.~\ref{fig:phasing_example}).

\begin{figure}
  \includegraphics[width=\columnwidth]{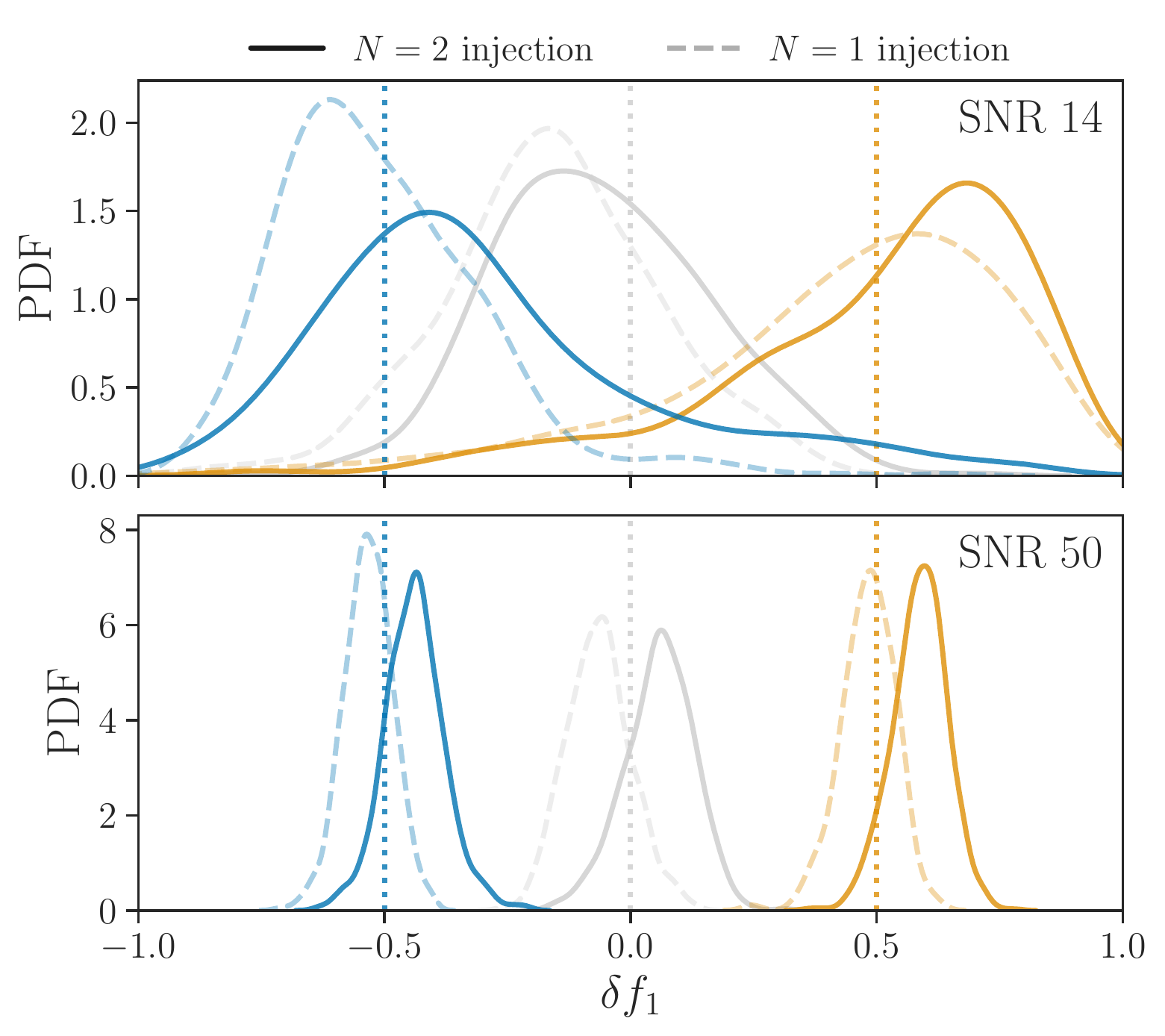}
  \caption{Measurement of the first overtone frequency deviation $\delta f_1$ in the presence of the second overtone, which is not accounted for in the recovery model (as in Fig.~\ref{fig:analytic_df_dtau_N1_Ninj2}).
  Color encodes the true injected value, as marked by the dotted lines: $\delta f_1 = -0.5$ ({blue}), $\delta f_1 = 0$ ({gray}), or $\delta f_1 = 0.5$ ({red}); $\delta \tau_1 = 0$ in all injections, although the model allowed this quantity to vary.
  For comparison, the dashed curves show the results obtained in the abscence of the second overtone, as in the top left panel of Fig.~\ref{fig:analytic_df_dtau}.
  The injected SNR was {14} (top) or {50} (bottom).
  }
  \label{fig:analytic_df_N1_Ninj2}
\end{figure}

Finally, we asses the potential effect of modeling systematics on this kind of measurement, as we did above for the case of a Kerr injection (Fig.~\ref{fig:analytic_df_dtau_N1_Ninj2}).
Here again, the results do not qualitatively change in the presence of the second overtone for moderate SNRs.
We show this by repeating the experiment from the end of Sec.~\ref{sec:analysis:pert:kerr} and injecting an $N=2$ signal without accounting for it in the recovery model; the fundamental mode and second overtone conform to the Kerr spectrum.
This time, however, we additionally perturb the frequency and damping time of the first overtone in the injection and repeat the analysis from Fig.~\ref{fig:analytic_df_dtau}.
Overall, the results in the presence of the $n=2$ mode are similar results to those in Fig.~\ref{fig:analytic_df_dtau} when the SNR is low.
As an example, Fig.~\ref{fig:analytic_df_N1_Ninj2} shows the $\delta f_1$ measurement for injections with $\mathrm{SNR} = 14$ (top) and 50 (bottom), $\delta \tau_1=0$ and $\delta f_1 = -0.5, 0, +0.5$.
Comparing to the distributions from the top left panel of Fig.~\ref{fig:analytic_df_dtau} (reproduced with a {dashed} trace in Fig.~\ref{fig:analytic_df_N1_Ninj2}), we conclude that bias is not significant at currently achievable SNRs.
Even though the effect of the $n=2$ mode starts becoming visible at SNR 50 (as it did in Fig.~\ref{fig:analytic_df_dtau_N1_Ninj2}), it is not significant at the 90\% credible level in this case.

\begin{figure*}
  \includegraphics[width=\columnwidth]{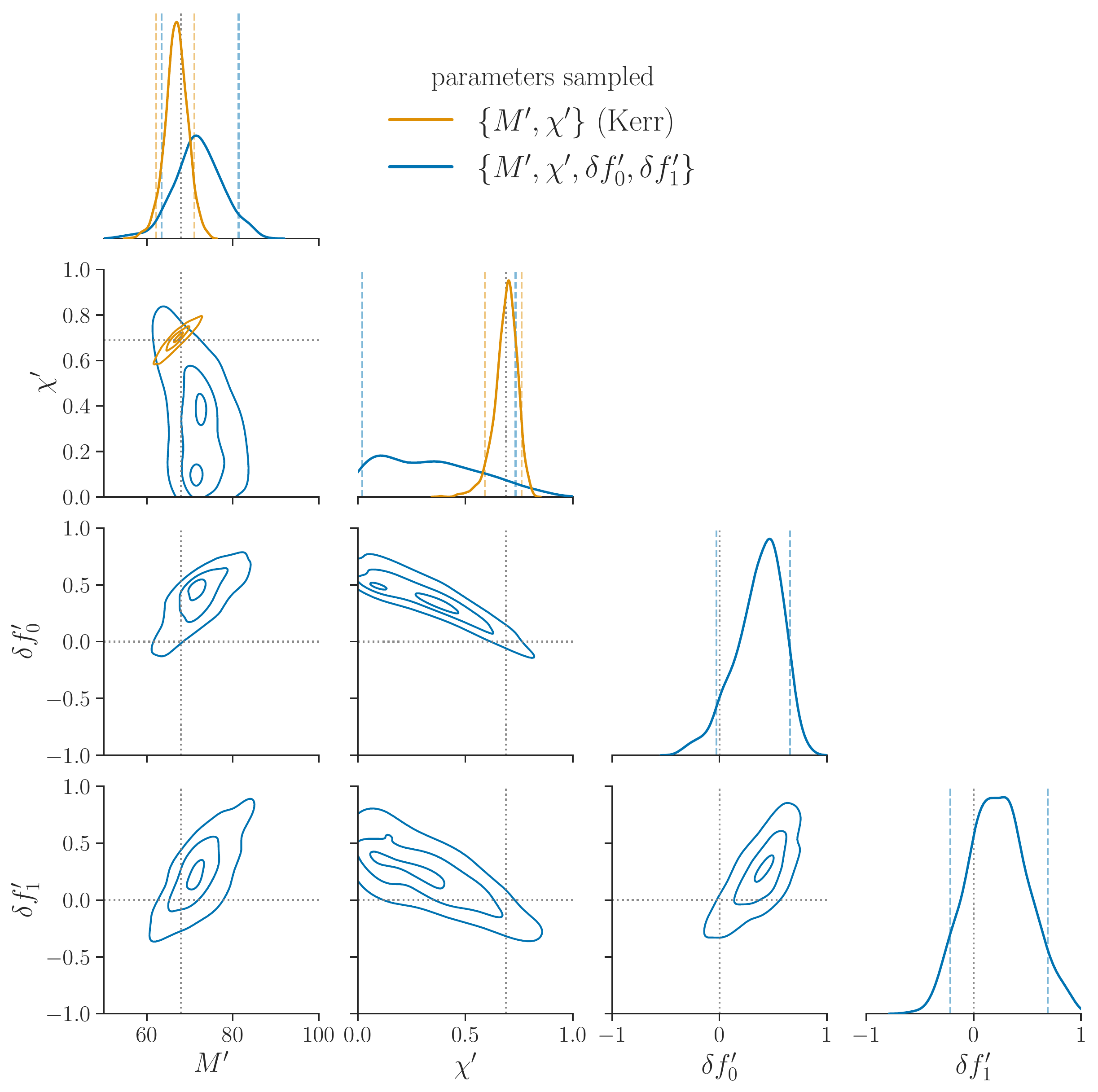}
  \includegraphics[width=\columnwidth]{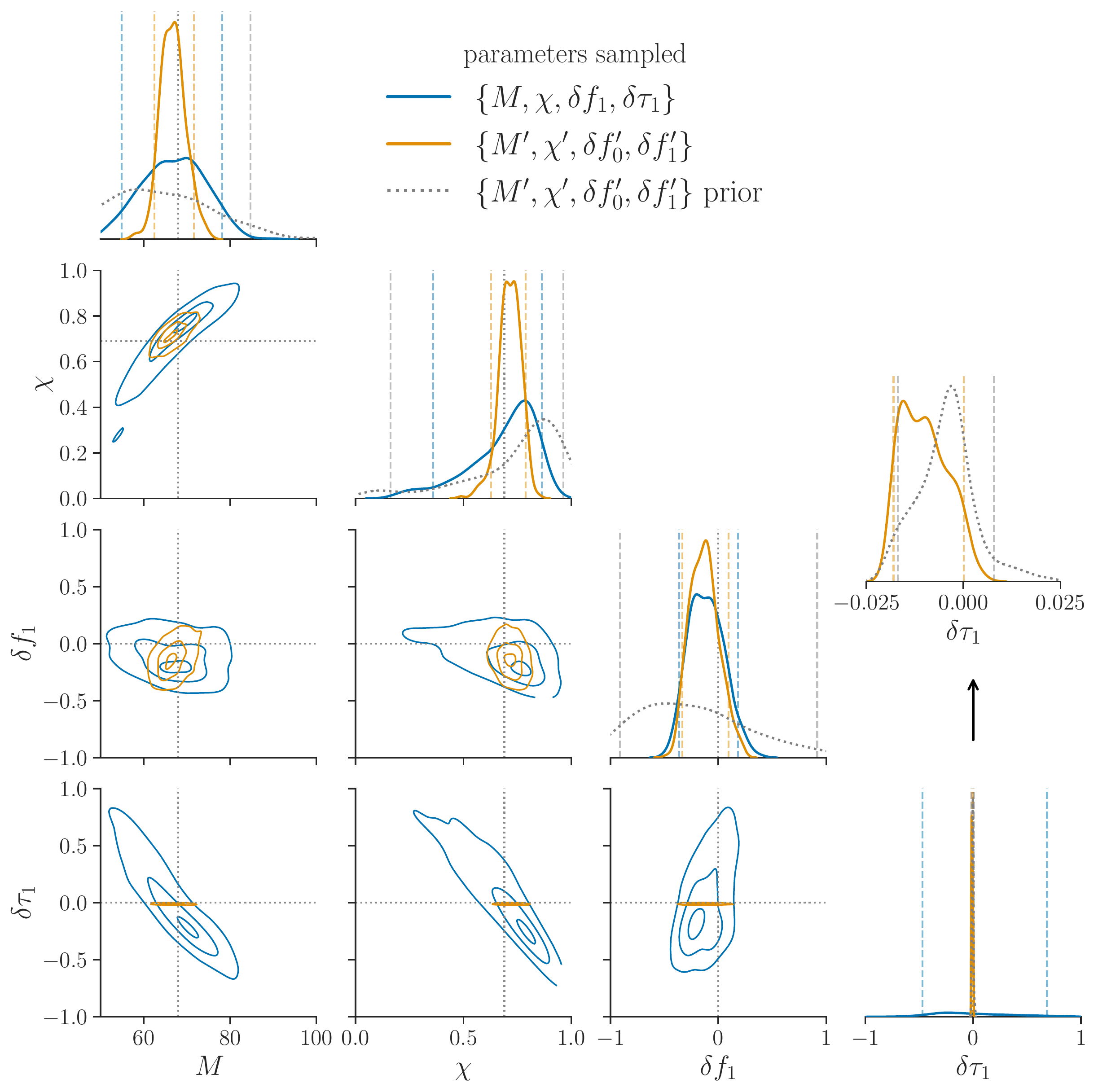}
  \caption{\label{fig:four-parameter-150914} Alternative parameterizations of deviations from the Kerr $N=1$ ringdown.
  On the left, we show the result of analyzing the Kerr $N=1$ signal in Fig.~\ref{fig:analytic_nongr_n1_models} (SNR 20) using the $\{M',\chi',\delta f_0', \delta f_1'\}$ parameterization of Eqs.~\eqref{eq:nongr_df0_df1} ({blue}), compared to a measurement assuming Kerr ({orange}).
  On the right, we show the same result transformed into our usual $\{M,\chi,\delta f_1, \delta \tau_1\}$ coordinates ({orange}), compared to the implied prior on this space ({dotted gray}) and a measurement carried out natively in the $\{M,\chi,\delta f_1, \delta \tau_1\}$ parameterization with flat priors on those quantities ({blue}).
  Contours enclose 90\%, 50\% and 10\% of the probability, while vertical dashed lines mark the symmetric 90\% credible interval.
  The effective prior
  induced on $\delta \tau_{1}$ (see Fig.~\ref{fig:four-parameter-posterior})
  by the inclusion of the $\delta f_{0}'$ parameter only allows for a very
  narrow range of values to be explored (around $\pm2\%$). Therefore, the addition of the
  $\delta f_{0}'$ degree of freedom does not meaningfully expand the range of
  parameters considered in the fit. }
\end{figure*}

\subsubsection{Alternative parameterizations}
\label{sec:analysis:pert:alt}

So far, we have explored departures from Kerr by parameterizing deviations from the $\ell=|m|=2,N=1$ model in terms of the overtone variables $\delta f_1$ and $\delta \tau_1$, as in Eqs.~\eqref{eq:nongr_df1_dtau1}.
However, as discussed in Sec.~\ref{sec:model:pert}, one could conceive of alternative parameterizations of the two-mode spectrum, in which the beyond-Kerr parameters are assigned to the fundamental mode quantities, as in Eqs.~\eqref{eq:nongr_df0_df1}.
In Sec.~\ref{sec:model:pert}, we argued that such alternative parameterizations are inferior to our default choice $(\delta f_1, \delta \tau_1)$ because they are effectively degenerate---that is, they have a reduced number of active degrees of freedom.
Here, we demonstrate this near degeneracy in a more concrete way by looking at simulated measurements.

To do this, we take the simulated $N=1$ Kerr signal studied above (e.g., Fig.~\ref{fig:analytic_nongr_n1_models}) at SNR 20, and analyze it using the primed parameterization of Eqs.~\eqref{eq:nongr_df0_df1}, $\{M',\chi',\delta f_0', \delta f_1'\}$.
This means that the model allows both frequencies to deviate from the Kerr values, rather than the frequency and damping time of the overtone as we had done so far.
We show the result in its native $\{M',\chi',\delta f_0', \delta f_1'\}$ parameterization in the left panel of Fig.~\ref{fig:four-parameter-posterior}.
As expected, allowing the fundamental to deviate from the Kerr prediction results in a highly degraded mass-spin distribution, which no longer resembles the much tighter posterior obtained assuming a Kerr spectrum (upper corner); this is unlike in our usual $\{M,\chi,\delta f_1, \delta \tau_1\}$ parameterization, wherein the mass-spin structure is broadened with respect to Kerr, but not destroyed.

The $\delta f_0'$ and $\delta f_1'$ marginals appear to return a measurement on the left of Fig.~\ref{fig:four-parameter-150914}.
However, in reality the four parameters $\{M',\chi',\delta f_0', \delta f_1'\}$ only encode three true degrees of freedom, as we argued in Sec.~\ref{sec:model:pert}.
The right panel in Fig.~\ref{fig:four-parameter-150914} illustrates this again for this measurement in particular by translating the $\{M',\chi',\delta f_0', \delta f_1'\}$ posterior on the right panel into the equivalent $\{M,\chi,\delta f_1, \delta \tau_1\}$ values, using Eqs.~\eqref{eq:Q220-transf}--\eqref{eq:dtau1_map}.
In this transformed space, the mass-spin posterior recovers its familiar correlation structure.
Furthermore, the result clearly shows that the $\{M',\chi',\delta f_0', \delta f_1'\}$ fit was equivalent to a three-parameter $\{M,\chi,\delta f_1\}$ fit with $\delta \tau_1 \approx 0$.
In fact, fitting $\{M',\chi',\delta f_0', \delta f_1'\}$ is functionally identical to to fitting $\{M,\chi,\delta f_1\}$ or $\{M',\chi',\delta f_0'\}$, in
spite of the apparent addition of a new parameter.
This again indicates that replacing $\delta \tau_{1}$ parameter with $\delta f_{220}'$ precludes meaningful exploration of the two-mode parameter space.

\begin{figure}
  \includegraphics[width=\columnwidth]{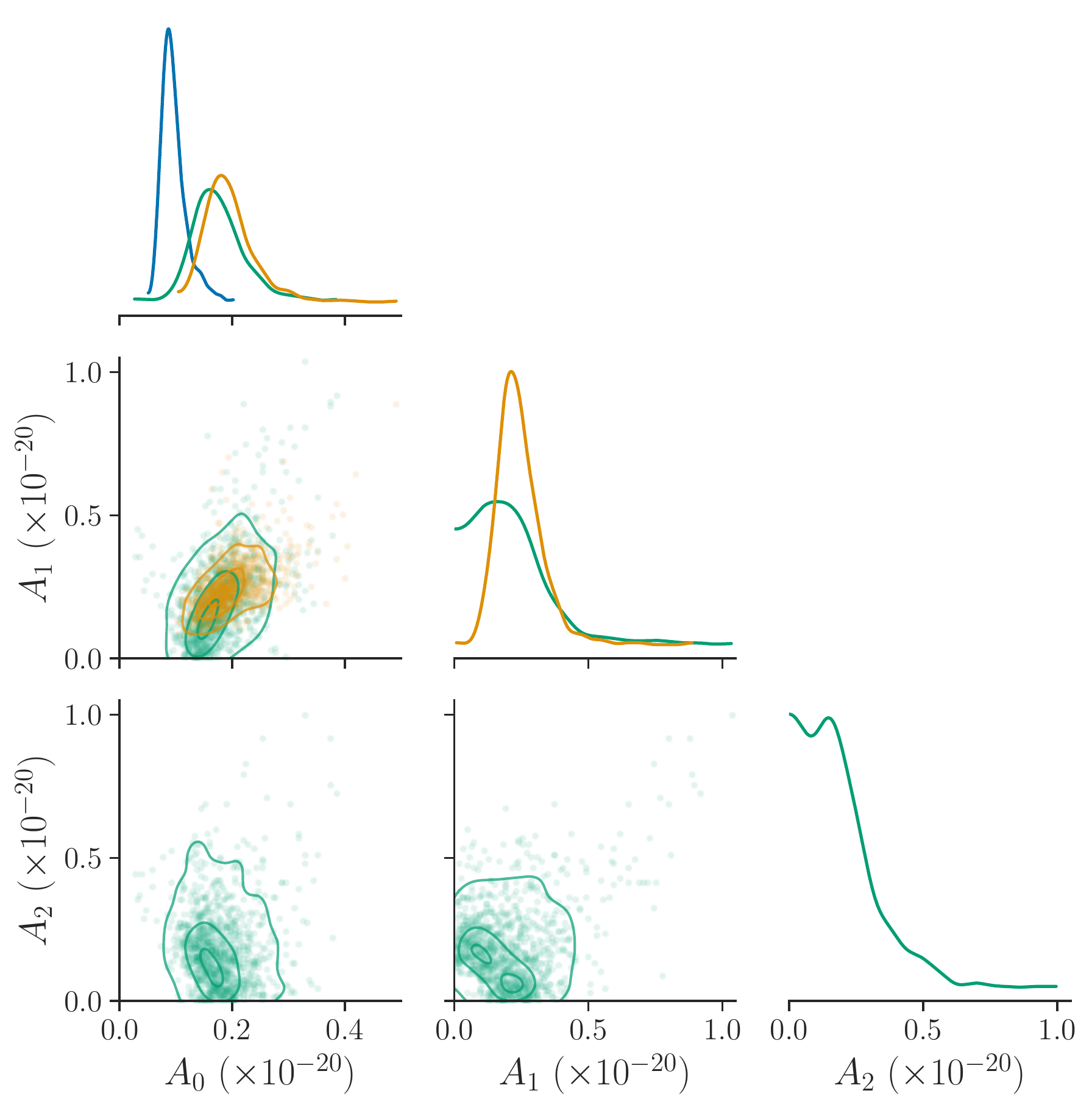}
  \caption{Amplitude posteriors for a GW150914-like numerical relativity simulation with a ringdown SNR of {12}, analyzed at peak strain with varying number of overtones (color) of the $\ell=|m|=2$ angular harmonic assuming a Kerr spectrum. The $N=1$ posterior (orange) indicates $A_0,A_1>0$, favoring the presence of two modes, while the $N=2$ posterior (green) displays an arch in the $A_1{-}A_2$ posterior that suggests the data do not require a third mode; following the procedure in Sec.~\ref{sec:ds:det}, this means we favor the $N=1$ model. Contours enclose 90\%, 50\% and 10\% of the posterior probability.}
  \label{fig:nr_amps}
\end{figure}

\subsection{Numerical relativity}
\label{sec:nr}

The damped-sinusoid studies in the preceding sections show how we can detect and characterize QNMs under different scenarios to learn about the source BH and verify agreement with the Kerr metric.
Because our method is insensitive to data before the truncation time $t_0$ (Secs.~\ref{sec:inference} and \ref{sec:ds:efficacy}), the general insights derived above must apply also to the ringdown portion of a binary black hole IMR signal.
To demonstrate this concretely, we inject a numerical relativity waveform into synthetic LIGO Hanford and Livingston noise, and operate as we would on real data.
The results of this exercise reproduce some of the key conclusions from the sections above, as well as with real and simulated data in \cite{Giesler:2019uxc,Isi:2019aib,Abbott:2020tfl,Abbott:2020mjq,Abbott:2020jks}

\begin{figure}
  \includegraphics[width=\columnwidth]{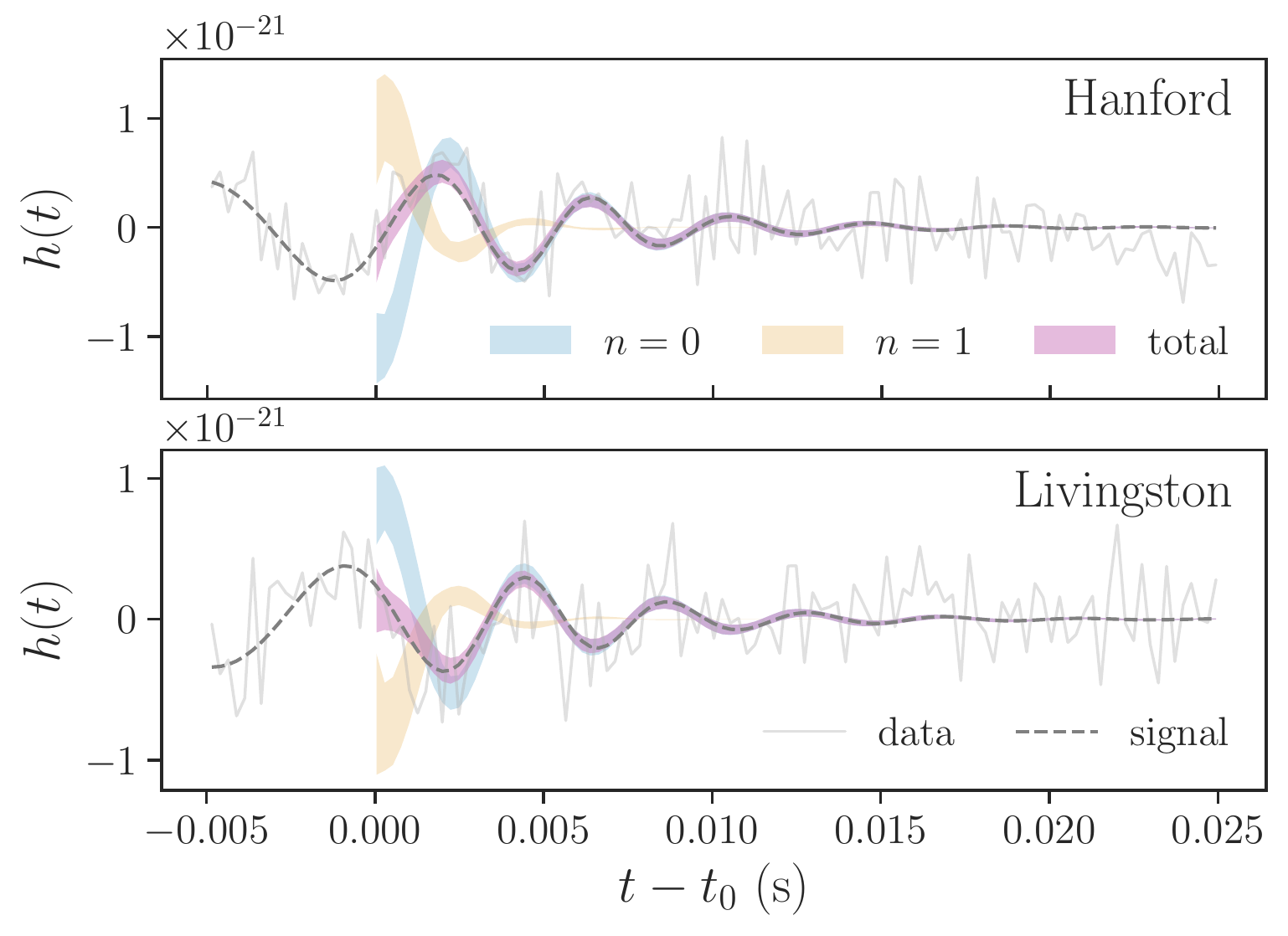}
  \caption{Reconstruction of numerical relativity ringdown strain using two tones ($N=1$) of the $\ell = |m|=2$ harmonic, assuming a Kerr spectrum.
  The $N=1$ model provides an accurate fit of the data after the strain peak at each detector ($t_0 = t_p + \delta t_{H/L}$), as seen by the agreement of the 90\%-credible reconstruction ({magenta} envelope) with the injected signal (dashed) for both LIGO Hanford (top) and Livingston (bottom).}
  \label{fig:nr_strain}
\end{figure}

\begin{figure}
  \includegraphics[width=\columnwidth]{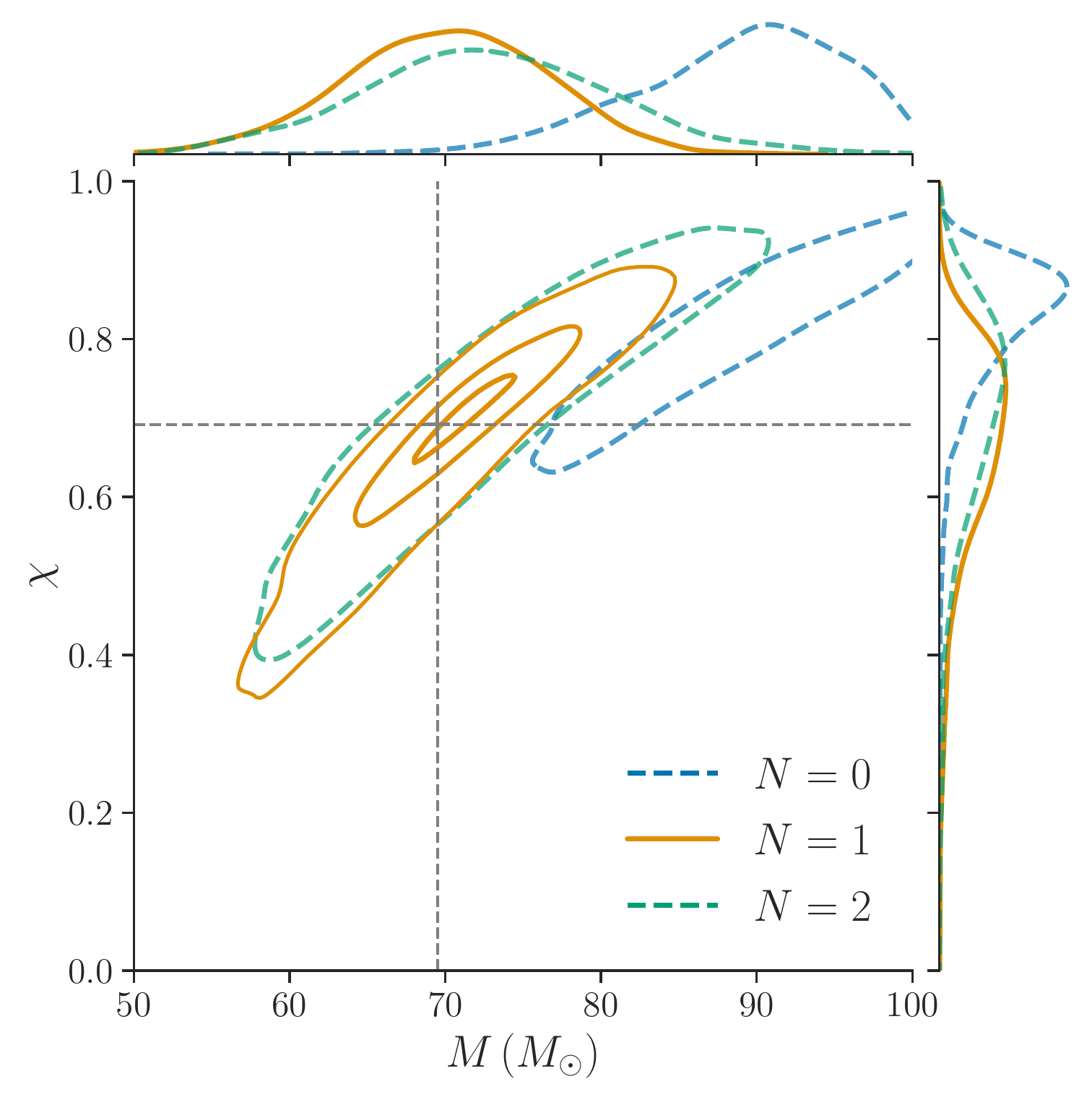}
  \caption{Remnant mass and spin measured from numerical relativity ringdown using a varying number of overtones of the $\ell=|m|=2$ angular harmonic: $N=0$ (blue), $N=1$ (orange) and $N=2$ (green). All analyses fit data starting at peak strain assuming a Kerr spectrum. As expected, $N=0$ is not an adequate model in this regime so it returns a measurement of $M$ and $\chi$ inconsistent with the true values (crosshairs); this is corrected by including overtones. Contours enclose 90\% (as well as, for $N=1$, 50\% and 10\%) of the posterior probability.}
  \label{fig:nr_mchi}
\end{figure}

We take as ground truth the simulation \nrsim \cite{kidder_larry_2019_3301877} from the SXS catalog  \cite{Boyle:2019kee}, which is consistent with GW150914 when scaled to a total (redshifted) binary mass of ${73}\, M_\odot$; we assume the source is oriented with its orbital angular momentum pointing to Earth (face on), and include all available angular harmonics using the infrastructure in \cite{Schmidt:2017btt,lalsuite}.
As in \cite{Giesler:2019uxc,Isi:2019aib}, we chose a truncation time corresponding to the peak of the injected complex strain, i.e., we set $t_\mathrm{start} = t_p$ for peak time $t_p$ such that
\begin{equation}
h_+^2(t_p) + h_\times^2(t_p) \equiv \max_t \left[ h_+^2(t) + h_\times^2(t) \right],
\end{equation}
with relative delays at each detector corresponding to the chosen sky location (Table \ref{tab:extrinsic}), as prescribed by Eq.~\eqref{eq:s}.
We assume the peak time is known a priori, rather than reconstruct it from the simulated data.

\begin{figure*}
  \includegraphics[width=1.5\columnwidth]{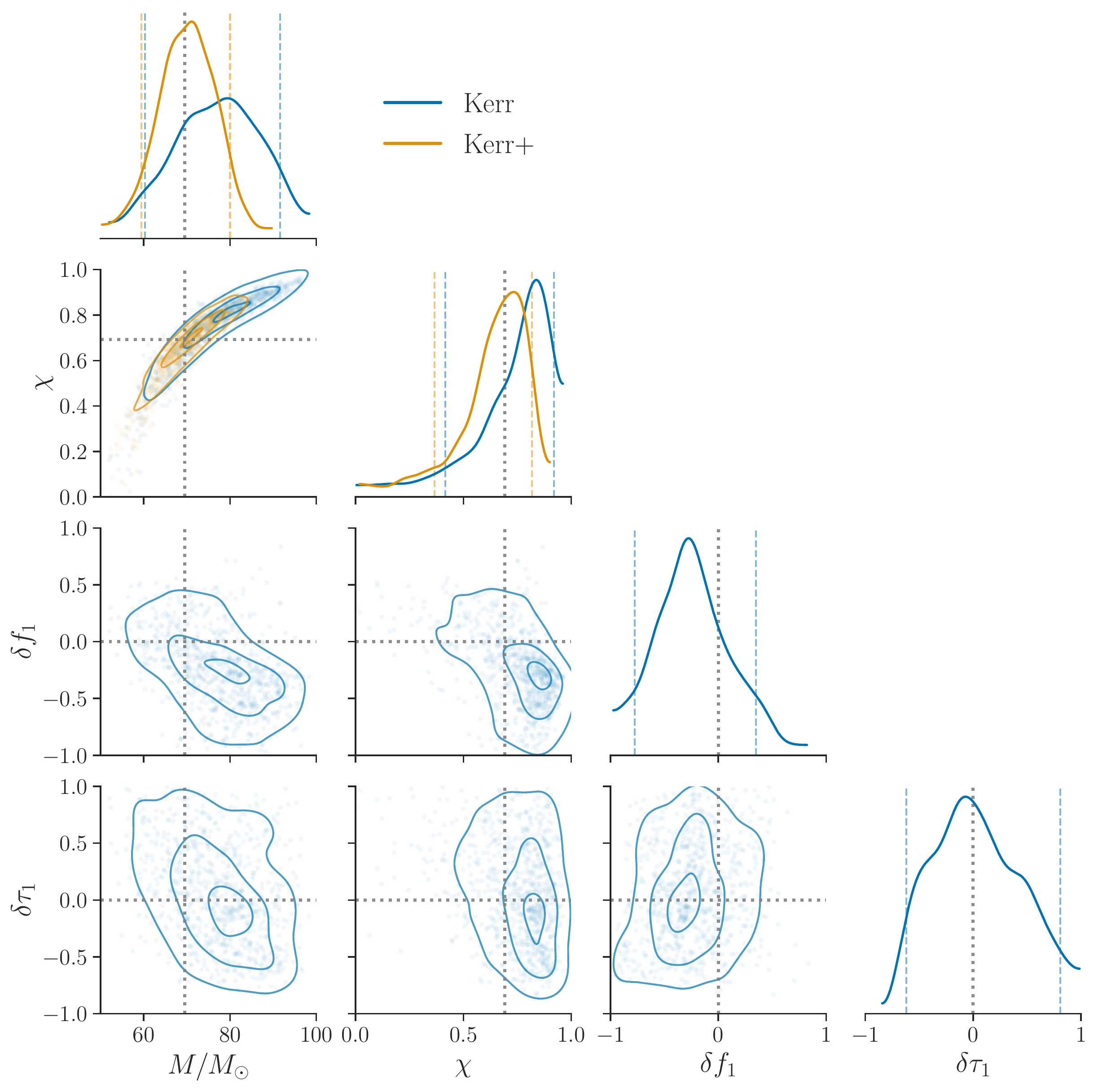}
  \caption{Results of a spectroscopic analysis applied to a GW150914-like numerical relativity ringdown in synthetic LIGO Hanford and Livingston noise, with an injected post-peak SNR of 12.
  We measure two tones ($N=1$) of the $\ell = |m|=2$ harmonic at peak strain, allowing for deviations from the Kerr spectrum via fractional deviations on the overtone frequency and damping time, $\delta f_1$ and $\delta \tau_1$.
  The distributions represent the posterior on those parameters, as well as the remnant mass $M$ and spin $\chi$, with the true values indicated by dotted gray lines.
  Besides the beyond-Kerr result ({blue}), we also show the corresponding Kerr measurement of the remnant parameters ({orange}), as obtained in Fig.~\ref{fig:nr_mchi}.
  Contours enclose 90\%, 50\% and {10\%} of the probability mass, while vertical dashed lines over the marginals mark the symmetric 90\% credible interval.
  The data are consistent with a Kerr spectrum ($\delta f_1 = \delta \tau_1=0$), finding $\delta f_1 = {-0.26^{+0.61}_{-0.51}}$ at 90\% credibility, which is comparable to actual GW150914 measurements \cite{Abbott:2020jks,GWTC2:TGR:release,Isi:2019aib} (see main text for details).
  }
  \label{fig:nr_dfdtau}
\end{figure*}

To start, we scale the overall signal amplitude (equivalently, the luminosity distance) to obtain an injected post-peak network SNR of {12}, comparable to the value recovered for GW150914.
Then, assuming a Kerr spectrum parameterized as in Eq.~\eqref{eq:params_gr}, we follow the procedure outlined in Sec.~\ref{sec:ds:det} to determine the number of detected modes.
Since we know the quadrupole to dominate for the injected signal, we focus on models with $N$ overtones ($\nmode=N+1$ tones, including the fundamental) of the $\ell=|m|=2$ harmonic as we have been doing so far.
Accordingly, we analyze the signal starting with $N=0$ and successively increase $N$ until the amplitude posterior no longer indicates a preference for that many modes.

We show the amplitude posterior for the case at hand in Fig.~\ref{fig:nr_amps}, for $N \leq 2$.
The distribution for of the $N=1$ model (orange) requires a nonzero contribution from both the fundamental mode and first overtone, $A_0, A_1 >0$, at 90\% credibility; however, the $N=2$ posterior (green) does not require both $A_1$ and $A_2$ to be simultaneously nonzero.
Since the presence of a third mode is not unequivocally established, this favors the two-tone model, $N=1$, over both $N=0$ and $N=2$.

Indeed, the $N=1$ model is sufficient to adequately reconstruct this signal, as we show in Fig.~\ref{fig:nr_strain}.
Notice that the $n=0$ and $n=1$ modes are inferred to be ${\sim}\pi~{\rm rad}$ out out phase, as in the analytic signals we studied above (c.f.~Figs.~\ref{fig:analytic_strain_M2} and \ref{fig:analytic_n1_wf}).
Those examples were constructed to emulate this feature, which was also seen in \cite{Giesler:2019uxc}.

\begin{figure}
  \includegraphics[width=\columnwidth]{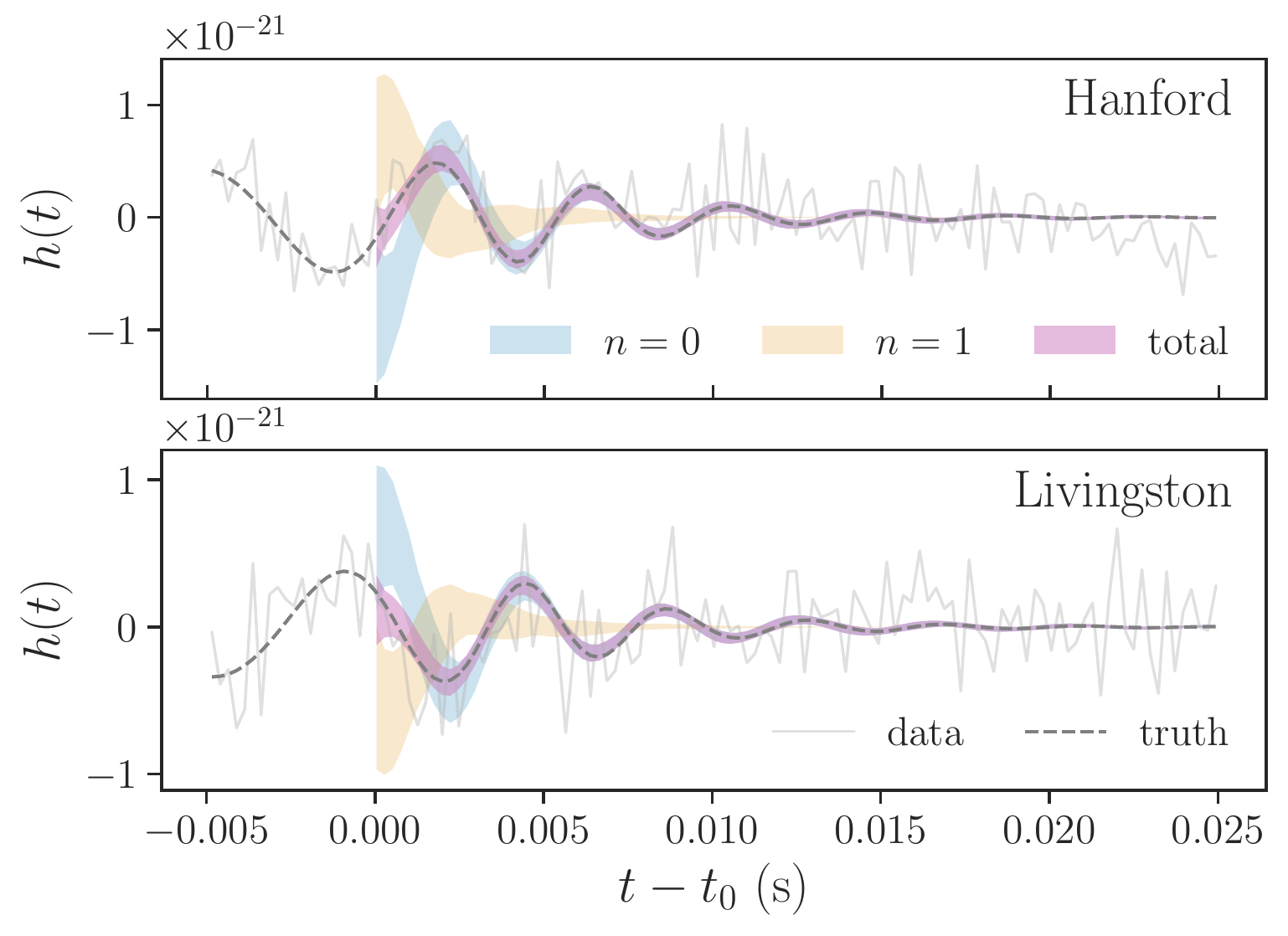}
  \caption{Reconstruction of numerical relativity ringdown strain using two tones ($N=1$) of the $\ell = |m|=2$ harmonic, allowing for deviations from Kerr spectrum through fractional deviations on the overtone parameters $\delta f_1$ and $\delta \tau_1$.
  The $N=1$ model provides an accurate fit of the data after the strain peak at each detector ($t_0 = t_p + \delta t_{H/L}$), as seen by the agreement of the 90\%-credible reconstruction ({magenta} envelope) with the injected signal (dashed) for both LIGO Hanford (top) and Livingston (bottom).}
  \label{fig:nr_strain_nongr}
\end{figure}

The $N=1$ and $N=2$ Kerr models both recover with high credibility the true remnant mass and spin known from the numerical relativity data (Fig.~\ref{fig:nr_mchi}).
The uncertainty is slightly greater for the $N=2$ case, which we expect from the fact that it introduces additional parameters that are not required to fit the data ($A_2, \epsilon_2, \theta_2, \phi_2$).
Although the $N=1$ model is known not to be a full description of the signal at the peak due to the presence of higher overtones, the systematic error from leaving those out is smaller than the statistical error at this SNR (e.g., Fig.~\ref{fig:analytic_df_dtau_N1_Ninj2}).
The same is not true for the $N=0$ model, which returns a visibly biased estimate of those quantities.
This was the behavior observed in \cite{Giesler:2019uxc,Isi:2019aib}, and agrees with the theoretical expectation that a template with insufficient modes should suffer from bias \cite{Berti:2007zu}.

Having established $N=1$ to be the most suitable model for these data, we can search for deviations from the Kerr spectrum through Eqs.~\eqref{eq:nongr_df1_dtau1}.
Figure \ref{fig:nr_dfdtau} shows the resulting posterior on the overtone deviation parameters $\delta f_1$ and $\delta \tau_1$ (lower right), together with $M$ and $\chi$ (upper left).
The joint posterior on the fractional deviations indicates that the data are consistent with a Kerr $N=1$ spectrum ($\delta f_1 = \delta \tau_1 = 0$) within 90\% credibility.
As we have come to expect from the analytic injections above (e.g., Fig.~\ref{fig:analytic_nongr_n1_models}), as well as the analyses of real data, the fractional deviation in the frequency is reasonably well-constrained to $\delta f_1 = {-0.26^{+0.61}_{-0.51}}$ at 90\% credibility; on the other hand, the fractional deviation in the damping time is only poorly bounded to $\delta \tau_1 = {0.03^{+0.78}_{-0.65}}$, spanning the full range of the prior.
For comparison, the GW150914 data were found to constrain $\delta f_1 = 0.13^{+0.59}_{-0.77}$ in \cite{Abbott:2020jks,GWTC2:TGR:release}, and $\delta f_1 = -0.08^{+0.44}_{-0.34}$ in \cite{Isi:2019aib}, both reported as symmetric 90\%-credible interals around the median.
\footnote{We computed the number for \cite{Abbott:2020jks} from the LIGO-Virgo samples publicly available in \cite{GWTC2:TGR:release}. We computed the result for \cite{Isi:2019aib} from the same samples used in that publication, which only explicitly reported the mean and standard deviation ($\delta f_1 = -0.05\pm0.2$).}
Those results appear consistent, especially considering they used slightly different models (see App.~\ref{app:slms} regarding the more constraining, circular-polarization model in \cite{Isi:2019aib}).
All these features also agree broadly with the analytic injections in Sec.~\ref{sec:analysis:pert:kerr}.

Finally, we show the signal reconstruction obtained with the beyond-Kerr model in Fig.~\ref{fig:nr_strain_nongr}.
Just as we had seen for the analytic injection in Fig.~\ref{fig:analytic_n1_wf}, uncertainty in the overall reconstruction is of a similar magnitude as in the Kerr case (compare {magenta} curve to that in Fig.~\ref{fig:nr_strain}), while the decomposition into the individual modes is more uncertain (notice broader envelopes for $n=0$ and $n=1$ in Fig.~\ref{fig:nr_strain_nongr} relative to Fig.~\ref{fig:nr_strain}).
This reflects the greater flexibility afforded by the $\delta f_1$ and $\delta \tau_1$ deviation parameters.

Since here we are studying a specific example, the results in this section should not be taken to be universally valid for all numerical relativity simulations; rather, as we have noted above, we expect the details at this SNR to vary somewhat with the specifics of the signal and noise configuration.
Nevertheless, this is sufficient to show that the observational results in \cite{Isi:2019aib,Abbott:2020jks} are consistent with simulations.
A systematic study of numerical relativity ringdowns over parameter space will be the subject of future work.

\section{Conclusion}
\label{sec:conclusion}

Ringdowns encode invaluable information about the structure of BH spacetimes, making them a crucial observable for fundamental physics and one of the primary targets of GW astronomy.
However, in spite of the vast literature on the topic, it is only recently that full-fledged, realistic techniques have been developed to tackle the unique data analysis challenges that ringdowns present.
Properly addressing these issues is not just a matter of technicalities: inadequate data analysis treatments can result in erroneous projections and biased measurements.
In this paper, we have aimed to rectify this by (1) providing an exhaustive overview of the challenges presented by ringdown analyses and how to overcome them; and, (2) using our robust analysis formalism to answer questions about mode detectatibility and resolvability, with a variety of realistic demonstrations on synthetic data focusing on tones of the $\ell = |m| = 2$ angular harmonic.
This contextualizes and supports recent observational results produced using variations of the techniques discussed here \cite{Isi:2019aib,Abbott:2020jks}.

The results of this paper can be split into three broad categories, corresponding to the three goals we outlined in the Introduction (Sec.~\ref{sec:intro}): provide a comprehensive account of the formalism behind ringdown-only analyses, examine the conditions for spectral characterization of ringdown signals, and demonstrate the use of our framework in answering questions about BH overtones.

The first set of results should be of interest to anyone attempting their own ringdown analyses, and comprises a discussion of how to construct a suitable ringdown template based on BH perturbation theory and how to set up a Bayesian likelihood that is blind to data before some designated start time.
Highlights of those results include the following observations regarding ringdown modeling (Sec.~\ref{sec:model}):
\begin{enumerate}
\item the most basic template suitable for data analysis treats each ringdown mode as an elliptically polarized damped sinusoid, with four degrees of freedom controlling the amplitude and phase of each mode;
\item the phenomenon of spherical-spheroidal mixing is irrelevant in generic spectroscopic studies, which are not informed by a specific model of the QNM excitation amplitudes;
\item to test the no-hair theorem, it is best to parameterize the frequency and damping of the best-measured mode as functions of $M$ and $\chi$, and only assign Kerr-deviation parameters to more loosely constrained modes (e.g., $n=1$ in an $N=1$ analysis)---doing otherwise precludes full exploration of the parameter space.
\end{enumerate}
They also include a number of statements about data analysis (Sec.~\ref{sec:inference}), among which we count:
\begin{enumerate}
\item it is not possible to fully isolate ringdown data without corruption using the traditional Fourier-based techniques prevalent in LIGO-Virgo analyses, because these are predicated on the ability to make the data and template periodic;
\item we can circumvent this issue by working fully in the time domain, but only if we use a \emph{noncirculant} covariance matrix and explicitly discard pre-ringdown data, as was first done in \cite{Isi:2019aib};
\item it is possible to construct an equivalent procedure in the frequency domain \cite{Capano2021}, but at greater computational cost.
\end{enumerate}
The time domain framework is useful not only for ringdown studies, but can also be repurposed to target other parts of the signal \cite{Isi:2020tac}.

The second set of results pertain to basic features of the detection and characterization of damped sinusoids in noisy data (Sec.~\ref{sec:ds}):
\begin{enumerate}
\item we can assert detection of $D$ ringdown modes by ensuring that the posterior unambiguously favors $D$ nonzero amplitudes, but not $D+1$---this avoids strong sensitivity to arbitrary prior choices;
\item it is incorrect to take the standard Rayleigh criterion of Eq.~\eqref{eq:rayleigh_f} as precondition for a frequency measurement, since a pair of damped sinusoids need not be separated in frequency to be resolved---Eq.~\eqref{eq:correctRayleigh} is a better heuristic;
\item being able to distinguish the damping times (or frequencies) of two modes does not imply high relative precision in measuring those parameters (i.e., resolving $\tau$ does not imply precision in $\delta\tau$).
\end{enumerate}

For the last set of results, we brought all the above insights together in the analysis of simulated BH ringdowns, focusing on spectroscopy with overtones of a GW150914-like remnant (Sec.~\ref{sec:analysis}).
Besides serving to demonstrate our whole framework in action end to end, this allowed us to conclude the following:
\begin{enumerate}
\item it is possible to detect and faithfully reconstruct the $\ell = |m|=2, n=1$ mode, and use it to constrain deviations from Kerr, with currently achievable SNRs;
\item although the $N=1$ model is known not to be a full description of IMR signals at the peak, we do not expect the associated systematic bias to play a role given current statistical uncertainties;
\item we can reproduce the $\delta f_1$ measurements for GW150914 in \cite{Isi:2019aib,Abbott:2020jks} with similar accuracy using a numerical relativity signal in synthetic LIGO noise.
\end{enumerate}
In summary, we have confirmed once more that ringdowns are detectable and resolvable with current detectors, and offer a comprehensive account of the framework necessary to do so.
We have also identified a number of points that demand further exploration.

Many of the examples discussed in this paper were intended to show that tones of the quadrupolar mode offer a viable avenue for BH spectroscopy.
This is motivated by the fact that the $\ell=|m|=2$ harmonic is almost always easier to detect.
It is certainly \emph{not} intended as an argument against the use of other angular angular harmonics: there is no question that, if discernible in the data, those additional modes would provide an exceptional tool for BH spectroscopy, perhaps probing different physics than $\ell=|m|=2$ overtones do.
Our formalism (as presented here, and as instantiated in the \textsc{ringdown} package) can seamlessly accommodate higher angular harmonics, but we leave a dedicated study of that use case for future work.

Additionally, the demonstrations in this paper focused on concrete examples, with a limited number of detector and signal configurations intended to illustrate specific points.
Future work will be required to fully explore the space of possible (or likely) signals.
This will allow us to draw global conclusions about the suitability of different signals for BH spectroscopy and to make robust projections for future detectors, which was not the goal of this paper.

The LIGO, Virgo and Kagra \cite{Aso:2013eba} detectors are currently undergoing major upgrades in anticipation of their fourth observing run, scheduled to begin in June 2022 \cite{Abbott:2020qfu,schedule}.
Once the instruments are back online, their enhanced sensitivity will allows us to detect large numbers of binary BHs with sufficient fidelity to carry out interesting ringdown analyses.
As the number and loudness of detections grows, our framework will make it possible to analyze the collection of sources without bias, and will enable increasingly powerful tests of the no-hair theorem and other predictions of GR.

\begin{acknowledgments}
We thank Leo C.~Stein and Saul A.~Teukolsky for insightful discussions.
We also thank Neil Cornish, Vitor Cardoso and Emanuele Berti for comments on the draft.
M.I.\ is supported by NASA through the NASA Hubble Fellowship
grant No.\ HST-HF2-51410.001-A awarded by the Space Telescope
Science Institute, which is operated by the Association of Universities
for Research in Astronomy, Inc., for NASA, under contract NAS5-26555.
The Flatiron Institute is a division of the Simons Foundation, supported through the generosity of Marilyn and Jim Simons.
This paper carries LIGO document number \dcc{}.
\end{acknowledgments}

\bibliography{refs}

\appendix
\section{Symmetry of the spin-weighted spheroidal harmonics}
\label{app:swsh_id}

Below Eq.~\eqref{eq:h_pt} we quoted the result that
\begin{equation} \label{eq:swsh_id}
{}_{\unaryminus 2} S_{[p]\ell \unaryminus m n}(\iota,\varphi) = (-1)^\ell {}_{\unaryminus 2} S^*_{[p]\ell m n}(\pi-\iota,\varphi)\, ,
\end{equation}
with the spin-weighted spheroidal harmonics ${}_{\unaryminus 2} S_{[p]\ell \unaryminus m n}(\iota,\varphi)$ as defined in Eq.~\eqref{eq:swsh}.
It is straightforward to derive this from the symmetries stated in Eq.~(48) of \cite{Cook:2014cta} (also see \cite{Press:1973zz}).
Starting from our definition in Eq.~\eqref{eq:swsh}, we can write
\begin{widetext}
\begin{align}
{}_{\unaryminus 2} S_{[p]\ell \unaryminus m n}(\iota,\varphi)
&= e^{-i m \varphi} {}_{\unaryminus 2} S_{\ell -m}\left(\chi t_M \tilde{\omega}_{[p]\ell \unaryminus m n}, \cos\iota\right) \label{eq:swsh_0}\\
&= (-1)^\ell e^{-i m \varphi} {}_{\unaryminus 2} S_{\ell m}\left(-\chi t_M \tilde{\omega}_{[p]\ell \unaryminus m n}, -\cos\iota\right) \label{eq:swsh_1}\\
&= (-1)^\ell e^{-i m \varphi} {}_{\unaryminus 2} S_{\ell m}\left(\chi t_M \tilde{\omega}_{[p]\ell m}^*, \cos(\pi-\iota)\right)\label{eq:swsh_2}\\
&=(-1)^\ell e^{-i m \varphi} {}_{\unaryminus 2} S^*_{\ell m}\left(\chi t_M \tilde{\omega}_{[p]\ell m}, \cos(\pi-\iota)\right) \label{eq:swsh_3}\\
&=  (-1)^\ell \left[e^{i m \varphi} {}_{\unaryminus 2} S_{\ell m}\left(\chi t_M \tilde{\omega}_{[p]\ell m n}, \cos(\pi-\iota)\right)\right]^*\label{eq:swsh_4}\\
&=  (-1)^\ell {}_{\unaryminus 2} S^*_{[p]\ell m n}(\pi-\iota, \varphi)\, .\label{eq:swsh_5}
\end{align}
\end{widetext}
In going from the first to the second line we applied Eq.~(48b) in \cite{Cook:2014cta} with $s=-2$.
We then took advantage of both the $\tilde{\omega}_{[p]\ell m n} =- \tilde{\omega}_{[p]\ell \unaryminus m n}^*$ symmetry and the fact that $\cos (\pi - \iota) = -\cos \iota$ to obtain Eq.~\eqref{eq:swsh_2}.
Next, in  Eq.~\eqref{eq:swsh_3}, we used Eq.~(48c) in \cite{Cook:2014cta}, and finally rearranged terms to arrive at Eq.~\eqref{eq:swsh_5}, which is our target as expressed in Eq.~\eqref{eq:swsh_id}.

This expression has often been presented without writing out the arguments of the spheroidal harmonics explicitly (e.g., \cite{Berti:2005ys,Buonanno:2006ui}), giving the wrong impression that ${}_{\unaryminus 2} S_{\ell \unaryminus m n}$ and ${}_{\unaryminus 2} S_{\ell m n}^*$ are directly interchangeable.

\section{Incorporating an amplitude model}
\label{app:slms}

We can narrow the scope of the ringdown template in Sec.~\ref{sec:template} by incorporating a model of the source of the BH perturbations.
This can be achieved by imposing symmetries on the unknown QNM amplitudes, or by re-parameterizing them in terms of fewer degrees of freedom (e.g., properties of the progenitors in a BH merger).
To do this, we must retain the angular factors in Eq.~\eqref{eq:h_pt} to obtain
\begin{align} \label{eq:h_pt_lmn_angles}
h_{j} &=  C_{\ell m n} e^{-i \tilde{\omega}_{\ell m n} t} {}_{\unaryminus 2} S_{\ell m n}(\iota,\varphi) \nonumber \\
&+ (-1)^\ell C_{\ell \unaryminus m n} e^{i \tilde{\omega}_{\ell m n} t} {}_{\unaryminus 2} S_{\ell m n}^*(\pi-\iota,\varphi) \, ,
\end{align}
instead of Eq.~\eqref{eq:h_pt_lmn}.
Here we have used the symmetry of spin-weighted spheroidal harmonics stated in Eq.~\eqref{eq:swsh_id}, which is derived in the previous appendix from properties shown in \cite{Press:1973zz,Cook:2014cta}.

Models for the amplitudes, as those provided in Ref.~\cite{London:2014cma}, can be plugged into Eq.~\eqref{eq:h_pt_lmn_angles} directly.
Alternatively, we may leave the amplitudes as free parameters but restrict them by enforcing symmetries.
For example, in the case of a non-precessing BH merger, we may expect the initial perturbations to be symmetric under equatorial reflections, satisfying $C_{\ell \unaryminus m n} = (-1)^\ell C_{\ell m n}^*$.
In that case, letting $C_{\ell m n} = |C_{\ell m n}| \exp{(i \phi_{\ell m n})}$, Eq.~\eqref{eq:h_pt_lmn_angles} can be written as
\begin{widetext}
\begin{align}
h_{j} &=  \left[ C_{\ell m n} e^{-i \omega_{\ell m n} t} {}_{\unaryminus 2} S_{\ell m n}(\iota,\varphi) + C_{\ell m n}^* e^{i \omega_{\ell m n} t} {}_{\unaryminus 2} S_{\ell m n}^*(\pi-\iota,\varphi) \right] e^{-t/\tau_{\ell m n}} \nonumber \\
 &=  \left[ C_{\ell m n} e^{-i (\omega_{\ell m n} t - m\varphi)} {}_{\unaryminus 2} S_{\ell m n}(\iota) +
C_{\ell m n}^* e^{i (\omega_{\ell m n} t - m\varphi)} {}_{\unaryminus 2} S_{\ell m n}^*(\pi - \iota) \right] e^{-t/\tau_{\ell m n}} \nonumber \\
 &=  \left| C_{\ell m n}\right| \left[ e^{-i \Phi_{\ell m n}(t)} {}_{\unaryminus 2} S_{\ell m n}(\iota) +
e^{i \Phi_{\ell m n}(t)} {}_{\unaryminus 2} S_{\ell m n}^*(\pi- \iota) \right] e^{-t/\tau_{\ell m n}} \, ,
\end{align}
\end{widetext}
with $\Phi_{\ell mn}(t) \equiv \omega_{\ell m n} t - \phi_{\ell m n} - m \varphi$. By comparison with Eq.~\eqref{eq:ellipticity}, we see that the ellipticity of this mode is fully determined by the inclination $\iota$,
\begin{equation}
\epsilon_{\ell|m|n} = \frac{\left|{}_{\unaryminus 2}S_{\ell m n}(\iota)\right| - \left|{}_{\unaryminus 2}S^*_{\ell m n}(\pi- \iota)\right|}{\left|{}_{\unaryminus 2}S_{\ell m n}(\iota)\right| + \left|{}_{\unaryminus 2}S^*_{\ell m n}(\pi- \iota)\right|} ;
\end{equation}
that is, $h_j$ would be circularly polarized were it not for the angular factors.
Approximating ${}_{\unaryminus 2}S_{\ell m n}(\iota) \approx {}_{\unaryminus 2}Y_{\ell m}(\iota)$ and noting that ${}_{\unaryminus 2}Y_{\ell m}(\iota)$ is real for all $\iota$, the complex strain simplifies to
\begin{align}
h_{j} = |C_j|\hspace{-1pt} \left[ Y^{(+)}_{j} \cos \Phi_j(t)
- i Y^{(\times)}_{j} \sin \Phi_j(t)\right] \hspace{-1pt} e^{-t/\tau_j} ,
\end{align}
using the shorthand $C_j \equiv C_{\ell m n}$, and defining the angular factors
\begin{equation}
Y^{(+/\times)}_{j} \equiv {}_{\unaryminus 2} Y_{\ell m}(\iota) \pm {}_{\unaryminus 2} Y_{\ell m}(\pi-\iota)\, ,
\end{equation}
which are algebraic functions of $\iota$.
For instance, the polarizations for the dominant $\ell = |m| = 2$ mode are
\begin{equation}
h_{j}^{(+)} =  \left(1 + \cos^2 \iota\right) \mathcal{A}_j\, e^{-t/\tau_{j}} \cos \Phi_j(t),
\end{equation}
\begin{equation}
h_{j}^{(\times)} =  \left(2\cos\iota\right) \mathcal{A}_j\, e^{-t/\tau_j} \sin \Phi_j(t),
\end{equation}
with $\mathcal{A}_j \equiv \sqrt{5/\pi} A_j$.
Up to an overall multiplicative constant, this was the model adopted in Ref.~\cite{Isi:2019aib}.

\section{Kerr spectrum coefficients}
\label{app:coeffs}

\begin{table}
\caption{Frequency coefficients for $\ell=|m|=2$ tones.}
\label{tab:fcoeff}
\resizebox{\columnwidth}{!}{\input{fcoeffs}}
\end{table}

\begin{table}
\caption{Damping rate coefficients for $\ell=|m|=2$ tones.}
\label{tab:gcoeff}
\resizebox{\columnwidth}{!}{\input{gcoeffs}}
\end{table}

\begin{figure}
  \includegraphics[width=\columnwidth]{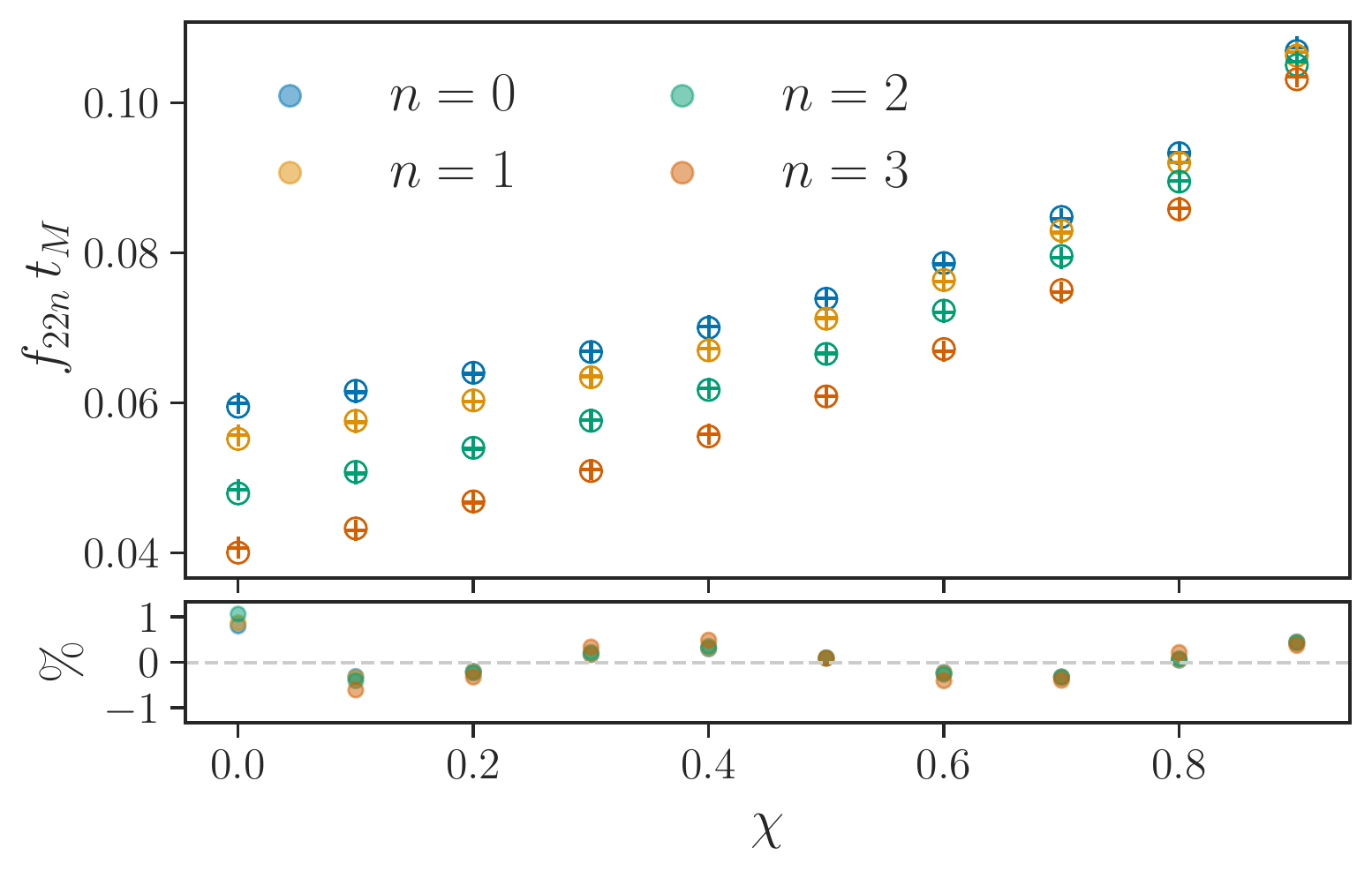}
  \includegraphics[width=\columnwidth]{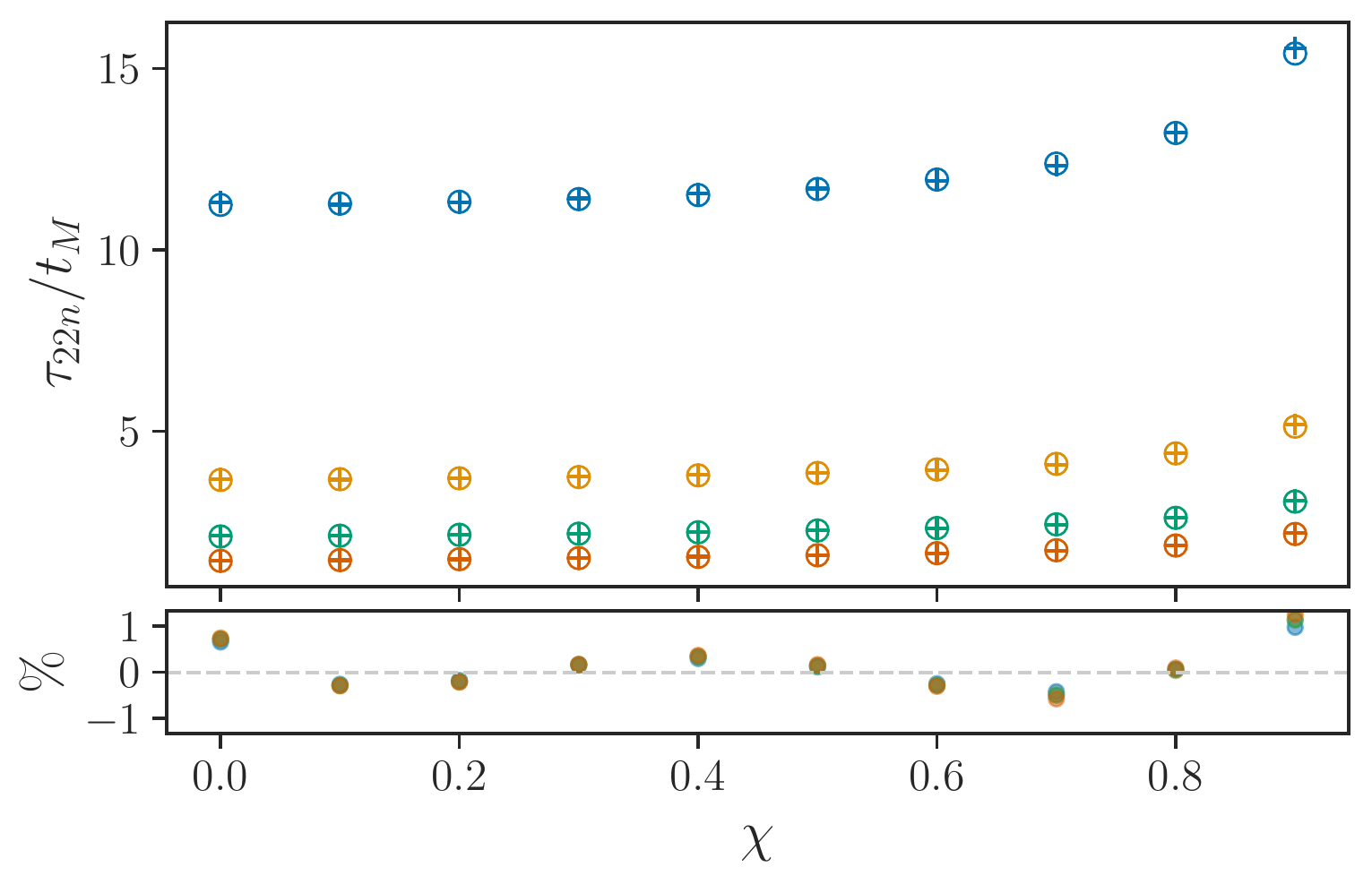}
  \caption{Analytic approximation to the dimensionless frequency (top) and damping time (bottom) as a function of spin $\chi$ for the first four tones of the $\ell=|m|=2$ harmonic (color).
  In the main panels, we show the approximation of Eq.~\eqref{app:eq:coeff} with the coefficients in Tables \ref{tab:fcoeff} and \ref{tab:gcoeff} (crosses) compared to the value obtained directly from the \textsc{qnm} package \citep{Stein:2019mop} (circles).
  We show the fractional difference between the approximation and the \textsc{qnm} value as a percentage in the secondary panels below each plot, with the \textsc{qnm} value as the baseline.
  The error of our approximation, Eq.~\eqref{app:eq:coeff}, remains roughly within $\pm 1\%$ irrespective of the overtone number $n$.}
  \label{fig:coeffs}
\end{figure}

As mentioned in Sec.~\ref{sec:model:kerr}, we have found that the spin-dependence of the $f$ and $\tau$ parameters for low
order Kerr modes in GR can be well approximated by a linear combination of $\log
(1 - \chi)$, and powers from $\chi^0$ to $\chi^4$: for each mode the dimensionless  $f\, t_M$ frequency satisfies
\begin{equation} \label{app:eq:coeff}
   f\, t_M \simeq c_l \log \left( 1 - \chi \right) + \sum_{i = 0}^4 c_i \chi^i,
\end{equation}
and similarly for the dimensionless damping rate $\gamma\, t_M$.
We find the approximating coefficients $c$ through standard least-squares fits to the complex mode frequencies computed using the \textsc{qnm} package \citep{Stein:2019mop}.
Our \textsc{ringdown} package does this automatically for any requested $\{\ell,|m|,n\}$.
As an example, we present the resulting values for the first eight tones of the $\ell=|m|=2$ modes in Table \ref{tab:fcoeff} for $f$ and Table \ref{tab:gcoeff} for $\gamma$.
We show the resulting fits to the frequency and damping rate for $0 \leq n \leq 3$ in Fig.~\ref{fig:coeffs}.

\section{Sampler-friendly parameterization}
\label{app:jacobian}

Unfortunately, although the elliptical QNM parameterization of Fig.~\ref{fig:ellipse} is conceptually clean, degenaricies in the $\{A_j, \epsilon_j, \theta_j, \phi_j\}$ make it hard to sample (Sec.~\ref{{sec:template}}).
We have found we can address this through by means of a reparameterization and a Jacobian.
For a more detailed discussion see \cite{Isi:inprep}.

Our expression for the linear polarizations, Eqs.~(\ref{eq:h_ellip_lmn_cross}--\ref{eq:h_ellip_lmn_cos}) can be rewritten as
\begin{equation}
h^{(+)}_j = A^{(+)}_j e^{-t/\tau_j} \cos(2\pi f_j t - \phi^{(+)}_j)\, ,
\end{equation}
\begin{equation}
h^{(\times)}_j = A^{(\times)}_j e^{-t/\tau_j} \cos(2\pi f_j t - \phi^{(\times)}_j)\, .
\end{equation}
The new amplitudes and phases $A^{(+/\times)}_j$ and $\phi^{(+/\times)}_j$ can be seen as the polar coordinates corresponding to Cartesian components
\begin{equation}
x^{(+/\times)}_j \equiv A^{(+/\times)}_j \cos \phi^{(+/\times)}_j\, ,
\end{equation}
\begin{equation}
y^{(+/\times)}_j \equiv A^{(+/\times)}_j \sin \phi^{(+/\times)}_j\, .
\end{equation}
In terms of these auxiliary quantities, the original amplitude parameters $\{A_j, \epsilon_j\}$ are
\begin{align}
A_j = \frac{1}{2}\left[\sqrt{\left(x^{(+)}_j + y^{(\times)}_j \right)^2 + \left(x^{(\times)}_j - y^{(+)}_j\right)^2} + \left(+ \leftrightarrow \times\right) \right]
\end{align}
\begin{align}
\epsilon_j =
\frac{\sqrt{\left(x^{(+)}_j + y^{(\times)}_j \right)^2 + \left(x^{(\times)}_j - y^{(+)}_j\right)^2} -  \left(+ \leftrightarrow \times \right)}{
\sqrt{\left(x^{(+)}_j + y^{(\times)}_j\right)^2 + \left(x^{(\times)}_j - y^{(+)}_j\right)^2}  + \left(+ \leftrightarrow \times\right)} ,
\end{align}
where $ \left(+ \leftrightarrow \times\right)$ stands for a term equal to the immediately preceding one except for a substitution of $+$ for $\times$, and viceversa.
The angles $\{\theta_j, \phi_j\}$ can also be given in terms of these quantities as
\begin{align}
\theta_j = -\frac{1}{2}&\left[\atantwo\left(-x^{(\times)}_j + y^{(+)}_j, y^{(\times)}_j + x^{(+)}_j\right) +\right. \nonumber \\ &\left.~\atantwo\left(-x^{(\times)}_j - y^{(+)}_j, -y^{(\times)}_j + x^{(+)}_j\right)\right]
\end{align}
and, lastly,
\begin{align}
\phi_j = \frac{1}{2}&\left[\atantwo\left(-x^{(\times)}_j + y^{(+)}_j, y^{(\times)}_j + x^{(+)}_j\right) -\right. \nonumber \\ &\left.~\atantwo\left(-x^{(\times)}_j - y^{(+)}_j, -y^{(\times)}_j + x^{(+)}_j\right)\right] .
\end{align}

With these expressions in hand, we can now sample in the $4j$ Cartesian quantities $x^{(+/\times)}_j$ and $y^{(+/\times)}_j$ by drawing them from an auxiliary normal distribution,
\begin{equation}
x^{(+/\times)}_j, y^{(+/\times)}_j \sim \mathcal{N}(0, \sigma_A)\, ,
\end{equation}
where $\sigma_A$ is some appropriate strain amplitude scale, e.g., $\sigma_A = 10^{-21}$.
This parameterization is free from any degeneracies and is easy to sample.

To enforce a prior uniform in $\{A_j, \epsilon_j\}$ we need only apply a Jacobian.
It may be shown that the desired correction to be added to the log probability $\ln P(d \mid s)$, Eq.~\eqref{eq:lnlike_td}, is
\begin{equation}
\Delta(\ln P)_j = \frac{1}{2} \sum_p \left(x^p_j + y^p_j\right) - 3 \ln A_j - \ln\left(1 - \epsilon_j^2\right)
\end{equation}
for each mode $j$ and for $p \in \{+,\times\}$.
The coordinate transformation is invertible everywhere except at the singular origin $x^p_j + y^p_j = 0$ for any given $p$ or $j$, which is a set of measure zero that has no practical relevance.

\section{Original LIGO-Virgo GW150914 ringdown analysis method}
\label{app:lvc}

The original GW150914 ringdown analysis in \cite{TheLIGOScientific:2016src} made use of the method detailed in \cite{Prix:T1500618}, which we replicated in Fig.~\ref{fig:lvc_performance}.
In that approach, we write the single-detector likelihood for data $d$ [cf.~Eqs.~\eqref{eq:lnlike_fd} and \eqref{eq:lnlike_td}] as
\begin{align}
\ln P(d \mid s) &= -\frac{1}{2}\left\langle d - s \mid d - s \right\rangle + \mathrm{const.}\nonumber \\
&= \left\langle d \mid s \right\rangle - \frac{1}{2}\left\langle s \mid s \right\rangle + \mathrm{const.},
\end{align}
where $\left\langle \cdot \mid \cdot \right\rangle$ is a noise-weighted inner product and the constant absorbs terms independent of the QNM template $s$, which is taken to be zero before the truncation time $t_0$ (as in the top panel of Fig.~\ref{fig:window}).
In standard LIGO-Virgo analyses, the inner product is computed in the frequency domain through an integral weighted by the PSD,
\begin{equation} \label{eq:ip_fd}
\left\langle x \mid y \right\rangle = 4 \Re \sum_{k=0}^{M-1} \frac{\tilde{x}_k^*\, \tilde{y}_k}{S(f_k)} \Delta f \, ,
\end{equation}
for a segment with $M$ data points, and where tilde indicates a Fourier domain quantity.
This is the expression used in \cite{TheLIGOScientific:2016src,Prix:T1500618} to evaluate the $\left\langle s \mid s \right \rangle$ term, with $\tilde{s}$ computed as a superposition of Lorentzians directly in the frequency domain.

On the other hand, in an attempt to avoid the windowing and cross-$t_0$ contamination issues described in Sec.~\ref{sec:inference}, the $\left\langle d \mid s \right \rangle$ term is evaluated in the time domain instead.
This is done in two steps, first starting from the continuoum-limit version of Eq.~\eqref{eq:ip_fd} and defining the overwhitened data series $\tilde{o}_k \equiv \tilde{d}_k / S(f_k)$,
\begin{align}
4\Re \int_{0}^{\infty} \tilde{o}^*(f)\, \tilde{s}(f)\, \infd f &= 2 \Re \int_{-\infty}^{\infty} \tilde{o}^*(f)\, \tilde{s}(f)\, \infd f \, , \nonumber \\
&= 2 \int_{-\infty}^{\infty} o(t)\, s(t)\, \infd t \, , \nonumber \\
&= 2 \int_{t_0}^{t_0+T} o(t)\, s(t)\, \infd t \,  ,
\end{align}
where we have used Parseval's theorem to bring the expression back to the time domain, assuming $s(t)$ only has support starting at $t_0$ and over some suitably long duration $T$, by which point the QNMs have decayed; $o(t)$ is the inverse Fourier transform of $\tilde{o}(f)$.
Returning to discrete quantities, the first term in the likelihood above is thus computed as
\begin{equation}
\left\langle d \mid s \right\rangle = 2 \sum_{k=0}^{M-1} o_k s_k\,  \Delta t\, .
\end{equation}
With this expression we avoid directly Fourier transforming $s(t)$, in favor of inverse Fourier transforming $\tilde{o}(t)$.

Unfortunately, the above derivation should make it clear that this approach is mathematically equivalent to a direct computation of the likelihood in the Fourier domain via Eq.~\eqref{eq:ip_fd}; therefore, it is therefore vulnerable to all the drawbacks associated with Eq.~\eqref{eq:lnlike_td}, as discussed in Sec.~\ref{sec:inference} and demonstrated concretely in Fig.~\ref{fig:lvc_performance}.

For the demonstration in Fig.~\ref{fig:lvc_performance}, we compute the likelihood as described above and condition the data following the process outlined in \cite{Prix:T1500618} as close as possible.
Unlike \cite{Prix:T1500618}, however, we place uniform priors on $f$, $\tau$, $A$ and $\phi$ for simplicity, rather than marginalizing over an unknown scale for $A$ by means of a hyperprior ($H$ in \cite{Prix:T1500618}).
As for our time-domain method in the main text, we sample the posterior using \textsc{Stan} \cite{Stan,Carpenter:2017}.

\end{document}

%% file: fcoeffs.tex
\begin{tabular}{lrrrrrr}
\toprule
$n$ &   $c_{l}$ &   $c_{0}$ &   $c_{1}$ &   $c_{2}$ &   $c_{3}$ &   $c_{4}$ \\
\midrule
0 & -0.008236 &  0.059950 & -0.001066 &  0.083542 & -0.151656 &  0.110213 \\[2.5pt]
1 & -0.008174 &  0.055662 &  0.001746 &  0.085319 & -0.154652 &  0.113264 \\[2.5pt]
2 & -0.008035 &  0.048425 &  0.005457 &  0.093005 & -0.169598 &  0.124871 \\[2.5pt]
3 & -0.007791 &  0.040673 &  0.004915 &  0.120850 & -0.222699 &  0.159911 \\[2.5pt]
4 & -0.007701 &  0.034185 & -0.008239 &  0.206435 & -0.376850 &  0.249180 \\[2.5pt]
5 &  0.003030 &  0.025584 &  0.067562 & -0.156557 &  0.367318 & -0.208803 \\[2.5pt]
6 & -0.009482 &  0.022091 & -0.006714 &  0.223895 & -0.363355 &  0.219673 \\[2.5pt]
7 & -0.009315 &  0.014293 &  0.033567 &  0.111958 & -0.205332 &  0.141090 \\
\bottomrule
\end{tabular}

%% file: gcoeffs.tex
\begin{tabular}{lrrrrrr}
\toprule
$n$ &   $c_{l}$ &   $c_{0}$ &   $c_{1}$ &   $c_{2}$ &   $c_{3}$ &   $c_{4}$ \\
\midrule
0 &  0.011807 &  0.088381 &  0.025283 & -0.090023 &  0.182455 & -0.121626 \\[2.5pt]
1 &  0.033605 &  0.271886 &  0.074607 & -0.313743 &  0.624993 & -0.411691 \\[2.5pt]
2 &  0.057548 &  0.474874 &  0.102760 & -0.524840 &  1.036581 & -0.672992 \\[2.5pt]
3 &  0.083005 &  0.700033 &  0.115212 & -0.770834 &  1.483327 & -0.933504 \\[2.5pt]
4 &  0.114385 &  0.940370 &  0.103270 & -0.899129 &  1.622337 & -0.963919 \\[2.5pt]
5 & -0.018886 &  1.204070 & -0.496516 &  1.047939 & -2.023199 &  0.881021 \\[2.5pt]
6 &  0.105308 &  1.438684 & -0.056218 & -1.383174 &  3.057700 & -2.259403 \\[2.5pt]
7 &  0.142801 &  1.690191 & -0.252107 & -0.670293 &  2.095130 & -1.825597 \\
\bottomrule
\end{tabular}